\documentclass [12pt]{article}
\usepackage{lscape}                                           %
\usepackage[centertags]{amsmath}
\usepackage{amsfonts} \usepackage{amssymb}\usepackage{eufrak}
\usepackage{graphicx}
\usepackage{a4}
\usepackage{cite}
\usepackage{lscape}

\newcommand{\COMMENTO}[1]{}

\newcommand{\hll}{{\hat l}}
\newcommand{\hmm}{{\hat m}}

\newcommand{\hmu}{{\hat \mu}}
\newcommand{\hnu}{{\hat \nu}}
\newcommand{\hlambda}{{\hat \lambda}}

\def\one{{\hbox{ 1\kern-.8mm l}}}

\newcommand{\oh}{\frac{1}{2}}

\newcommand{\dcalA}{{\dot{\cal A}}}

\newcommand{\hAA}{{\hat A}}

\newcommand{\dal}{{\dot{\alpha}}}

\begin{document}
\begin{titlepage}
\rightline{DFTT/9/2010}
 \rightline{NORDITA-2010-115} \vskip 3.0cm
\centerline{\LARGE \bf Open strings in the system D5/D9 } \vskip
1.0cm \centerline{\bf P. Di Vecchia$^{a,b}$, R. Marotta$^c$, I.
Pesando $^d $ and F. Pezzella$^c$\footnote{Partially supported by
INFN-MIT  ``Bruno Rossi" Exchange Program.}} \vskip .6cm
\centerline{\sl $^a$ The Niels Bohr Institute, Blegdamsvej 17,
DK-2100 Copenhagen \O, Denmark} \vskip .4cm \centerline{\sl $^b$
Nordita, Roslagstullsbacken 23, SE-10691 Stockholm, Sweden} \vskip
.4cm \centerline{\sl $^c$ Istituto Nazionale di Fisica Nucleare,
Sezione di Napoli} \centerline{ \sl Complesso Universitario di Monte
S. Angelo ed. 6, via Cintia,  80126 Napoli, Italy} \vskip 0.4cm
\centerline{ \sl $^d$ Dipartimento di Fisica Teorica,
  Universit\`a di Torino and INFN, Sezione di Torino,} \centerline{\sl
 via P. Giuria 1,  10125 Torino, Italy}
 \vskip 1cm

\begin{abstract}
We construct   the six-dimensional Lagrangian for the massless twisted open strings
with one  end-point ending on  a stack of D5 and the other on a  stack of D9 branes, interacting with the gauge multiplets living respectively on the D5 and D9 branes.
It is  first obtained by uplifting to six
dimensions  the four-dimensional Lagrangian of  the ${\cal{N}}=2$ hypermultiplet and manifestly exhibits an  $SU(2)$  symmetry.
We  show by an explicit calculation that it is ${\cal{N}}=1$ supersymmetric  in six dimensions
and
then we check  various terms of this Lagrangian by computing string amplitudes on the disk.
Finally, starting from this  Lagrangian and assuming the presence of
non-zero magnetic fluxes along the extra compact  dimensions, we determine the
spectrum of the Kaluza-Klein  states which agrees with the corresponding one obtained
from  string theory in the field theory limit.

\end{abstract}

\end{titlepage}

\newpage

\tableofcontents       %
\vskip 1cm             %

\section{Introduction}
\label{intro}

A very powerful aspect of string theory consists in its being able to view
different field theories as different limits of the same string model.  An example of
this is provided by open strings attached to  magnetized D9-branes
 on the background $R^{3,1} \times T^2 \times T^2 \times T^2$.
In this case the spectrum of open strings ending on branes with different magnetizations (called in the following {\em twisted} open strings)
 and their interaction depends on the
difference of magnetizations $\nu_i$  in the three tori $T^2$. In particular, in the limit
where the D branes have the same magnetization, i.e.  $\nu_i =0$  with  $i=1,2,3$,
the open strings stretched
between them are described by the ten dimensional ${\cal{N}}=1$ super Yang-Mills theory
on $R^{3,1} \times T^2 \times T^2 \times T^2$. On the other hand, by taking
$\nu_1 =0, \nu_{2,3} = \frac{1}{2}$,
one has effectively a system of D5/D9 branes with different kinds of open strings. There are the open strings attached to two D9 branes that are again described by  the  ten-dimensional ${\cal{N}}=1$ super Yang-Mills on $R^{3,1} \times T^2 \times T^2 \times T^2$.
Then, there are open strings attached to two  D5 branes that are described by a six-dimensional
gauge theory obtained by dimensionally reducing the ten-dimensional
${\cal{N}}=1$ super Yang-Mills compactified  on $R^{3,1} \times T^2$. Finally, there are
the open strings stretched between a D9 and a D5 that are described by a six-dimensional
theory whose Lagrangian is that of   the ${\cal{N}}=1$ hypermultiplet in  six dimensions.

Therefore,  in string theory,
one has a complete   description of the open strings ending on two D branes with
different magnetizations \cite{FT1985,T1988, ACNY, 9503030,0512067,9209032} and it is possible to interpolate between the two field theories discussed above by suitably changing the magnetization.
Magnetized branes allow
one to construct semirealistic extensions of the Standard Model~\footnote{See for instance Refs. \cite{0902.3251, 0610327, 0005067} and References therein.}
 and therefore it can be very helpful to derive the low-energy effective actions of the open strings attached to them. In this perspective, if one is not interested in string corrections, it is
 easier to derive such effective  actions in a purely field theoretical context rather than to extract them from the string amplitudes.
 In fact,
as shown in Refs.~\cite{0404229,0807.0789,0810.5509} (see also \cite{0812.2247,0812.3534,0904.0910,0906.3033, 1003.5805}), for the open strings attached to two unmagnetized D9 branes,
one can start from ${\cal{N}}=1$ super Yang-Mills in ten dimensions and, introducing
background magnetic fields in the extra dimensions, one  obtains
a description  of the  twisted open strings
in the limit $\alpha' \rightarrow 0$. It is then natural to expect that, if one starts instead
from the low-energy effective action describing the open strings attached to a D9 and a D5, and
introduces background magnetic fields in the extra dimensions,
then a description of the twisted D5/D9 strings is obtained  in the limit  $\alpha' \rightarrow 0$. This description is
T-dual to the one given in Ref.~\cite{0408036,1002.0006}. Naturally, it is expected that the field theory description of magnetized branes has to coincide with the stringy one in the limit $\alpha' \rightarrow 0$.

In this paper we show that this is indeed the case also  for the D5/D9 twisted open  strings.
In order to do that, we need to construct the complete six-dimensional action
describing  the  massless D5/D9 twisted open strings attached to unmagnetized D branes interacting with the untwisted
ones, corresponding to the gauge multiplets living respectively in the world-volume
of the D5 and D9 branes. The new feature of this action, with respect to
${\cal{N}}=1$ super Yang-Mills, is that, while the D5/D9 and D5/D5 open strings live
in six dimensions, the D9/D9 live instead in the entire ten dimensional space-time.

In this paper we construct this six-dimensional action  by two different methods.
The first one consists in starting from the four-dimensional Lagrangian of  the ${\cal{N}}=2$ hypermultiplet interacting with both  gauge multiplets living respectively on the D5 and D9 branes   and then by uplifting it to six dimensions. In this way we get a six-dimensional theory
where not only the fields corresponding to the twisted open strings, but also both   gauge multiplets living on the D5 and D9 branes are six  dimensional. This approach is straightforward because the four-dimensional Lagrangian for the ${\cal{N}}=2$ hypermultiplet can be easily constructed  using a superfield formalism and also its uplift to six dimensions is straightforward. A nice and new  feature of this six dimensional action is the fact that
the $SU(2)$ R-symmetry of the original four-dimensional theory is still manifest
although only  an ${\cal{N}}=1$ supersymmetry survives in six dimensions. During the uplift we may, in general,  lose  the original supersymmetry and therefore we also
check, through a direct calculation,  that the six-dimensional action  is supersymmetric.
In this approach, however, the gauge multiplet living on the D9 brane is still treated as  six- and not as  ten-dimensional.
Furthermore, the six-dimensional Lagrangian so constructed still contains the auxiliary fields of the gauge multiplet living on the D9 branes and, in order to eliminate them, we should consider it together with the Lagrangian of
the D9/D9 open strings. When we eliminate them, by using their algebraic equation of motion,
we get in  the  six-dimensional Lagrangian for the open strings D5/D9  quartic terms for the
twisted scalars. We stress, however, that all fields, including those that live in the
world-volume of the D9 branes, are treated as six-dimensional fields in this field theoretical approach.

In order to  have an independent check of this six-dimensional Lagrangian, we construct the vertex operators
corresponding to the massless open strings D9/D9, D5/D5 and D5/D9 and we use them to
derive some of the  terms of the D5/D9 and D5/D5  Lagrangians by  computing string amplitudes.  In string theory, however,
the fields of the gauge multiplet living on the D9 are treated  as ten dimensional  fields.
This means that in string theory we are not only able to check the previously constructed six-dimensional Lagrangian, but we can also  determine  the couplings where the fields corresponding
to the twisted open strings  D5/D9  are six-dimensional,
while those corresponding to the open strings D9/D9 are ten-dimensional.
In conclusion, by using these two combined methods, we can construct the complete
low energy six-dimensional Lagrangian for the system D5/D9 and we get various hints on how to extend it to a Lagrangian where the fields of the gauge multiplet living on the D9 branes
are treated as ten-dimensional.

We face, however,  a problem. The uplift of the  six-dimensional
supersymmetric theory to  ten dimensions in general does not insure that the
uplifted theory still remains supersymmetric. This problem may be connected
with the fact that, in principle  in string theory, we should treat  all fields, including those
living on the D5 brane, as ten dimensional fields and that then
the six-dimensional action is obtained by integrating over their wave function in the
four extra dimensions. We do not analyze this problem in this paper and we hope to
be able to discuss it in a future publication.

Finally, starting from the complete D5/D9 Lagrangian where the fields living on
the world-volume of
the D5 and D9 branes are respectively six- and ten-dimensional, and assuming the presence of
non-zero magnetic fluxes along the six extra dimensions, we determine the spectrum of Kaluza-Klein states  and we show that it is identical to the one that one gets directly from string theory in the corresponding field theory limit.  This procedure generalizes to the system D5/D9 what has already been   done for the system D9/D9.

In conclusion, the most important and new results of this paper are the construction of the
complete six-dimensional Lagrangian describing the interaction of the massless fields, corresponding to the twisted D5/D9 open strings, with the gauge multiplets living respectively on the
D5 and D9 branes and the proof that, by introducing in this
Lagrangian background magnetic fields, one gets the spectrum of Kaluza-Klein states that one obtains from that of the
open strings attached to magnetized D branes in the field theory limit ($\alpha ' \rightarrow 0$).
It is also new, as far as we know,  that the Lagrangian has a manifest $SU(2)$ symmetry that is  the remnant
of the $SU(2)$ R-symmetry of the four-dimensional Lagrangian of the hypermultiplet.

The paper is organized as follows. In Sect. \ref{2} we give the spectrum of the twisted
open strings  attached to two magnetized D9 branes and we perform two different
field theory limits.

In Sect. \ref{595559}
we construct the six-dimensional Lagrangian describing the
interaction of the massless twisted D5/D9 open strings interacting with the gauge multiplets
living respectively on the D5 and D9 branes and  in Sect. \ref{susys} we show
that this six-dimensional Lagrangian is supersymmetric.

Sect. \ref{vertexope} is devoted to the construction of the vertex operators associated to all
massless open strings of the system D5/D9 and  Sect. \ref{striampli} to compute some
string amplitudes checking some terms of the previously constructed Lagrangians.

Finally, in Sect. \ref{KKredu} we introduce  background magnetizations and we recover
the spectrum of the Kaluza-Klein excitations that agrees with the one obtained directly from string theory in the field theory limit.

This paper contains also seven Appendices where many technical details are presented.
In App. \ref{nota} we discuss our notations for the spinors in four, six and ten dimensions
and for the $\Gamma$-matrices used, while  in App. \ref{magne} we give some more detail
about magnetized D branes.  In Appendices  \ref{uplift55}) and   \ref{uplift59}
 we discuss the uplift from four to six dimensions respectively for the Lagrangian of the massless
open strings 55 and that of the massless open strings 59 and in App. \ref{Uplift99} we consider
the uplift from six to ten dimensions of the Lagrangian describing the massless open strings 99.
In App. \ref{susyap} we show in detail that the six-dimensional Lagrangian for the strings 59 is supersymmetric. Finally, in App. \ref{correla} we compute some correlators in string theory
to have an independent check of the various Lagrangians.

\section{Open strings attached to magnetized D branes and the field theory limit}
\label{2}

In Sect. \ref{2.1}  we give the spectrum of open strings attached to two magnetized D branes
having  different magnetizations and we perform two kinds of field theory limits. The first one,
presented in Sect. \ref{2.2},  provides the spectrum of open strings attached to two magnetized D9 branes, while the other, presented in Sect. \ref{2.3},  gives the spectrum of open strings attached to a D9 and a D5 brane,  both in the limit $\alpha' \rightarrow 0$.

\subsection{Open string spectrum for magnetized D branes}
\label{2.1}

The spectrum of open strings attached  to two magnetized D9 branes living on
$R^{3,1} \times T^2 \times T^2 \times T^2$ is given by:
\begin{eqnarray}
M^2 = \frac{1}{\alpha'} \left[  N^{X}  + N^{\psi} + \sum_{r=1}^{3} \left( N^{\cal{Z}}_{r}  +
N^{\Psi}_{r}
\right)  - \frac{x}{2}
+ \frac{x}{2} \sum_{r=1}^{3} \nu_r      \right]
\label{M2}
\end{eqnarray}
where $x=1\, (0)$ for the NS (R) sector, the number operators
\begin{eqnarray}
N^X = \sum_{n=1}^\infty  n \, a^{\dagger}_{n}  \cdot  a_n~~;~~N^{\psi} =
\sum_{n= 1 - \frac{x}{2}}^{\infty}  n  \, \psi_{n}^{\dagger}  \cdot \psi_n
\label{NXNpsi}
\end{eqnarray}
contain the harmonic oscillators along the non-compact directions,
while the number operators
\begin{eqnarray}
N^{\cal{Z}}_{r} =
\sum_{n=0}^{\infty} \left[ (n + \nu_r ) a^{(r)\dagger}_{n+\nu_r}  a_{n+\nu_r}^{(r)} +
 (n+1 - \nu_r ) \bar{a}^{(r) \dagger}_{n+1 -\nu_r} \bar{a}_{n+1-\nu_r}^{(r)}\right]
\label{NcalZ}
\end{eqnarray}
\begin{eqnarray}
N^{\Psi}_{r} =     \sum_{n=\frac{x}{2} }^\infty (n+\nu_r)
\Psi^{\,\dagger \,(r)}_{n+  \nu_{r}} \Psi^{\, (r)}_{n+  \nu_{r}}
 +     \sum_{n=1- \frac{x}{2}}^\infty (n-\nu_r)
{\overline \Psi}^{\,\dagger \,(r)}_{n-\nu_r}{\overline \Psi}^{\, (r)}_{n-\nu_r}
\label{NPsi}
\end{eqnarray}
correspond to the contribution of the oscillators along the three tori $T^2$.

Let us denote by $a$ and $b$ the two D9 branes with different magnetization. Their
magnetization  is encoded  in the quantities   $\nu_{r}^{(a)}$ and $\nu_{r}^{(b)}$ $ (r=1,2,3)$
along the three tori that are given by:
\begin{eqnarray}
\tan\pi\nu^{(a,b)}_{r} =
\frac{ I^{(r)}_{(a,b)} }{ n^{(r)}_{(a,b)} { T}_2^{(r)}}~~;~~
{ T}_{2}^{(r)} \equiv \frac{V_{T^2_{r} }}{(2 \pi \sqrt{\alpha'})^2}
\label{tana}
\end{eqnarray}
where   $ V_{T^2_{r}} $ is the physical volume of the $r$-th torus,
while $I_{(a,b)}^{(r)}$ and   $n_{(a,b)}^{(r)}$ are respectively the
integer  magnetic flux and  the number of times that the D brane is
``wrapped" on the $r$-th torus. They are related to the Chern class
by:
\begin{eqnarray}
\int \frac{{\rm Tr_{n^{(r)}_{(a,b)}}} \left( F^{(r)}_{(a,b)} \right)}{2 \pi} = I^{(r)}_{(a,b)}
\label{Chern}
\end{eqnarray}
The quantity that gives  the shift in the frequency of the oscillators is given by
\begin{eqnarray}
\nu_r \equiv \nu^{(a)}_{r} - \nu^{(b)}_{r}
\label{tan1}
\end{eqnarray}
 corresponding to the difference of the magnetizations of the two D branes  along the three tori.

The previous expressions  for the number operators in the six compact directions
are valid  for   $0 \leq \nu_r \leq \frac{1}{2}$.  They can be extended
to the interval $ - \frac{1}{2}  \leq \nu_r \leq \frac{1}{2}$  that is the natural range for the NS
sector.  In this range the first equation in  (\ref{tana}) shows that $\tan \pi \nu_r$ varies monotonically from $- \infty$ to $+ \infty$. When $\nu_r$ is negative  the
number operators along the  compactified directions change their form in terms of the harmonic oscillators.  In particular, the number operators for the fermionic coordinate of the NS sector are left unchanged, while in those for the fermionic coordinate in the R sector and for the bosonic coordinate we have to exchange $\Psi$ with ${\bar{\Psi}}$ and $a$ with ${\bar{a}}$.
This is explained in detail in Appendix \ref{magne}.  Here we limit ourselves
to give the mass spectrum of open strings to include both the case of positive and negative
$\nu_r$. One gets:
\begin{eqnarray}
M^2 = \frac{1}{\alpha'} \left[  N^{X}  + N^{\psi} + \sum_{r=1}^{3} \left(
N^{\cal{Z}}_{r}  +
N^{\Psi}_{r}
\right)  - \frac{x}{2}
+ \frac{x}{2} \sum_{r=1}^{3} |\nu_r |      \right]
\label{M2b}
\end{eqnarray}
where the  number operators ${\cal{N}}^{\cal{Z}}_{r}$ and ${\cal{N}}^{\Psi}_{r}$
are defined above if $\nu_r$ is positive, while  they are given in Appendix \ref{magne} when $ \nu_r $ is negative.

Having introduced different  magnetizations on the two D branes, the previous formula
describes both the spectrum of open strings ending on two magnetized D9 branes and
the spectrum of the open strings with one end-point on a magnetized D9 and the
other end-point attached to a magnetized D5 brane. All this is true in string theory.
The situation changes when we take the field theory limit that consists in sending $\alpha' \rightarrow 0$, while keeping the physical volume $V_{T^2}$ of the torus $T^2$ fixed.
In the following two subsections we take two different field theory limits corresponding to open string excitations attached to two D9 branes or to a D9 and a D5 brane.

\subsection{Field theory limit for D9/D9}
\label{2.2}

In the case of two D9 branes, as it can be seen from Eq. (\ref{tana}), the field theory limit as defined above corresponds to
values of $\nu_r$ proportional to $\alpha'$. The states left in this limit are those for which
the factor $\alpha'$ in the denominator of Eq. (\ref{M2}) is cancelled by a similar factor in the numerator. This happens for the following states in the NS sector:
\begin{eqnarray}
&&{\bar{\Psi}}_{\frac{1}{2} \mp | \nu_r |}^{(r) \dagger} |0 \rangle~~~;~~~
\Psi^{(r) \dagger}_{\frac{1}{2} \pm |\nu_r |} | 0 \rangle
~~~;~~~r=1,2,3 \nonumber\\
&& \Psi_{\frac{1}{2}}^{(i)} |0 \rangle~~;~~~i=2,3  \,\, .
\label{lows}
\end{eqnarray}
For the scalar states in the first line of the previous equation the upper (lower) sign is valid if
$\nu_r \geq0 (\nu_r \leq 0)$.
We work in the light-cone gauge and therefore the index $i$ runs over the transverse
directions $2$ and $3$.
The mass of the previous eight physical states is given respectively by:
\begin{eqnarray}
&&\alpha' M^{2}_{r} = \frac{1}{2}  \sum_{s=1}^{3}  |\nu_s| \pm |\nu_r|~~~;~~~r=1,2,3
\nonumber \\
&&\alpha' M^{2}_{i} =  \frac{1}{2}  \sum_{s=1}^{3} |\nu_s| ~~~;~~~i=2,3 \,\, .
\label{mass2}
\end{eqnarray}
 They become all massless in the limit of  $\nu_r \rightarrow 0$ reproducing the eight on-shell components of a massless gauge field. By using Eq. (\ref{tana}) it is easy to see that their mass remains finite in the field theory limit and is equal to~\cite{0901.4458}:
\begin{eqnarray}
&&M_{r}^{2} = 2 \pi \left[
 \sum_{s=1}^{3}  \frac{ | { \tilde{I} }_{ab}^{(s)} |  }{  V_{T_{s}^{2}} } \pm 2
 \frac{ | {\tilde{I}}_{ab}^{(r)} |    }{ V_{T_{r}^{2}}} \right]~~~;~~~r=1,2,3 \nonumber \\
&&M^{2}_{i}  =  \sum_{s=1}^{3} | { \tilde{I} }_{ab}^{(s)} | \frac{ 2 \pi }{V_{T_{s}^{2}} }
\label{ftlim73a}
\end{eqnarray}
where
\begin{eqnarray}
{\tilde{I}}_{ab}^{(r)} = \frac{I^{(r)}_{a} n^{(r)}_{b} -  I^{(r)}_{b} n^{(r)}_{a} }{n^{(r)}_{a}
n^{(r)}_{b}}    \,\, .
\label{mn}
\end{eqnarray}
We have a massless scalar if there is a value of $r$ for which one of the
previous masses is zero.

They are not, however, the only states that survive in  the field theory limit. We can also
 include  an additional bosonic oscillator that we call $a^{(r)}$ and that corresponds to $a^{(r)}_{\nu_r}$ or to ${\bar{a}}^{(r)}_{\nu_r}$ depending on the sign of $\nu_r$.
By including these extra excitations one finally gets the complete field theoretical result:
\begin{eqnarray}
&&M_{r}^{2} =  2 \pi \left[  \sum_{s=1}^{3}   \frac{ | {\tilde{I}}_{ab}^{(s)} |   }{V_{T_{s}^{2}}}
( 2 N^{(s)}   +1)
\pm  2  \frac{  | {\tilde{I}}_{ab}^{(r)} |    }{V_{T_{r}^{2}}} \right]~~~;~~~r=1,2,3 \nonumber \\
&&M^{2}_{i}  =  2 \pi   \sum_{s=1}^{3}   \frac{  | {\tilde{I}}_{ab}^{(s)} | }{V_{T_{s}^{2}}}
( 2 N^{(s)}  +1 )~~~;~~~i=2,3
\label{ftlim73}
\end{eqnarray}
where $N^{(s)} = a^{(s)\dagger}  a^{(s)}$ is the oscillator number.

The fermionic mass spectrum of the limiting field theory can be easily obtained
from the R sector proceeding as in the NS sector and it results to be:
\begin{eqnarray}
M^2 = 4 \pi  \sum_{s=1}^{3}  \frac{  | {\tilde{I}}_{ab}^{(s)} |    }{V_{T_{s}^{2}}}
(  N^{(s)}  + N^{(s)}_{f})
\label{M2R}
\end{eqnarray}
where $N^{(s)}_{f} = b^{ (s) \dagger} b^{(s)}$ is the oscillator number corresponding to the fermionic oscillators $\Psi_{\nu_s}$ or ${\bar{\Psi}}_{\nu_s}$ depending on the sign of $\nu_s$.
The  lowest state is the vacuum that, because of the GSO projection, is a four-dimensional chiral spinor.

The mass spectrum obtained in the field theory limit agrees completely, as already noticed in Ref. \cite{0901.4458},  with the spectrum of
states  found in Refs. \cite{0404229} and \cite{0810.5509} starting
from the ten-dimensional ${\cal{N}}=1$
super Yang-Mills after the introduction of  background magnetic fields in the six extra dimensions.

\subsection{Field theory limit for D5/D9}
\label{2.3}

In this subsection we perform a different field theory limit that  gives  the spectrum of open strings
attached to a D5 and a D9 in the limit $\alpha' \rightarrow 0$. If we assume that the world-volume of the D5 branes is along the four non-compact dimensions and along the
first torus $T_{2}^{(1)}$, then, requiring that $\nu_r$ lies in the interval
between $-\frac{1}{2}$ and $\frac{1}{2}$ implies the following
three values of $\nu_r$ in the mass formula:
\begin{eqnarray}
&&\nu_1 \equiv \nu_{1}^{(5)} - \nu_{1}^{(9)} \nonumber \\
&&\nu_{2,3} =\left\{\begin{array}{cc} \frac{1}{2} - \nu_{2,3}^{(9)}~~&if~~~  \nu_{2,3}^{(9)} >0
 \\
 - \frac{1}{2} - \nu_{2,3}^{(9)}~~&if~~~  \nu_{2,3}^{(9)} <0\end{array} \,\, \right..
\label{nur4}
\end{eqnarray}
By inserting the previous expressions in Eq. (\ref{M2b}) one gets:
\begin{eqnarray}
M^2 =\frac{1}{\alpha'} \left[  N^{X}  + N^{\psi} +
 \sum_{r=1}^{3} \left( N^{\cal{Z}}_{r}  +
N^{\Psi}_{r}
\right)  +  \frac{x}{2} \left( |\nu_{1}|  - |\nu_{2}^{(9)}| -  |\nu_{3}^{(9)}|
\right)      \right]
\label{M2mb}
\end{eqnarray}
where $N^X$ and $N^{\psi}$ are given in Eqs. (\ref{NXNpsi}), $N_{1}^{\cal{Z}}$ is given in Eqs. (\ref{NcalZbcx}) and (\ref{NcalZbd})
with analogous expressions for $N^{\Psi}_{1}$ in  the Ramond sector, while for
the NS sector $N^{\Psi}_{1}$ is given in Eq. (\ref{NPsi})
that is valid in the entire interval $- \frac{1}{2}\leq \nu_1 \leq \frac{1}{2}$.
The number operators for the second and third torus are listed in Appendix \ref{magne}.

We want now to perform the field theory limit taking care of Eq. (\ref{tana}). Let us start from the NS sector. In the field theory limit we are left with the following mass formula:
\begin{eqnarray}
M^2 = \frac{1}{\alpha'} \left[ |\nu_1| \left(  N^{(1)}   + \frac{1}{2} \right) +
| \nu_{2}^{(9)} | \left( N^{(2)}_{f}  - \frac{1}{2} \right)  +
| \nu_{3}^{(9)} | \left(  N^{(3)}_{f}  - \frac{1}{2} \right)  \right]
\label{M2we}
\end{eqnarray}
where
\begin{eqnarray}
&&
N^{(1)}  \equiv a^{(1) \dagger}_{\nu_1} a^{(1)}_{\nu_1} ~~~\mbox{if}~~~
\nu_1  >0
\nonumber \\
&&
N^{(1)}  \equiv {\bar{a}}^{(1) \dagger}_{|\nu_1|}
{\bar{a}}^{(1)}_{|\nu_1|}~~~\mbox{if}~~~\nu_1 <0
\label{aa}
\end{eqnarray}
\begin{eqnarray}
&&
N^{(2,3)}_f   =
{\bar{\Psi}}^{(2,3) \dagger}_{\nu_{2,3}^{(9)}} {\bar{\Psi}}^{(2,3) }_{\nu_{2,3}^{(9)}}
~~\mbox{if}~~\nu_{2,3}^{(9)} >0
\nonumber \\
&&
N^{(2,3)}_{f}  = {{\Psi}}^{(2,3) \dagger}_{|\nu_{2,3}^{(9)}|} {{\Psi}}^{(2,3) }_{|\nu_{2,3}^{(9)}|}~~\mbox{if}~~\nu_{2,3}^{(9)} <0 \,\, .
\label{bb}
\end{eqnarray}
In the following, for simplifying the notation, we name $N^{(1)} = a^{\dagger} a$ and $N^{(2,3)}_{f} =
b^{\dagger}_{2,3} b_{2,3}$.
Therefore, the following four states survive in the field theory limit. Two of them have
an odd number of fermions:
\begin{eqnarray}
(a^{\dagger})^m b_{2}^{\dagger} | 0 \rangle ~~~~~;~~~~~ (a^{\dagger})^m   b_{3}^{\dagger} | 0 \rangle
\nonumber
\end{eqnarray}
with a mass given  respectively by:
\begin{eqnarray}
M^2 =\frac{1}{\alpha'} \left[ |\nu_1| \left( m + \frac{1}{2} \right) \pm \frac{1}{2}
 \left( |\nu_{2}^{(9)}| -  |\nu_{3}^{(9)}| \right)\right] \label{twos} \,\, .
  \end{eqnarray}
 The other two states have an even number of fermions:
 \begin{eqnarray}
(a^{\dagger})^m| 0 \rangle ~~~~;~~~~
(a^{\dagger})^m b_{2}^{\dagger}  b_{3}^{\dagger} | 0 \rangle
\label{ios}
\end{eqnarray}
with a mass given respectively by:
\begin{eqnarray}
M^2 =
 \frac{1}{\alpha'} \left[ |\nu_1| \left( m + \frac{1}{2} \right) \mp \frac{1}{2}
 \left( |\nu_{2}^{(9)}| +  |\nu_{3}^{(9)}| \right)\right]  \,\, .
\label{twost}
\end{eqnarray}
By performing the field theory limit through Eq. (\ref{tana}) we can group together the two
expressions in Eqs. (\ref{twos}) and (\ref{twost}) respectively as follows:
\begin{eqnarray}
M^{2}_{GSO} = \frac{2\pi}{ \left(2\pi \sqrt{\alpha'} \, \right)^2}
\left[ \frac{ | {\tilde{I}}^{(1)}_{59}|}{ T_{2}^{(1)}  }\left( 2 N^{(1)}  + 1 \right) \pm
 \left(   \frac{ | I_{9}^{(2)} |}{ T_{2}^{(2)} n_{9}^{(2)} } -
  \frac{| I_{9}^{(3)} |}{ T_{2}^{(3)} n_{9}^{(3)} }
 \right)
   \right]
\label{M259}
\end{eqnarray}
and
\begin{eqnarray}
M^{2}_{OGSO} = \frac{2\pi}{ \left(2\pi \sqrt{\alpha'} \, \right)^2} \left[
\frac{ | {\tilde{I}}^{(1)}_{59}| }{ T_{2}^{(1)}  } \left( 2 N^{(1)} + 1 \right) \pm
 \left(   \frac{| I_{9}^{(2)}| }{ T_{2}^{(2)} n_{9}^{(2)} } +
  \frac{| I_{9}^{(3)}| }{ T_{2}^{(3)} n_{9}^{(3)} }
 \right)\right]  \,\, .
\label{M259b}
\end{eqnarray}
The mass spectrum in Eq. (\ref{M259}) corresponds to the one
obtained by means of the usual GSO projection, while the spectrum in
Eq. (\ref{M259b}) corresponds to the one with the opposite GSO
projection which keeps the states written in Eq.(\ref{ios}).

The previous  analysis  can be easily extended to the R sector. In this case the mass
spectrum, in the field theory limit, is given by:
\begin{eqnarray}
M^2 = \frac{1}{\alpha'} \left[ | \nu_1 |  a^{\dagger} a + |\nu_1| b_{1}^{\dagger} b_1    \right]  \,\, .
\label{M2R1}
\end{eqnarray}
The lowest state, the vacuum, is a four-dimensional massless spinor because the presence of
the magnetization in the first torus and the Neumann-Dirichlet boundary conditions on the other two tori allow fermionic zero modes only along the four non-compact directions.
The first excited level is obtained by applying on this vacuum
the fermionic creation operator $b_{1}^\dagger$. Two towers of
Kaluza-Klein states are generated by acting on these two fermionic
states with the bosonic creation operators:
\begin{eqnarray}
&&(a^{\dagger})^m |0 \rangle ~~~;~~~(a^{\dagger})^m b_{1}^{\dagger} |0 \rangle~~~;~~~m=0,1 \dots
\label{Rspec}
\end{eqnarray}
These two towers differ by the number of fermionic oscillators,
having respectively an even and an odd number of such operators and
their masses are:
\begin{eqnarray}
M^2_{1} = \frac{ 2\pi  }{ (2\pi \sqrt{\alpha'} )^2 }    \frac{ | I^{(1)}_{59} | }{ T_{2}^{(1)} } 2 N^{(1)} ~~~~~;~~~~~~
M_{2}^{2} =\frac{ 2\pi  }{ (2\pi \sqrt{\alpha'} )^2 }    \frac{ | I^{(1)}_{59} | }{ T_{2}^{(1)} }  \left( 2 N^{(1)} +2\right)
\label{spectM2}
\end{eqnarray}
where we have used Eq.(\ref{tana}).

In order to have a consistent string model  the
GSO projection has to  be imposed. The standard GSO projection selects, in the
NS sector, states with an odd number of fermionic
oscillators, while in the R sector fixes  the chirality  of the
vacuum. In particular, the vacuum becomes a chiral massless state.
The remaining states of the two towers have in pair the same mass but opposite chirality.
They can be combined together to form a single tower of Kaluza-Klein Dirac fermions.
Supersymmetry is achieved if the following condition is imposed:
\begin{eqnarray}
|\nu_1 | = |\nu_{2}^{(9)} | - |\nu_{3}^{(9)} | \,\, .
\label{cond12}
\end{eqnarray}
In this case the states in Eq. (\ref{twos}) have, respectively, the
following masses:
\begin{eqnarray}
 M^2 =
 \frac{1}{\alpha'} \left[ |\nu_1| \left( m + 1 \right)  \right]~~;~~ M^2 =
 \frac{1}{\alpha'}  |\nu_1|  m
\label{twosbc}
\end{eqnarray}
which coincide with the ones of the R sector in Eq. (\ref{Rspec}).

One gets the same conclusion by  imposing,  instead of the condition in Eq.
(\ref{cond12}), the following one:
\begin{eqnarray}
|\nu_1 | = |\nu_{3}^{(9)} | - |\nu_{2}^{(9)} |  \,\, .
\label{cond12b}
\end{eqnarray}
The masses of the states in Eq. (\ref{twosbc}) are just exchanged.

It is also interesting to notice that, considering both in the NS and
R-sector the opposite GSO condition, we have Bose-Fermi degeneracy
if the following condition is imposed:
\begin{eqnarray}
|\nu_1 | = |\nu_{3}^{(9)} | + |\nu_{2}^{(9)} | \,\, .
\end{eqnarray}
In the untwisted sector of the theory the opposite GSO projection is never
taken in consideration because, in absence of magnetic fluxes, it does not  project out
the tachyon and, therefore, leads to inconsistent string models. In the
unmagnetized twisted D5/D9 open string sector, instead, the tachyon
is not present in the spectrum and
 both GSO projections give a supersymmetric spectrum.
Therefore,
from this point of view both projections are allowed.
This is an interesting aspect of the D5/D9 twisted sector and should be further analyzed.

So far we have derived the spectrum of  the open strings D5/D9 that one obtains from string
theory in the  field theory limit. Can the same spectrum be obtained directly from
a field theoretical calculation? In the case of the strings D9/D9 it  has been
shown~\cite{0404229,0810.5509,0901.4458}
that it follows from the low-energy effective action for the massless open strings D9/D9
that is ${\cal{N}}=1$ super Yang-Mills. Therefore, in this case, we expect that the spectrum
obtained above from string theory can also be obtained from the low-energy
effective Lagrangian for the massless D5/D9 strings.
That is the reason why  we are going now to construct this Lagrangian.

 \section{Lagrangians for the open strings 99, 55 and 59}
 \label{595559}

 In this section we construct the  six-dimensional  Lagrangian describing the
 massless open strings attached to a stack of  D9 and a stack of D5 branes
 interacting with the gauge multiplets  living respectively on the D5 and the D9 branes.
 As  shown in Appendix \ref{uplift59}, we start from the four-dimensional Lagrangian
describing the ${\cal{N}}=2$  hypermultiplet    interacting  with the
  gauge multiplets and we uplift it  to six dimensions by
explicitly keeping  the  original $SU(2)$ R-symmetry of the four-dimensional Lagrangian.
This may seem strange at the first sight because the six-dimensional Lagrangian has only
an ${\cal{N}}=1$ supersymmetry, but string calculations, which will be discussed later,
confirm  that such a symmetry is indeed present in  six dimensions.
It turns out  that one can keep an $SU(2)$ symmetry
provided that some of the fields as the gaugino that are  $SU(2)$ doublets satisfy some
constraints.

 In this way we get a Lagrangian in which all fields, including those living on the D9 branes,
 are six dimensional.
  On the other hand, the Lagrangian so constructed still contains the auxiliary fields of the gauge multiplet living on the D9 branes. In order to eliminate them
 we  have to consider the Lagrangian of the open strings D5/D9 together with
 that of the open strings D9/D9.  The elimination of the auxiliary fields living on the D9 branes
 generates, in the Lagrangian for the strings 59, a four-scalar interaction involving the twisted scalar of the hypermultiplet and a trilinear term  containing two twisted scalar fields and the field strength  of the gauge field living on the D9 branes.  After eliminating the auxiliary fields, we can finally uplift to ten dimensions that part of the Lagrangian containing the gauge multiplet on the D9 branes.

 The six-dimensional Lagrangian of the ${\cal{N}}=2$
 gauge multiplet living on a  stack of D5 branes
 is  derived in Appendix \ref{uplift55} and is given by:
\begin{eqnarray}
{\cal{L}}_{55} &=& \,\,\, 2  {\rm Tr} \Bigg[
- \frac{1}{4} F_{ {\hat{\mu}}  {\hat{\nu}} }^{(5)}  F_{ {\hat{\mu}}   {\hat{\nu}}}^{(5)}
 + \frac{1}{2} \sum_{a=1}^{3} ({\cal{D}}^{(5)}_{c+5} )^{2} -
 \frac{i}{2}   {\bar{\Lambda}}_{i}^{(5)} \Gamma^{\hat{\mu}}D_{\hat{\mu}}^{(5)}
 (\Lambda^{i})^{(5)}
\Bigg] \nonumber \\
&&+ \,
 2  {\rm Tr} \Bigg[   - {( D_{\hat{\mu}}^{(5)} { ({Z}}^{(5)} )^i)}^{\dagger}
 ( D_{\hat{\mu}}^{(5)} (Z^{(5)})^{i} )
+ g_5 \bar{Z}_{i}^{(5)} \sum_{c=1}^{3} ({\cal D}_{c+5}^{(5)}\tau^{c})^i_{\,j}(Z^{(5)})^{j}
\nonumber \\
&& - \, g_{5} Z_{i}^{(5)}
\sum_{c=1}^{3}({\cal D}_{c+5}^{(5)}\tau^{c})^i_{\,j} ({\bar{Z}}^{(5)})^{j}
 -  i   {\bar{\Psi}}^{(5)}\Gamma^{\hat{\mu}} D_{\hat{\mu}}^{(5)} \Psi^{(5)}
   \nonumber \\
 && + \,
  i \sqrt{2} g_5 b \left( [{\bar{\Psi}}^{(5)}, Z^{j}_{(5)}] \epsilon_{ij} (\Lambda^{(5)})^{i} -
  {\bar{\Lambda}}_{i}^{(5)} [ \Psi^{(5)} , {\bar{Z}}_{j}^{(5)}] \epsilon^{ij} \right) \Bigg]
\label{L55}
\end{eqnarray}
where $b=\pm 1$, the indices $i,j$ label the $SU(2)$  symmetry and are lowered and raised
by means of the $\epsilon$ tensor as described in Appendix \ref{nota}.
The trace is over the fundamental matrices of the group $U(N_5)$.
In the previous Lagrangian the six-dimensional gaugino is described by an $SU(2)$ doublet  $\Lambda^i$ subject, in our notation,  to the symplectic Majorana condition\cite{9811101}
\begin{eqnarray}
B_6\Lambda^i=ac \epsilon_{ij}(\Lambda^j)^*
\label{Lambdai}
\end{eqnarray}
where  $a = \pm 1$ and $c$ is a phase factor. $B_6$ is the operator which relates the six-dimensional Dirac-matrices $\Gamma_{(6)}$ with their complex cojugates\cite{GSO}. More details on the properties of such operator are given in appendix \ref{nota} and its relation with  the  $\Gamma_{(6)}$'s is: $B_6= c\Gamma^2_{(6)} \Gamma^4_{(6)}$.
The constraint in  Eq. (\ref{Lambdai}) reduces by one half the number of independent fermions giving, as expected, the correct number of degrees of freedom for a gaugino.

Before proceeding further, let us discuss the three factors $a, c, b $ equal to $\pm1$
that appear in
the Lagrangian (\ref{L55}) and in the constraint for the gaugino in Eq. (\ref{Lambda}).
The six-dimensional     fermions $\Lambda^{(5)}$ and $\Psi^{(5)}$ come from a Weyl-Majorana
ten-dimensional spinor that satisfies Eq. (\ref{Weyl10}) and this implies that they
satisfy Eqs. (\ref{weylcondi}).  As a consequence, we get a factor of $b$ in the last two  terms
of Eq. (\ref{L55}). In the string calculations we have taken the ten-dimensional spinor to be anti-chiral ($b= 1$), but here in the field-theoretical formulation we leave $b$ to be arbitrary.
The two other factors $c$ and $a$ appear respectively in the definition
of $B_6$ in Eq. (\ref{B6})  and in Eq. (\ref{ans1}).
For the sake of simplicity, we will take them to be equal to $a= c = 1$.

The six-dimensional Lagrangian in Eq. (\ref{L55}) is obtained by dimensional
reduction from ${\cal{N}}=1$ super Yang-Mills in ten dimension and should be invariant under
the R-symmetry group $SO(4) \equiv SU(2) \times SU(2)$. Since, however, only one of the two $SU(2)$ is  maintained in the Lagrangian for the strings D5/D9, we write it in a form that shows
only a manifest invariance under this $SU(2)$.

In Appendix \ref{uplift59} we also construct  the six-dimensional
Lagrangian corresponding to the uplift of the ${\cal{N}}=2$ hypermultiplet from four to six dimension.  It is given in Eq. (\ref{Lfi}). When we include both the gauge theory living on the D5
and that on the D9 branes we get the following six-dimensional Lagrangian:
\begin{eqnarray}
{\cal{L}}_{59}   &=&  \,\, \epsilon^{i j} ( D_{ {\hat{\mu}} }  {\bar{w}}_{i })^{u}_{\,\,\,a}
(D^{{\hat{\mu}}}  {{w}}_{j})^{a}_{\,\,\,u} -
g_5 {\bar{w}}^{u}_{i\,a}  \sum_{c=1}^{3}(\tau^c)^{i}_{\,\,\,j} ( {\hat{ {\cal{D}}} }_{c+5}^{(5)})^{a}_{\,\,\,\,\,b}
w^{jb}_{\,\,\,u}
\nonumber \\
&& - \,
i  {\bar{\mu}}^{u}_{\,\,a} \Gamma^{\hat{\mu}} (D_{\hat{\mu}}  \mu)^{a}_{\,\,u} +
 \sqrt{2} g_5 i b  \left[  {\bar{\mu}}^{u}_{\,\,a}  ({{\Lambda}}^{(5)i})^{a}_{\,\,b}
 \epsilon_{ij}(w^j)^{b}_{\,\,u}    +
({\bar{w}}_i )^{u}_{\,\,a} \epsilon^{ij}   ({\bar{\Lambda}}_{j}^{(5)})^{a}_{\,\,b} \mu^{b}_{\,\,u} \right] \nonumber \\
&& + \,
{\tilde{g}}_9 {\bar{w}}_{ia}^{u}  \sum_{c=1}^{3}(\tau^c)^{i}_{\,\,\,j}
({\hat{{\cal{D}}}}_{c+5}^{(9)})^{v}_{\,\,\,\,\,u}   w^{ja}_{\,\,v}
\nonumber \\
&& - \,
 \sqrt{2} {\tilde{g}}_{9} i b  \left[  {\bar{\mu}}^{u}_{\,\,a} ( {\hat{ \Lambda}}^{i}_{(9)})^{v}_{\,\,u}
  \epsilon_{ij}(w^j)^{a}_{\,\,v}    +
({\bar{w}}_i)^{u}_{\,\,a} \epsilon^{ij}   ({{{\bar{{\hat{\Lambda}}}}}}_{j}^{(9)})^{v}_{\,\,u} \mu^{a}_{\,\,v} \right]
\label{59}
\end{eqnarray}
where $\bar{ w}_j=(w^j)^\dag$ and  ${\tilde{g}}_9$ is the gauge coupling constant of the gauge theory
living on the D9 branes properly rescaled with the volume factors in order to have
the same dimension as  $g_5$:
\begin{eqnarray}
\frac{1}{ {\tilde{g}}_{9}^{2} } =
\frac{1}{ g_{9}^{2} } \,\, {(2 \pi \sqrt{\alpha'})^2} \,\,T_{2}^{(2)} \,\, T_{2}^{(3)}
\label{gtilde9}
\end{eqnarray}
and
\begin{eqnarray}
&&(D_{\mu} w^i)^{a}_{u}  = \partial_{\mu} w^{ia}_{u} + ig_5 (A_{\mu}^{(5)})^{a}_{\,\,\,b} w^{ib}_{u}
- i {\tilde{g}}_9  ( {\hat{A}}_{\mu}^{(9)} )^{v}_{\,\,\,u} w^{ia}_{v}  \nonumber \\
&&(D_{\mu} {\bar{w}}_i )_{a}^{u}  = \partial_{\mu} {\bar{w}}_{ia}^{u}  -ig_{5}
(A_{\mu}^{(5)})^{b}_{\,\,\,a}  {\bar{w}}_{i b}^{u} + i{\tilde{g}}_{9} ( {\hat{A}}_{\mu}^{(9)} )^{u}_{\,\,\,v}
{\bar{w}}_{ia}^{v}\nonumber \\
&&(D_{\hat{\mu}} \mu )^{a}_{u}  = \partial_{{\hat{\mu}}} \mu^{a}_{u}  + i g_5
(A_{\hat{\mu}}^{(5)})^{a}_{\,\,\,b}\mu^{b}_{u}  -
i {\tilde{g}}_{9} ({\hat{A}}_{\hat{\mu}}^{(9)})^{v}_{\,\,\,u}\mu^{a}_{v} \,\, .
\label{DDeri}
\end{eqnarray}
Here, $T_2^{(2,3)}$ are the imaginary part of the K\"ahler moduli of the two tori associated to the directions $6\dots 9$.
We have put a hat on the fields living on the D9 to remember that they are six-dimensional.
The previous Lagrangian has a manifest $SU(2)$ symmetry that is the remnant of the
$SU(2)$ R-symmetry of the Lagrangian of the hypermultiplet in four dimensions.
We are able to keep it in six dimensions imposing
 the constraint in Eq. (\ref{Lambdai})  that eliminates the redundant components of the gaugino.

 The Lagrangian in Eq. (\ref{59}) has been obtained by uplifting to six dimensions a four-dimensional ${\cal{N}}=2$ supersymmetric Lagrangian. Although the original Lagrangian was supersymmetric, it is not clear that the uplifted Lagrangian is still  supersymmetric as the original one.
In Sect. \ref{susys} we show, with an explicit calculation, that indeed the uplifted Lagrangian has
the same amount of  conserved supersymmetry charges as the original four-dimensional one.

Finally, we have also the six-dimensional Lagrangian of the fields living on the D9 branes:
\begin{eqnarray}
{\cal{L}}_{99} &= &2  {\rm Tr} \left[  - \frac{1}{4} {\hat{F}}_{ {\hat{\mu}}  {\hat{\nu}} }^{(9)}
{\hat{F}}_{ {\hat{\mu}}   {\hat{\nu}}}^{(9)}
 + \frac{1}{2} \sum_{c=1}^{3} \left( ( {\hat{{\cal{D}}}_{c+5}^{(9)}} )^{2} +  i {\tilde{g}}_9
  {\hat{{\cal{D}}}}_{c+5}^{(9)}    \eta_{(c+5)mn} [ {\hat{A}}_{m}^{(9)},  {\hat{A}}_{n}^{(9)}]  \right)
 \right.
\nonumber \\
&&
\left.  - \frac{1}{2} \sum_{m=6}^{9} D_{\hat{\mu}} {\hat{A}}_{m}^{(9)}
  D_{\hat{\mu}} {\hat{A}}_{m}^{(9)}
-  \frac{i}{2}   {\bar{\hat{\Lambda}}}_{i}^{(9)} \Gamma^{\hat{\mu}}D_{\hat{\mu}} (
 \hat{\Lambda}^{(9)})^{i} -
  i   {\bar{\Psi}}^{(9)}\Gamma^{\hat{\mu}} D_{\hat{\mu}} \Psi^{(9)} \right.
  \nonumber \\
  &&  +
\left.  i \sqrt{2} {\tilde{g}}_9  b
\left( [{\bar{\hat{\Psi}}}^{(9)}, {\hat{Z}}^{(9)\,j}] \epsilon_{ij} ({\hat{\Lambda}}^{(9)})^{i} -
  {\bar{\hat{\Lambda}}}_{i}^{(9)} [ {\hat{\Psi}}^{(9)} , {\bar{\hat{Z}}}_{j}^{(9)}] \epsilon^{ij} \right)
 \right]    \,\, .
\label{L99}
\end{eqnarray}
The trace is over the fundamental matrices of the group $U(N_9)$ and
$\eta_{ {m} {n} }^{c+5}$ are the 't Hooft symbols   defined in Eq.
(\ref{eta}) \cite{thooft}.

It is important to stress that each of the three previous Lagrangians  is supersymmetric
independently from the others.
In particular, the Lagrangians of the open strings 55 and 99 are invariant under
16 supercharges, while that of the strings 59 is invariant under 8 supercharges.
They still contain the auxiliary fields of the gauge multiplets. In the following we
will eliminate them by using their equation of motion obtaining a complete Lagrangian
for the strings 59  and a Lagrangian for the strings 99 that can be uplifted from six to ten dimensions.

Let us consider the terms in Eqs. (\ref{59}) and (\ref{L99}) that contain the auxiliary
field ${\hat{\cal{D}}}^{(9)}$. They are:
\begin{eqnarray}
{\cal{L}}_{\cal D} &=& i {\tilde{g}}_9
\sum_{{c}=1}^{3} ({\hat{{\cal{D}}}}^{(9)} _{c+5} )^{u}_{\,\,\,v}
\eta_{(c+5)mn} ([ {\hat{A}}_{m}^{(9)},  {\hat{A}}_{n}^{(9)}]  )^{v}_{\,\,\,u}  +
\sum_{c=1}^3     ({\hat{{\cal D}}}_{c +5}^{(9)} )^{u}_{\,\,\,v}
({\hat{{\cal D}}}_{c +5}^{(9)} )^{v}_{\,\,\,u} \nonumber \\
 &&  + \,\, {\tilde{g}}_9 \,  \bar{w}_{ia}^{u} \sum_{c=1}^3
(\tau^{c})^i_{~j}\left( {\hat{{\cal D}}}_{c+5}^{(9)} \right)^v_{~u}w^{aj}_{v}  \,\, .
\label{calD4}
\end{eqnarray}
The equation of motion for ${\hat{\cal{D}}}^{(9)}$ is given by:
\begin{eqnarray}
({\hat{{\cal{D}}}}^{(9)}_{c+5})^{u}_{\,\,\,v}= \frac{1}{2} \left(- {\tilde{g}}_9
\bar{w}_{ia}^{u}
(\tau^{c})^i_{~j}w^{aj}_{v}-  i {\tilde{g}}_9 \eta_{(c+5)mn}
\left(   [ {\hat{A}}_{m}^{(9)},  {\hat{A}}_{n}^{(9)}]
\right)^u_{~v} \right)
\label{moti}
\end{eqnarray}
which, inserted back in Eq. (\ref{calD4}), yields:
\begin{eqnarray}
{\cal{L}}_{\cal D} & = &
\,\,\, \frac{1}{4} \, {\tilde{g}}_{9}^{2}
\sum_{c=1}^3 \eta_{(c+5)mn} \eta_{(c+5)pq}
\left( [ {\hat{A}}_{p}^{(9)},  {\hat{A}}_{q}^{(9)}]  \right)^u_{~v}
\left( [ {\hat{A}}_{m}^{(9)},  {\hat{A}}_{n}^{(9)}]  \right)^v_{~u}
\nonumber \\
&&-
\frac{i}{2} \, {\tilde{g}}_{9}^{2} \,  \bar{w}_{ia}^{u}  \sum_{c=1}^3  (\tau^{c})^i_{~j} \eta_{(c+5)mn}
 \left([ {\hat{A}}_{m}^{(9)},  {\hat{A}}_{n}^{(9)}]     \right)^v_{~u} w^{aj}_{v}
 \nonumber \\
&& - \frac{1}{4}  \, {\tilde{g}}_{9}^{2} \sum_{c=1}^3 \bar{w}_{ib}^{u}
(\tau^{c})^i_{~j}w^{aj}_{v}  \bar{w}_{hb}^{v}
(\tau^{c})^h_{~k}w^{bk}_{u}  \,\,.
\label{Dterm7}
\end{eqnarray}
The last two terms  of the previous equation go into the Lagrangian of the strings 59
that becomes:
\begin{eqnarray}
{\cal{L}}_{59}   &=&  \epsilon^{i j} ( D_{ {\hat{\mu}} }  {\bar{w}}_{i })_{a}^{u}
(D^{{\hat{\mu}}}   {{w}}_{j})^{a}_{u} -
g_5 {\bar{w}}_{ia}^{u}  \sum_{c=1}^{3}(\tau^c)^{i}_{\,\,\,j} ({\cal{D}}_{c+2}^{(5)})^{a}_{\,\,\,\,\,b}
w^{bj}_{u}
\nonumber \\
&-&
i  {\bar{\mu}}^{u}_{a} \Gamma^{\hat{\mu}} (D_{\hat{\mu}}  \mu)^{a}_{u} +
 \sqrt{2} g_5 i b  \left[  {\bar{\mu}}^{u}_{a} (\Lambda^{i}_{5})^{a}_{\,\,\,b} \epsilon_{ij}(w^j)^{b}_{u}
   +
({\bar{w}}_i)^{u}_{a} \epsilon^{ij}   ({\bar{\Lambda}}_{j}^{5})^{a}_{\,\,\,b} (\mu)^{b}_{u} \right] \nonumber \\
&-&     \frac{1}{4} {{g}}_{5}^{2} \sum_{c=1}^3 \bar{w}_{ia}^{u}
(\tau^{c})^i_{~j}w^{aj}_{v} \bar{w}_{hb}^{v}
(\tau^{c})^h_{~k}w^{bk} _{u}
\nonumber \\
&-&
 \sqrt{2} {{g}}_{9} i b  \left[  {\bar{\mu}}^{u}_{a} ( \Lambda^{i}_{(9)})^{v}_{\,\,\,u}
 \epsilon_{ij} (w^j )^{a}_{\,\,v}    +
({\bar{w}}_i)^{u}_{a}  \epsilon^{ij}   ({\bar{\Lambda}}_{j}^{(9)})^{v}_{\,\,u} (\mu)^{a}_{v} \right] \nonumber \\
&-&    \frac{i}{2}{\tilde{g}}_9 \bar{w}_{ia}^{u}  \sum_{c=1}^3  (\tau^{c})^i_{~j} \eta_{(c+5)mn}
  \left([ {\hat{A}}_{m}^{(9)},  {\hat{A}}_{n}^{(9)}]     \right)^v_{~u}
 w^{aj}_{v}  \,\, .
\label{59c}
\end{eqnarray}
This   Lagrangian  is  six-dimensional and  also the  fields of the gauge multiplet
living on the D9 branes are six-dimensional fields.  In the Lagrangian for the strings 55 and the one for the strings 59 there is still an auxiliary field
living on the world-volume of the D5 branes. By eliminating it one gets additional quartic
terms involving the fields $w$ and $Z$.

The first term in the right-hand side of Eq. (\ref{Dterm7}) goes instead together with
the others terms in Eq. (\ref{L99}) to give:
\begin{eqnarray}
{\cal{L}}_{99} & = & \,\, \, 2  {\rm Tr} \left[  - \frac{1}{4} {\hat{F}}_{ {\hat{\mu}}  {\hat{\nu}} }^{(9)}
{\hat{F}}_{ {\hat{\mu}}   {\hat{\nu}}}^{(9)}
 + \frac{{\tilde{g}}_{9}^{2}}{4} \sum_{c=1}^3 \eta_{(c+5)mn} \eta_{(c+5)pq}
[ {\hat{A}}_{m}^{(9)},  {\hat{A}}_{n}^{(9)}]    [ {\hat{A}}_{p}^{(9)},  {\hat{A}}_{q}^{(9)}]
\right.
\nonumber \\
&&
\left.  - \frac{1}{2} \sum_{m=6}^{9} D_{\hat{\mu}} {\hat{A}}_{m}^{(9)}
  D_{\hat{\mu}} {\hat{A}}_{m}^{(9)} -
 \frac{i}{2}   {\bar{{\hat{\Lambda}}}}_{i}^{(9)} \Gamma^{\hat{\mu}}D_{\hat{\mu}}
 ({\hat{\Lambda}}^{(9)})^{i} -
  i   {\bar{\Psi}}^{(9)}\Gamma^{\hat{\mu}} D_{\hat{\mu}} \Psi^{(9)} \right.
  \nonumber \\
  &&  +
\left.  i \sqrt{2} {\tilde{g}}_9  b
\left( [{\bar{\Psi}}^{(9)}, (Z^{(9)})^{j}] \epsilon_{ij} (\Lambda^{(9)})^{i} -
  {\bar{\Lambda}}_{i}^{(9)} [ \Psi^{(9)} , {\bar{Z}}_{j}^{(9)}] \epsilon^{ij} \right)
 \right]  \,\, .
\label{L99a}
\end{eqnarray}
This Lagrangian can be uplifted from six to ten dimensions. This is easy to perform for
the purely bosonic part of the action, while it requires some attention for the fermionic part. The point is that the six-dimensional gaugino is an SU(2) doublet and in ${\cal N
}=1$ SYM in ten dimensions that symmetry is lacking.  The presence of such a doublet is a consequence of the ten-dimensional Majorana-Weyl condition, as it is explained in Appendix \ref{uplift55}. In this Appendix  it is also shown that the uplifting from six to ten dimensions yields
${\cal{N}}=1$  SYM:
\begin{eqnarray}
{\cal{L}}_{99} = 2 {\rm Tr} \left[  - \frac{1}{4} F_{MN} F^{MN}  - \frac{i}{2} {\bar{\lambda}}
\Gamma^{M} D_M \lambda  \right]   \,\, .
\label{99}
\end{eqnarray}
In  Appendices \ref{uplift55} and \ref{uplift59} the Lagrangians (\ref{L55}, \ref{59}, \ref{L99a}) have been obtained by uplifting four-dimensional supersymmetric Lagrangians to six dimensions. While the uplift to ten dimensions for the strings D9/D9 is straightforward and gives rise to a Lagrangian that is supersymmetric in ten dimensions, this is not quite so for the strings D5/D9. The reason is that, while the fields corresponding to the open strings D5/D9 and
D5/D5 are six-dimensional, those living on the D9 branes are ten-dimensional. In order to
get an intuition on how this is going to work,  in Sect. \ref{vertexope}
we will  go back to string theory that treats the
fields living on the D9 branes as ten-dimensional, and we compute various terms of the D5/D9
action, checking, on the one hand, the previously constructed Lagrangian and seeing, on the
other hand, how the ten-dimensional fields appear in it.

\section{Supersymmetry  invariance of the action  for the strings D5/D9}
\label{susys}

The six-dimensional   Lagrangian   in Eq. (\ref{59}), describing the interaction of
 the massless open strings stretched between the D9 and the D5-branes
(twisted-matter) with the gauge multiplets living respectively on the D5 and D9 branes,
has been obtained from  the four-dimensional supersymmetric
${\cal N}=2$    Lagrangian of the hypermultiplet.
It  is expected to preserve ${\cal N}=1$ supersymmetry in six dimensions.
The uplifting procedure, however, in general does not generate all the terms
of the six-dimensional theory; for example it does not  give terms
depending on the derivative with respect to the compact dimensions. The
requirement of  the gauge invariance
allows one to obtain many of the missing terms but it
does not  ensure, in general,
that the uplifted action is complete.
The explicit proof of  the   invariance  of the six-dimensional
uplifted theory under ${\cal{N}}=1$ supersymmetry transformations, which we
are going now to discuss, is a strong check  that
the uplifted   Lagrangian is correct.

The six-dimensional Lagrangians for the massless D5/D5 and D9/D9
open strings contain a gauge multiplet and a hypermultiplet, transforming  in the
adjoint representation of the gauge group, and are invariant under
${\cal N}=2$ supersymmetry, while that of the  twisted matter
coupled to the two previous gauge multiplets preserves only half of
the previous supersymmetry and is therefore only invariant under
${\cal N}=1$ supersymmetry.  For the sake of simplicity, in the analysis of
this section we neglect the gauge multiplet living on the D9 branes, that, however,
can  be trivially included.  The ${\cal{N}}=1$ supersymmetry
transformations of the gauge multiplet can be found  from the requirement
that  they leave invariant the first line of Eq.  (\ref{L55}) corresponding to the
Lagrangian of a ${\cal{N}}=1$ gauge multiplet in six dimensions\cite{NPB121}
given by:
\begin{eqnarray}
&&{\cal S}_{g}= 2 \int d^6x {\rm Tr}\left[ -\frac{1}{4}F_{\hat{\mu}\hat{\nu}}^2+\frac{1}{2}\sum_{c=1}^3 {\cal D}_c^2-\frac{i}{2}\bar{\Lambda}_i \Gamma^{\hat{\mu}}D_{\hat{\mu}}\Lambda^i\right] \,\,.
\label{as1x}
\end{eqnarray}
It is easy to see that the following supersymmetry transformations
leave the  previous action  invariant:
\begin{eqnarray}
&\delta A^{\hat{\mu}}=\frac{i}{2}\left( \bar{\epsilon}_i
\Gamma^{\hat{\mu}} \Lambda^i-\bar{\Lambda}_i \Gamma^{\hat{\mu}} \epsilon^i \right)~~;~~\delta \Lambda^i=\frac{1}{2} F_{\hat{\mu}\hat{\nu}}\Gamma^{\hat{\mu}\hat{\nu}}\epsilon^i+ i { {\cal{D}}}^i_{~j}\epsilon^j&\nonumber\\
&
\delta {\cal D}^c= \frac{1}{2}(\tau^c)^i_{~j} \left( D_{\hat{\mu}}\bar{\Lambda}_i\Gamma^{\hat{\mu}}\epsilon^j+
\bar{\epsilon}_i\Gamma^{\hat{\mu}}D_{\hat{\mu}}\Lambda^j\right)&
\label{susy1f}
\end{eqnarray}
where $\Gamma^{\hat{\mu}\hat{\nu}}=\frac{1}{2} [\Gamma^{\hat{\mu}},\,\Gamma^{\hat{\nu}}]$ and
${\cal{D}}^{i}_{\,\,j} \equiv {\cal{D}}^c ( \tau^{c} )^{i}_{\,\,j}$.
The parameters of the supersymmetry transformations  $\epsilon^i$ are two six-dimensional spinors having,  for consistency
with Eqs. (\ref{susy1}), the same chirality as the gaugino and, as the gaugino,
they have  to satisfy Eq. (\ref{Majoco}). More details about the properties satisfied  by the $\epsilon^i$'s can be found in Appendix \ref{susyap}.
We are not going to show here explicitly that the action in Eq. (\ref{as1x}) is invariant under the transformations in Eq. (\ref{susy1}), but  we refer to Ref.~\cite{NPB121}  where the proof has been given. For more details see also  Sect. 9 of Ref.~\cite{9803026}. There is, however, a slight
difference with the action and the transformations given in Ref.~\cite{NPB121}, namely the presence of the auxiliary fields ${\cal{D}}^c$. In Appendix \ref{susyap} we show that the extra terms coming from the presence of the auxiliary fields, also give rise to a total derivative that
leave the action invariant.

Having determined the supersymmetry transformations of the fields of the gauge multiplet,
we are going now to study the supersymmetry  invariance of the Lagrangian for the twisted strings
D5/D9 given in Eq. (\ref{59}) without including, for the sake of simplicity,  the gauge multiplet living on the D9 branes. The Lagrangian is equal to:
\begin{eqnarray}
{\cal{L}}_{59}   &=& \,\,  \epsilon^{i j} ( D_{ {\hat{\mu}} }  {\bar{w}}_{i })_{a}
(D^{{\hat{\mu}}}  {{w}}_{j})^{a}
-
i  {\bar{\mu}}_{a} \Gamma^{\hat{\mu}} (D_{\hat{\mu}}  \mu)^{a}
-
g {\bar{w}}_{ia}  \sum_{c=1}^{3}(\tau^c)^{i}_{\,\,\,j}
({\hat{{\cal{D}}}}_{c+2})^{a}_{\,\,\,\,\,b}   w^{jb}
\nonumber \\
&& + \,
 \sqrt{2} g i b  \left[  {\bar{\mu}}_{a} ( {\hat{ \Lambda}}^i )^{a}_{~b}
  \epsilon_{ij}(w^j)^{b} +
({\bar{w}}_i)_{a} \epsilon^{ij}   ({{{\bar{{\hat{\Lambda}}}}}}_{j})^{a}_{~b} \mu^{b} \right]  \,\, .
\label{59ax}
\end{eqnarray}
In Appendix \ref{susyap} we show that  it transforms as a total derivative under the action of
the transformations in Eq. (\ref{susy1}) that act on the fields of the gauge multiplet,  together
with the following supersymmetry transformations acting on the twisted fields:
\begin{eqnarray}
\delta w^{ia}= - \sqrt{2} \, b \, \epsilon^{ij} \bar{\epsilon}_j\mu^a~~&;&~~\delta\mu^a =-
i \, \sqrt{2} \, b
\, \Gamma^{\hat{\mu}}\epsilon^i\epsilon_{ij}(D_{\hat{\mu}} w^j)^a.
\label{susy}
\end{eqnarray}
These transformations are similar to the ones involving the hypermultiplets in Ref. \cite{TOINE} with the main difference that they preserve ${\cal N}=1$ in six dimensions instead of ${\cal{N}}=2$ in five dimensions.  The doubling of the susy parameters, constrained by Eq. ({\ref{as6}), is necessary
to manifestly  keep the SU(2) invariance of the action.
The proof of the supersymmetric invariance of the twisted Lagrangian is  long and tedious.
The details are again given in  Appendix  \ref{susyap}. Here we quote just the final result
of the supersymmetry transformations given by:
\begin{eqnarray}
 \delta {\cal{L}}_{59} & =  & \,\,\,  \partial_{\hat{\mu}} \left\{ - \sqrt{2} \, b\,
  \bar{w}_{ja}\epsilon^{ji}\bar{\epsilon}_i [\Gamma^{\hat{\mu}},\,\Gamma^{\hat{\nu}}](D_{\hat{\nu}} \mu)^a     +
\sqrt{2} \, b \,   {\bar{\mu}}_a \epsilon^i \epsilon_{ij} (D^{\hat{\mu}} w^j)^a \right. \nonumber\\
&& \left. + \, {g} \,  \bar{w}_{ja}\epsilon^{jk}\left( \bar{\epsilon}_k \Gamma^{\hat{\mu}}(\Lambda^i)^a_{~b}+(\bar{\Lambda}_k)^a_{~b}\Gamma^{\hat{\mu}}\epsilon^i\right)\epsilon_{ik}w^{kb}\right\}  \,\, .
\label{totder}
\end{eqnarray}
In conclusion, we have shown that the Lagrangian for the open strings D5/D9 is invariant under
the supersymmetry transformations given in Eqs. (\ref{susy1f}) and (\ref{susy}).

\section{Vertex operators for open strings 99, 55 and 59}
\label{vertexope}

In this section we write the vertex operators of the massless open string states in the system D5/D9. We have three kinds of open strings:  those with the
two end-points attached to a system of $N_9$  parallel D9 branes, those
with the two end-points
attached to a system of $N_5$  parallel D5 branes
and the mixed open strings having  one end-point attached to a D9 brane
and the other end-point attached to a D5 brane.
All the following vertices are in a one-to-one correspondence with those written in
Ref~\cite{Billo:2002hm} due to the T-duality between  the system D9/D5 and the  system D3/D(-1) discussed in that paper. The main difference consists  in the momentum dependence in all sectors  of our vertices and in the $SU(2)$ structure that is  manifest in the expression of the string vertex
operators.

\subsection{$D9/D9$ open strings}
\label{D9D9}
The massless fields living on the system of $N_9$ $D9$ branes consist in a gauge field
$(A_{M}^{(9)} )^u_v (x^\hmu,y^m)$ and a
ten-dimensional  Majorana-Weyl spinor with negative chirality~\footnote{This is a
consequence of the fact that we have chosen the  GSO projection in the
ten-dimensional  $II B$ string theory such as  to keep spinors
with negative chirality.}
$(\Lambda_{\cal{\dot{A}}}^{(9)} )^u_v (x^\hmu,y^m)$.
All the fields carry a ten-dimensional momentum with $k_Mk^M=0$ and their vertex  operators  are:
\begin{eqnarray}
&&V_{A^{(9)}}^{(-1)}  (z) =ig_9
\sqrt{2 \pi\alpha'}\, (A_{M}^{(9)} )^u_v (k_M)
~ e^{-\phi (z)} ~\psi^M  (z) ~ e^{i \sqrt{2 \pi \alpha'}
k_M X^M (z) } \nonumber \\
\label{-1pict}
\end{eqnarray}
for the gauge field in the picture $(-1)$ and
\begin{eqnarray}
&&\!\!\!\!V_{A^{(9)}}^{(0)} (z) =
i g_9\sqrt{ 2\pi\alpha'}
(A_{M}^{(9)} )^{u}_{v} (k)\left[ \partial_z  X^{M} (z)
+ i \sqrt{2\pi\alpha'} k_N \psi^N (z)  \,  \psi^M (z) \right]
 {\rm e}^{i \sqrt{2 \pi \alpha'} k_M  X^M}    \nonumber \\
\label{vertA}
\end{eqnarray}
in the picture $0$. For both of them the trasversality condition $k_MA^M=0$ holds. Here  $M,N=0,\dots 9$; $\dcalA=1,\dots 16$ and $u,v=1,\dots N_9$ are the indices of the  gauge group $U(N_9)$
living on the set
of the $N_9$  $D9$ branes.

The vertex for the gaugino is:
\begin{eqnarray}
V_{\Lambda^{(9)}} ^{(-{1/}{2})} (z) =g_9
(\pi\alpha')^{\frac 3 4}\,
(\Lambda_{\dcalA}^{(9)})^{u}_{v} (k_M)~ e^{-\oh \phi (z)} ~
S^\dcalA  (z) ~ e^{i \sqrt{2 \pi \alpha'}
~k_M X^M (z)} \,\, .
\label{gaugi}
\end{eqnarray}
The presence of the $D5$ branes breaks the original ten-dimensional Lorentz group $SO(1,9)$ into $SO(1,5) \otimes SO(4)$.  Hence,  it is convenient to split the ten-dimensional
Weyl index $\dcalA$ as follows
\begin{eqnarray}
S^\dcalA = (S^\hAA S^\dal, S_\hAA S_\alpha)
\label{deco}
\end{eqnarray}
where  the upper (lower) index $\hAA$ is the one of  a Weyl spinor in six dimensions
with positive (negative) chirality
and similarly  $\alpha$ (${\dot{\alpha}}$) is the index of a four-dimensional Weyl spinor
with positive (negative) chirality.
Introducing the  decomposition  (\ref{deco}) in the vertex in Eq. (\ref{gaugi}),
one gets:
\begin{eqnarray}
V_{ \Lambda^{(9)} _{ {\hat{A}} {\dot{\alpha}} } }^{(-\frac{1}{2}) }(z) =\sqrt{2} g_{9}
(\pi\alpha')^{\frac 3 4}\,
(\Lambda_{  {\hat{A}} {\dot{\alpha}}    }^{(9)} )^{u}_{v}(k_M)~ e^{-\oh \phi (z)} ~
S^{\hat{A}}  (z) S^{ \dot{\alpha} } (z) ~ e^{i \sqrt{2 \pi \alpha'}
~k_M X^M (z)}
\label{Vla}
\end{eqnarray}
The latter expression has an index $\dot{\alpha}$ that for simplicity,  in the previous section,
we have denoted by the indices $i$ and $j$, and that corresponds to the $SU(2)$  symmetry previously
discussed. In string theory the role of the internal symmetry is thus played by one of the two $SU(2)$ factors  of  the $SO(4)$ group  associated to the directions 6\dots 9.

The normalization factor of the vertex operator in Eq. (\ref{Vla}) has been determined by requiring
that the three-point function computed in string theory  and involving two gauginos and a gauge field agrees with the corresponding coupling in field theory.  In particular, the factor $\sqrt{2}$
in Eq. (\ref{Vla}) is a direct consequence of the constraint for the gaugino field  in Eq. (\ref{Lambdai}).

In principle one should also introduce a vertex for the field $\bar\Lambda_i$ besides the one for $\Lambda^i$, but in this case this is not necessary because the two fields are not independent being related by the symplectic Majorana condition.

The last vertex coming from the decomposition of the ten-dimensional spinor is:
\begin{eqnarray}
V_{\Psi^{(9)\hat{A} \alpha}}^{ (- \frac{1}{2} ) }  (z) = \sqrt{2}g_{9}
(\pi\alpha')^{\frac 3 4}\,
(\Psi^{\hat{A} \alpha })^{u}_{v} (k_M)~ e^{-\oh \phi (z)} ~
S_{\hat{A}}  (z) S_{\alpha} (z) ~ e^{i \sqrt{2 \pi \alpha'}
~k_M X^M (z)}  \,\,
\label{Vlab}
\end{eqnarray}
which  again  contains an  $SU(2)$-doublet corresponding to the $SU(2)$ that is a symmetry of
the Lagrangian of the open strings D5/D5, but not of the Lagrangian of the open
strings D5/D9. In the six-dimensional Lagrangian  of the open strings D5/D5 in Eq. (\ref{L55})
such a doublet is, however, missing because this $SU(2)$ symmetry is not
manifestly realized  and we
have a field $\Psi$ instead of a doublet. It turns out that the vertex operator corresponding to this field is given  in Eq. (\ref{Vlab}) where the six-dimensional spinor is  taken to be:
\begin{eqnarray}
\Psi^{ {\hat{A}}\alpha}=\left(\begin{array}{c}
                        \Psi^{\hat{A}} \\0\end{array}\right)\label{psi9}
\end{eqnarray}
Notice that the ten-dimensional spinor index is dimensionally reduced,
but the four-momentum of the vertices is  still ten-dimensional.

\subsection{$D5/D5$ open strings}
\label{D5D5}
The vertex operators corresponding to open strings of the gauge multiplet living on  the $D5$ branes are obtained through a trivial dimensional reduction from those living  on  a
$D9$ brane.
They are a gauge field $(A_{\hmu}^{(5)} )^a_b  (x^\hmu)$ (with $\hmu,\hnu=0,\dots 5$ and $a,b$ running over the number of $D5$ branes, namely
$a,b=1,\dots N_5$), four real  scalars $(\phi_{m}^{(5)})^a_b (x^\hmu)$, a couple of
six-dimensional  Weyl spinors $(\Lambda^{(5)\hAA \dal})^a_b (x^\hmu)$
with positive six-dimensional  chirality and a couple of
six-dimensional  Weyl spinors $( { \Psi_{\hAA \alpha}^{(5)})}^a_b
(x^\hmu)$  with negative  six-dimensional  chirality.  The components of the latter spinor are  given as in Eq. (\ref{psi9}).

The corresponding  vertex operators, carrying a six-dimensional momentum with $k^2=0$,
are:
\begin{eqnarray}
&&V_{A^{(5)}}^{(-1)} (z) = i (2\pi\alpha')^{\frac12} g_{5}\,
(A_{\hmu}^{(5)})^a_b (k_\hnu)~ e^{-\phi (z)} ~\psi^\hmu (z) ~ e^{i \sqrt{2 \pi \alpha'}
~k_\hnu X^\hnu (z)}
\label{Verta1}
\end{eqnarray}
for the gauge field in picture $(-1)$,
\begin{eqnarray}
V_{A^{(5)}}^{(0)} (z) &=& i
(2\pi\alpha')^{\frac 1 2}  g_{5}\,
(A_{\hmu}^{(5)})^a_b (k)~
\left[ \partial_z  X^{\hmu} (z)
+ i (2 \pi \alpha')^{\frac{1}{2}} k_\hnu \psi^\hnu  (z) \,  \psi^\hmu (z) \right]
\nonumber \\
&\times&  {\rm e}^{i  \sqrt{2 \pi \alpha'}
k_\hnu  X^\hnu (z)}
\label{verta0}
\end{eqnarray}
for the gauge field in the picture $0$, and for both of them
the transversality condition $k_\hmu  A^{(5)\hmu} =0$ holds.

For the scalars in the picture $(-1)$ we have:
\begin{eqnarray}
V_{\phi^{(5)}_{{m}}}^{(-1)} =i  g_{5}
(2\pi\alpha')^{\frac12}\,
(\phi_{ m}^{(5)})^a_b (k_\hnu)
~ e^{-\phi(z)} ~\psi^m (z)~ e^{i \sqrt{2 \pi \alpha'}
~k_\hnu X^\hnu}
\label{ verta2}
\end{eqnarray}
while for the scalars in the picture $0$ we have:
\begin{eqnarray}
V_{ \phi^{(5)}_m }^{(0) } &=& i g_{5} (2\pi\alpha')^{\frac 1 2}
(\phi_{m}^{(5)} )^a_b (k_\hnu)~
\left[ \partial_z  X^{m} (z)
+ i  (2 \pi \alpha')^{\frac{1}{2}}k_\hnu \psi^\hnu  \,  (z) ,  \psi^m (z) \right]
\nonumber \\
&\times&{\rm e}^{i  \sqrt{2 \pi \alpha'} k_\hnu  X^\hnu (z)} .
\label{verta3}
\end{eqnarray}
The vertex for the gaugino with negative six-dimensional chirality is:
\begin{eqnarray}
V_{ \Lambda^{(5)}_{ {\hat{A}} \dal } } (z) = \sqrt{2} g_{5}
(\pi\alpha')^{\frac 3 4}\,
(\Lambda_{\hAA \dal}^{(5)})^{a}_{b}
(k_\hmu)~ e^{-\oh \phi (z)} ~
S^{\hAA} (z) S^{ \dal} (z) ~ e^{i  \sqrt{2 \pi \alpha'} ~k_\hnu X^\hnu}
\label{verta4}
\end{eqnarray}
and for $\Psi^{(5)}$, which has the same $SU(2)$ structure given in Eq. (\ref{psi9}),
we have:
\begin{eqnarray}
V_{\Psi^{ (5) {\hat{A}} \alpha }}= \sqrt{2} g_{5}
(\pi\alpha')^{\frac 3 4}\,
(( { \Psi}^{(5)} )^{\hAA \alpha})^{a}_{b}  (k_\hmu)~
e^{-\oh \phi (z)} ~S_{\hAA} (z)
S_{ \alpha} (z)~ e^{i  \sqrt{2 \pi \alpha'} ~k_\hnu X^\hnu (z)} .
\label{verta5}
\end{eqnarray}
Finally, we add also the vertex operator of the  auxiliary fields  whose introduction is convenient
for computing  only the  cubic couplings instead of the more difficult
quartic ones:
\begin{eqnarray}
V^{(5)}_{{\cal{D}}_{c+5}} = i (\pi \alpha' ) g_{5} ({\cal{D}}_{c+5}^{(5)})^{a}_{b} \eta^{c+5}_{{m} t{n}} \psi^{{m}} \psi^{{n}} e^{-\phi(z)} e^{i \sqrt{ 2 \pi \alpha' } k_{\hat{\mu}} X^{\hat{\mu}} (z) }
\label{aus}
\end{eqnarray}
where $c=1,2,3$; $ {m},  {n} = 6,7,8,9$.

\subsection{$D9/D5$ and $D5/D9$ open strings}
\label{D9D5}

In the system $D5/D9$, along  the four directions of
the last two tori, there are Ramond (Neveu-Schwarz) boundary conditions
in the NS (R) sector. This means that the massless state of the
NS sector is an
$SO(4)$ spinor. Due to the GSO projection it is a chiral
spinor,  $w_{\dot{\alpha}}$. In the R sector, the massless state is
a  $SO(6)$  spinor that, because again of the GSO projection, is a chiral spinor
in  six dimensions:  the four non-compact directions and the
two compact ones belonging to the first torus.  In conclusion, the
massless $D9/D5$ sector contains a  doublet of scalars
${( w_{{\dot\alpha}})^a_u} (x )$ under $SU(2)$
and a  six-dimensional
Weyl spinor $(\mu^A )^a_u (x )$.

Their vertex operators are ~\cite{Billo:2002hm}:
\begin{eqnarray}
\label{vertw}
V_{w}^{(-1)}(z)&=&
g_{5} (\pi\alpha')^{\frac12}\,{(w_{\dot\alpha})^a_u} (k)~~
\Delta(z) S^{\dot\alpha}(z)\,{\rm e}^{-\varphi(z)}  {\rm e}^{i  \sqrt{2 \pi \alpha'}
k_\hmu  X^\hmu (z)}
\nonumber\\
V_{\mu}^{(- \frac{1}{2})} (z)
&=&
g_{5}(\pi\alpha')^{\frac 3 4}\, (\mu^{\hat{A}})^a_u(k)~~
\Delta(z) S_{\hat{A}}(z)\,{\rm e}^{-\phi/2}  {\rm e}^{i \sqrt{2 \pi \alpha'}
k_\hmu  X^\hmu (z)}
\end{eqnarray}
where
$\Delta$ is the twist operator with conformal weight $1/4$ which changes
the boundary conditions of the four compact coordinates  from Neumann to
Dirichlet. $k_{\hmu}$ is the six-dimensional momentum. The vertex operators
in Eq. (\ref{vertw}) describe the strings D9/D5. The strings D5/D9 are described
instead by:
\begin{eqnarray}
\label{vertwb}
V_{\bar{w}}^{(-1)} (z) &=&
g_{5} (\pi\alpha')^{\frac12}\,
( {\bar{w}}_{\dot\alpha} )^u_a  (k)~~
{\bar{\Delta}}(z) S^{\dot\alpha}(z)\,{\rm e}^{-\varphi(z)}  {\rm e}^{i \sqrt{2 \pi \alpha'}
k_\hmu  X^\hmu (z)}
\nonumber\\
V_{\bar{\mu}}^{(- \frac{1}{2})} (z)
&=&
g_{5} (\pi\alpha')^{\frac 3 4}\, ({\bar{\mu}}^{\hat{A}})^u_a(k)~~
{\bar{\Delta}}(z) S_{\hat{A}}(z)\,{\rm e}^{-\phi/2}  {\rm e}^{i \sqrt{2 \pi \alpha'}
k_\hmu  X^\hmu (z)}
\end{eqnarray}
with $\Delta(z)$ replaced by the anti-twist
$\bar\Delta(z)$ operator, corresponding to DN (instead of ND) boundary conditions along
the directions orthogonal to the world-sheet of the D5 brane. The upper index  in brackets of each vertex denotes, as usual,  the corresponding picture. All previous states transform according to the
bifundamental of the group $U (N_9 ) \times U (N_5 )$.

\section{Computing string amplitudes}
\label{striampli}

Having identified  the massless states of the three  open string sectors
and their vertex operators, we can now perform string disk
calculations in order to determine the various terms of their low-energy Lagrangian. This
approach  is complementary to that presented  in Sect. [\ref{595559}], where we
have constructed the supersymmetric low-energy Lagrangians in six dimensions  by
uplifting   to six dimensions the four-dimensional ones.

The calculation of various three-point couplings  provides an additional check  of the six-dimensional Lagrangians already
derived in Sect.[\ref{595559}].  An example of this is
the coupling of two adjoint fermions and a gauge boson living on the D5 branes.
This coupling has been computed in Eq. (\ref{ppl}) and agrees with the one in Eq. (\ref{L55}).
Other three-point couplings can also be easily  computed and one can see that they
reproduce the various trilinear couplings  of the Lagrangian in Eq. (\ref{L55}).
In particular, one can   use the vertex operator of the auxiliary fields to determine also their couplings with the propagating fields. In this way all trilinear couplings can be computed.
The quartic terms present  in the Lagrangian (\ref{L55}) can be obtained from the cubic one
by gauge invariance. In this way one can derive the three six-dimensional Lagrangians for
the strings D5/D5, D9/D9 and D5/D9.

 On the other hand, string theory treats  the fields of the gauge multiplet living on the D9 branes
as ten-dimensional and therefore one can determine their couplings with the twisted fields living on the D5 branes. In this way one can obtain a six-dimensional Lagrangian for the massless
open strings D5/D9 interacting with the gauge multiplet living on the D9 branes, where
the latter fields are ten-dimensional and not six-dimensional as in Eq. (\ref{59}).

Examples are provided by the three-point amplitudes involving a twisted scalar, a twisted fermion and a D5/D9 or D5/D5 gaugino. These are computed in Appendix~ \ref{correla}. Particularly interesting
 is the three-point amplitude
involving two twisted scalar fields $w$ and ${\bar{w}
}$ and a gauge boson living on the D9 branes, given by:
\begin{eqnarray}
{\cal{A}}^{ {\bar{w}} A_{M}^{(9)}w} = C_0 \int \frac{d x_1 dx_2 dx_3}{dV}
\langle  0| V_{\bar{w}} ^{(-1)}(x_1)\,V_{A_{M}^{(9)}}^{(0)}(x_2)\, V_{w }^{(-1)}(x_3)
|0 \rangle
\label{corre}
\end{eqnarray}
where $C_0$ is the  normalization of the disk which is related to the five dimensional gauge coupling   $g_{5}$ through the relation $C_{0} =  g_{5}^{-2} ( \pi \alpha')^{-2}$ and $dV$ is the projective invariant  M{\"{o}}bius volume.
This amplitude, which can be read from Eq.(\ref{fin3a}), is given
by:
\begin{eqnarray}
{\cal{A}}^{ {\bar{w}} A^{(9)}w } & = & \,\,\, g_9 \,
( {\bar {w}}_{\dot\alpha})^{v}_{ a } (p_{\hat\mu},\,y_0)
({w}_{{\dot{\beta}}})^{a}_{ u}(q_{\hat{\mu}},\,y_0) \left\{  \epsilon^{ \dot{\alpha} \dot{\beta}}
(A^{(9)}_{ {\hat{\mu}} })^{u}_{v} (k_{\hat m},\,y_0)  ( p^{\hat{\mu} }  -q^{\hat{\mu} } )
\right.
\nonumber \\
&&
\left.
- \frac{1}{2} \left(( A^{(9)}_{ {{m}} })^{u}_{v} (k_{\hat{m}},\,y_0)
k_{{n}}  - (A^{(9)}_{ {{n}} })^{u}_{v} (k_{\hat m},\,y_0)
k_{{m}} \right) (\bar\sigma^{n m} ){}^{\dot \alpha  \dot \beta}
\right\}
\label{fin4b}
\end{eqnarray}
where $y_0$ is the position of the $D5$-branes in the last two tori.
The first term of the previous equation reproduces the coupling of the twisted scalars
with the D9 gauge fields that  can be read from the first term of the  Eq. (\ref{59c}).
Having computed a three point coupling implies that the second term contains only the  derivative part
of the gauge field strength living on the D9 branes. It can be easily made  gauge invariant by adding the term with the commutator obtaining:
\begin{eqnarray}
&&{\cal{A}}^{{\bar{w}} A^{(9)} w}_{(2)} =
-\frac{i}{2}
( {\bar {w}}_{i})^{v}_{ a } (p_{\hat\mu},\,y_0)
({w}^{j})^{a}_{ u}(q_{\hat\mu},\,y_0)
(F_{{m}{n}})^u_{~v}(k,\,y_0)\sum_{c=1}^{3} \eta_{c+5}^{~~~mn} (\tau^{c})^{i}_{\,\,\,j}
\label{gn}
\end{eqnarray}
where  the identity
\begin{eqnarray}
(\bar{\sigma}^{mn})^{i}_{\,\,\,j}=i\sum_{c=1}^{3} \eta_{c+5}^{~~~mn} (\tau^{c})^{i}_{\,\,\,j}
\label{barsigma}
\end{eqnarray}
has been used and the indices $(\dot{\alpha},\,\dot{\beta})\equiv (i,\,j)$ have been redefined. The quantities  $\eta$'s are again the 't~Hooft symbols, previously introduced.

The amplitude in Eq. (\ref{gn}) is the string result of the last term in the Lagrangian  (\ref{59c}). It differs from the corresponding one obtained in field theory  because it depends on the whole field strength of the internal components of the gauge field instead of  only the commutator. Furthermore, the field strength
of the D9 gauge field is  ten-dimensional and is computed at the point $y_0$ of the four-dimensional space where the D5 branes are located.

In conclusion, we have shown that, by performing string disk calculations, we can uplift the six-dimensional Lagrangian in (\ref{59}) to contain the ten-dimensional  fields   of the gauge
multiplet living on the D9 branes (and not just
their reduction to six dimensions as obtained in the field theoretical  approach).  In practice, one obtains the same Lagrangian
as in Eq. (\ref{59c}), where now all the ten-dimensional fields are computed at $y_0$ with the last term in Eq. (\ref{59c}) that becomes:
\begin{eqnarray}
- \frac{i}{2}  {\bar{w}}^{u}_{ia} ({\bar{\sigma}}^{mn})^{i}_{\,\,j}
 ( F_{mn})^{v}_{\,\,u} w^{aj}_{v}  \,\, .
\label{59cd}
\end{eqnarray}
The Lagrangian in Eq. (\ref{59c}), with the modifications discussed above, will be used in the next section to compute  the spectrum of the Kaluza-Klein states in the presence of a background magnetization in the six extra-dimensions.

\section{The Kaluza-Klein reduction}
\label{KKredu}

After having  determined the six-dimensional Lagrangian
describing the interaction of the massless  fields corresponding to the
open strings  D5/D9 with the gauge multiplets living respectively on the D5 and D9 branes,
in this section  we perform the Kaluza-Klein reduction of the six-dimensional
Lagrangian to $R^{3,1} \times T^2$  with a non-zero magnetization on the D9 branes
and we show that the spectrum of states agrees with the one
obtained from string theory in the field theory limit given in Sect. \ref{2.3}.
Since we are interested only in the spectrum of Kaluza-Klein states, we can limit ourselves
to the following part of the Lagrangian in Eq. (\ref{59c}):
\begin{eqnarray}
{\cal L}_{59} & = & \,\,\,
\epsilon^{ \dot\alpha \dot \beta}
\left(    {D^{\hat{\mu}} {\bar w}_{\dot\alpha }} \right)^{a}_{\,\,u}
\left( D_{\hat{\mu}} w_{\dot\beta }\right)^{a}_{\,\,u}  - i {\bar{\mu}}^{u}_{\,\,a}\Gamma^{\hat{\mu}}
\left( D_{\hat{\mu}} \mu \right)^{a}_{\,\,u} \nonumber \\
&&- \frac{i}{2}
 ({\bar{w}}_{\dot{\alpha} })^{u}_{\,\,a}
({\bar\sigma}^{m n} )^{\dot  \alpha}_{~ \dot  \beta } F_{mn}^{(9)}(x,y_0)
\epsilon^{\dot \beta \dot \gamma}
( w_{\dot\gamma } )^{a}_{\,\,u}  \,\, .
\label{LAG}
\end{eqnarray}
Let us consider the bosonic part of this Lagrangian that, after a partial integration,
 is equal to:
\begin{eqnarray}
{\cal L}^{bos.}_{59} & =  & \,\,\, \epsilon^{ \dot\alpha \dot \beta}
\left(   {D^{{\mu}} {\bar w}_{\dot\alpha }} \right)^{a}_{\,\,u}
\left( D_{{\mu}} w_{\dot\beta }\right)^{a}_{\,\,u} -
\epsilon^{ \dot\alpha \dot \beta}
\left(  {\bar w}_{\dot\alpha }  \right)^{a}_{\,\,u}
\left( {\bar{D}}_i  D_i  w_{\dot\beta }\right)^{a}_{\,\,u} \nonumber \\
&&- \frac{i}{2}
 ({\bar{w}}_{\dot{\alpha} })^{u}_{\,\,a}
({\bar\sigma}^{m n} )^{\dot  \alpha}_{~ \dot  \beta } F_{mn}^{(9)}(x,y_0)
\epsilon^{\dot \beta \dot \gamma}
( w_{\dot\gamma } )^{a}_{\,\,u}
 \label{LAGB}
\end{eqnarray}
where $\mu=0 \dots 3$, $i=4,5$ and ${{m, n }}=6, 7, 8, 9$ and the
indices ${{m}}, {{n}}$ are contracted with the flat metric.
Furthermore:
\begin{eqnarray}
(\bar\sigma^{{m} {n}} )^{\dot  \alpha}_{~ \dot  \gamma}=\frac{1}{2}\left(
(\bar{\sigma}^{ {m}})^{\dot\alpha\alpha}(\sigma^{{n}})_{\alpha\dot \gamma}-
(\bar{\sigma}^{{n}})^{\dot\alpha\alpha}(\sigma^{{m}})_{\alpha\dot \gamma}\right)
\label{sig}
\end{eqnarray}
with $\sigma^{{m}}\equiv(-i\vec{\tau},\,\mathbb{I})$ and
$\bar{\sigma}^{{m}}\equiv(i\vec{\tau},\,\mathbb{I})$\footnote{The identity matrix corresponds to the index $9$, while the matrix $\vec{\sigma}$ corresponds to the indices $6, 7, 8$.}.
In the following, we restrict ourselves to a six-dimensional compact
manifold that is the product of three tori $T^2$ with a background magnetic field $F\equiv (F_{45}^{(9)},\,F_{67}^{(9)}, \, F_{89}^{(9)})$   turned on
only on the D9-branes.  The three background magnetic fields must satisfy the constraint that the corresponding Chern class is an integer.  For the sake of simplicity, we assume that the magnetic fields lie in the $U(1)$ gauge group. This implies that they have the following form:
\begin{eqnarray}
F_{45} =\frac{2\pi I^{(1)}}{(2\pi\sqrt{\alpha'})^2T_2^{(1)}}~ ~~;~~F_{67}^{(9)}=\frac{2\pi I^{(2)}}{(2\pi\sqrt{\alpha'})^2T_2^{(2)}}~~;~~F_{89}^{(9)}=\frac{2\pi I^{(3)}}{(2\pi\sqrt{\alpha'})^2T_2^{(3)}}
\label{3mag}
\end{eqnarray}
where $ I^{(1)}, I^{(2)}, I^{(3)}$ are three integers. We take the D5 branes unmagnetized.

The extension to the case of magnetized D5-branes is easily obtained by writing $I^{(1)}=I^{(1)}_5-I^{(1)}_9$, while the extension to a non abelian bundle is obtained by the substitution  $I \rightarrow
\frac{I}{n}$ where $I$  is the first Chern-class and $n$ is the
wrapping number.

Because of the background magnetizations the quantity appearing in the last term of Eq. (\ref{LAGB}) is equal to:
\begin{eqnarray}
(\bar\sigma^{{m} {n}} )^{\dot  \alpha}_{~ \dot  \gamma}F_{{m} {n}}^{(9)}(x,y_0)=2 i
\left(\begin{array}{cc}
                                     F_{67}^{(9)}+F_{89}^{(9)}&0\\
                                    0&-F_{67}^{(9)}-F_{89}^{(9)}
                                    \end{array}\right) \,\, .
\label{EFFE}
\end{eqnarray}
Furthermore,  according to the notation of Ref. \cite{0810.5509}, we have:
\begin{eqnarray}
- \bar D^i D_i=\frac{2\pi | I^{(1)} |}{(2\pi\sqrt{\alpha'})^2T_2^{(1)}} (2a^{(1)\dag} \, a^{(1)}+1) \,\, . \label{DDF}
\end{eqnarray}
By inserting Eqs. (\ref{EFFE}) and (\ref{DDF}) in Eq. (\ref{LAGB}) we get the following expression for
the quadratic term involving the twisted field $w$:
\begin{eqnarray}
{\cal L}^{bos} =
-\epsilon^{\dot\alpha \dot \beta}
\bar D^\mu \bar w_{\dot\alpha }
D_\mu w_{\dot\beta }+\bar w_{\dot\alpha }\left( \frac{4\pi |I^{(1)} |}{(2\pi\sqrt{\alpha'})^2T_2^{(1)}}
N^{(1)}
\delta^{\dot \alpha}_{~\dot\gamma}+M^{\dot \alpha}_{~\dot\gamma} \right)
\epsilon^{\dot\gamma \dot \beta}w_{\dot\beta }
\label{LAGB2}
\end{eqnarray}
being $N^{(1)} = a^{(1)\dagger} a^{(1)}$ and
\begin{eqnarray}
M^{\dot \alpha}_{~\dot\gamma}=\frac{2\pi}{(2\pi\sqrt{\alpha'})^2}\left(\begin{array}{cc}
                 \frac{|I^{(1)}|}{T_2^{(1)}}+ \left(\frac{I^{(2)}}{T_2^{(2)}}+
                 \frac{I^{(3)}}{T_2^{(3)}} \right)&0\\
0&\frac{|I^{(1)} |}{T_2^{(1)}}- \left( \frac{I^{(2)}}{T_2^{(2)}}+ \frac{I^{(3)}}{T_2^{(3)}}\right)
  \end{array} \right)  \,\, .
\label{EMMEa}
\end{eqnarray}
The previous matrix is diagonal with eigenvalues given by:
\begin{eqnarray}
\lambda_{\dot \alpha}= \frac{2\pi}{(2\pi\sqrt{\alpha'})^2}\left[\frac{|I^{(1)}|}{T_2^{(1)}}
\pm \left( \frac{I^{(2)}}{T_2^{(2)}} + \frac{I^{(3)}}{T_2^{(3)}} \right)\right]  \,\, .
\label{EMME}
\end{eqnarray}
The massless state is obtained by choosing $N^{(1)}=0$ and imposing
the constraint $\frac{|I^{(1)}|}{T_2^{(1)}} \pm \left( \frac{I^{(2)}}{T_2^{(2)}}+
\frac{I^{(3)}}{T_2^{(3)}} \right) =0$ where the sign $\pm$ to be chosen depends on the sign
of the round bracket.
The corresponding wave function can be obtained by solving the Eq.
$a^{(1)} \,w_{\dot \alpha}=0$ with suitable boundary conditions. These are given by the identification, up to a gauge transformation, of the fields under a translation on the torus \cite{ 0404229}. The solution to this problem is provided in Ref.\cite{0810.5509}. Therefore,
we conclude that the wave-function is again a theta-function depending
just on the moduli of the first torus.
The general expression for the mass is then:
\begin{eqnarray}
m^2_n=\frac{2\pi }{(2\pi\sqrt{\alpha'})^2}\left[\frac{ | I^{(1)} |}{T_2^{(1)}}(2N^{(1)}+1)
\pm \left( \frac{ I^{(2)}}{T_2^{(2)}} + \frac{ I^{(3)}}{T_2^{(3)}} \right)\right]  \,\, .
\label{spect5}
\end{eqnarray}
Let us  now show  that the  spectrum of
Kaluza-Klein states in Eq. (\ref{spect5}) agrees
with the one obtained from string theory in Eqs. (\ref{M259}) and (\ref{M259b}).
In Eq. (\ref{spect5}) $I^{(2)}$ and $I^{(3)}$ can be both positive and negative. When they have
opposite sign, the previous mass formula becomes:
\begin{eqnarray}
m^2_n=\frac{2\pi }{(2\pi\sqrt{\alpha'})^2}\left[\frac{ | I^{(1)} |}{T_2^{(1)}}(2N^{(1)}+1)
\pm \left( \frac{| I^{(2)}|}{T_2^{(2)} } - \frac{ |I^{(3)}|}{T_2^{(3)}} \right)\right]
\label{spect5a}
\end{eqnarray}
that is equal to Eq. (\ref{M259}). On the other hand, when they have the same sign, we get:
\begin{eqnarray}
m^2_n=\frac{2\pi }{(2\pi\sqrt{\alpha'})^2}\left[\frac{ | I^{(1)} |}{T_2^{(1)}}(2N^{(1)}+1)
\pm \left( \frac{| I^{(2)}|}{T_2^{(2)} } + \frac{ |I^{(3)}|}{T_2^{(3)}} \right)\right]
\label{spect5ab}
\end{eqnarray}
that is equal to Eq. (\ref{M259b}). In conclusion, one obtains the string theory spectrum
in the field theory limit.

The Kaluza-Klein spectrum of the fermionic states is obtained by
expanding the six dimensional fermionic field in terms of its
internal wave-functions associated to the compact directions of the
first torus. These latter are determined by solving the eigenvalue
equation of the two dimensional Dirac-equation \cite{0404229}:
\begin{eqnarray}
\Gamma^iD_i \eta_n(x^i)=m_n \eta_n
\end{eqnarray}
Squaring  the Dirac equation and
using the six dimension $\Gamma$-matrices  given in the Appendix \ref{nota}, one
easily arrives to write\cite{0810.5509}:
\begin{eqnarray}
\left(-D_iD^i\mathbb{I}_2+\frac{2 \pi I^{(1)}}{(2\pi \sqrt{\alpha'})^2T_2^{(1)}}\tau^3\right)\eta_n=m_n^2\eta_n \,\, .
\end{eqnarray}
Using Eq. (\ref{DDF}) we see  that we have again two towers  of Kaluza-Klein fermionic states with  masses:
\begin{eqnarray}
m_n^2= \frac{2\pi |I^{(1)}|}{(2\pi\sqrt{\alpha'})^2T_2^{(1)}}(2N^{(1)}+1)\pm\frac{2\pi  I^{(1)}}{(2\pi \sqrt{\alpha'})^2T_2^{(1)}}  \,\, .
\end{eqnarray}
{From} the previous expression we see that, once we fix the sign of the first
Chern-class, the spectrum contains
 two towers of fermionic Kaluza-Klein states having opposite  internal two-dimensional
chirality. Being the six-dimensional spinors
also chiral spinors, we conclude that the two sets of states have also
opposite four dimensional chirality.
This is exactly what happens in string theory and the previous expression
coincides with the string result given in eq. (\ref{spectM2}). This agreement is another check on the correctness of the interaction term given in eq. (\ref{59cd}).

\section{Conclusions and outlook}
\label{conclu}

In this paper we have constructed the six-dimensional Lagrangian for the open strings D5/D9
interacting with the gauge multiplets living on the world-volumes of respectively the D5 and D9 branes. We have shown with an explicit calculation that, when we treat the fields of the gauge multiplet living on the D9 branes as six-dimensional,  this Lagrangian is ${\cal{N}}=1$ supersymmetric.
In order to check various couplings of this Lagrangian and to extend it to the case in which the fields of the gauge multiplet living on the D9 branes are treated as ten-dimensional, we construct the vertex operators corresponding to the massless states of our system D5/D9 and we use them to compute three-point couplings. In this way
we obtain a six-dimensional Lagrangian where the open strings D5/D5 and D5/D9 are treated as six-dimensional, while the open strings D9/D9 are treated as ten-dimensional.
Finally, by introducing background magnetizations in the extra dimensions both on D5 and D9 branes, we compute the spectrum of the open strings attached with one end-point to a D5 and the other end-point to a D9 brane where the two D branes have different magnetizations.
The spectrum of states completely agrees with that obtained in string theory in the field
theory limit $(\alpha' \rightarrow 0)$.

Actually, string theory requires that all states, including the open strings D5/D9 and D5/D5,
should be treated as ten-dimensional. We hope to be able to come back to this point in a future publication.

\vspace{2mm}
\noindent {\large \textbf{Acknowledgements} }

\vspace{2mm}We thank  E.  Brynjolfsson, L.  Thorlacius, A.  Wijns and T. Zingg for an early participation in this paper and M. Berg, P. Camara, G. Ferretti, A. Van Proeyen and G. Villadoro for useful discussions. F. P. thanks the Simons Center for Geometry
and Physics at Stony Brook and the MIT Center for Theoretical Physics for hospitality during different stages of this work.

\appendix

\section{Notations}
\label{nota}

We start writing the various Lagrangians in four dimensions
in terms of Weyl spinors $\psi^{\alpha}$ and
$\chi^{\dot{\alpha}}$
 that are lowered or raised by means of the antisymmetric $\epsilon$ tensor given by:
\begin{eqnarray}
\epsilon^{\alpha \beta} = \left( \begin{array}{cc} 0 &1 \\
-1   & 0 \end{array} \right)~~;~~\epsilon_{\alpha \beta} = \left( \begin{array}{cc} 0 & -1 \\
 1   & 0 \end{array} \right)~~;~~\epsilon^{\alpha \gamma} \epsilon_{\gamma \beta} =
 \delta^{\alpha}_{\,\,\, \beta}
\label{epsi93}
\end{eqnarray}
Analogous formulas are valid for the dotted indices.

We can define a Dirac spinor and a Majorana spinor in terms of two Weyl spinors  as follows:
\begin{eqnarray}
\Psi_{D} = \left( \begin{array}{c}     \psi_{\alpha}  \\
{\bar{\chi}}^{ \dot{\alpha}}   \end{array} \right)~~;~~~
\Psi_{M} = \left( \begin{array}{c}     \psi_{\alpha}  \\
{\bar{\psi}}^{ \dot{\alpha}}   \end{array} \right)
\label{DM}
\end{eqnarray}
In the case of a Dirac spinor the two Weyl spinors are different, while in the case of Majorana spinor the two are one the complex conjugate of the other.

The four-dimensional $\gamma$-matrices are given by:
\begin{eqnarray}
\gamma^{\mu} = \left(  \begin{array}{cc} 0 & \sigma^{\mu} \\
{\bar{\sigma}}^{\mu} & 0 \end{array} \right)~~;~~\sigma^{\mu} = (1, \tau^i )~~;~~
{\bar{\sigma}}^{\mu} = (1, - \tau^i )
\label{ga}
\end{eqnarray}
in terms of the two-dimensional Pauli matrices $\tau^i$ and the identity matrix.
The previous notations are the same as those in Ref.~\cite{Wessbagger}
except that in their case
$\sigma^0 =-1$, while in our case $\sigma^0 =1$.

In order to go to  from four to six dimensions we   introduce the following representation
of the six-dimensional gamma matrices  in terms of
the four-dimensional ones:
\begin{eqnarray}
\Gamma^{\mu} = \gamma^{\mu} \otimes 1~~;~~
\Gamma^4 = \gamma_5 \otimes i \tau^1~~;~~
\Gamma^5 = \gamma_5 \otimes i \tau^2~~;~~
\Gamma_7 \equiv \gamma_5 \otimes \tau^3
\label{Gamma}
\end{eqnarray}
where
\begin{eqnarray}
\Gamma_7 \equiv \Gamma^0 \Gamma^1 \Gamma^2 \Gamma^3 \Gamma^4 \Gamma^5~~;~~
\gamma_5 \equiv -i \gamma^0 \gamma^1 \gamma^2 \gamma^3 ~~;~~
\{ \Gamma^{{\hat{\mu}}}, \Gamma^{\hat{\nu}}\} = - 2 \eta^{\hat{\mu} \hat{\nu}}
\label{Gamma7}
\end{eqnarray}
where we are using, in any space-time number of dimensions, the mostly plus metric $(-, +, \cdots, +)$.
A chiral spinor in six dimensions satisfies the condition:
\begin{eqnarray}
(1 - b \Gamma_7) \zeta =0
\label{6chi}
\end{eqnarray}
where $\Gamma_7$ is defined in Eq. (\ref{Gamma}) and $\zeta$
has eight  components.
The most general solution of the previous equation is given by:
\begin{eqnarray}
\zeta = \left( \begin{array}{c}  \left( \frac{1 + b \gamma_5}{2} \right)  \beta \\
 a \left( \frac{1 - b \gamma_5}{2} \right) \beta \end{array} \right)
= \zeta_1 \otimes  \left( \begin{array}{c} 1 \\ 0 \end{array} \right) +
\zeta_2 \otimes \left( \begin{array}{c} 0 \\ 1 \end{array} \right)
 \label{geneso1}
 \end{eqnarray}
 where
 \begin{eqnarray}
\zeta_1 =\frac{1 + b \gamma_5}{2}   \beta~~~;~~~
\zeta_2 =a\frac{1 - b \gamma_5}{2}   \beta
\label{zeta12}
\end{eqnarray}
and
 \begin{eqnarray}
&&  {\bar{\zeta}}\equiv \zeta^{\dagger} \Gamma^0  =  \left[\zeta_{1}^{\dagger}
\otimes  \left( \begin{array}{cc} 1 & 0 \end{array} \right) +  \zeta_{2}^{\dagger}
 \otimes
 \left( \begin{array}{cc} 0 & 1 \end{array} \right) \right] \left(\gamma^0 \otimes 1\right)
  \nonumber \\
 &&=\zeta_{1}^{\dagger} \gamma^0 \otimes
  \left( \begin{array}{cc} 1 & 0 \end{array} \right) +  \zeta_{2}^{\dagger} \gamma^0
  \otimes
  \left( \begin{array}{cc} 0 & 1 \end{array} \right)= \left( \begin{array}{cc}   {\bar{\beta}}  \left( \frac{1 - b \gamma_5}{2} \right) &
a {\bar{\beta}}  \left( \frac{1 + b \gamma_5}{2} \right) \end{array} \right)
\label{geneso}
\end{eqnarray}
where $ a= \pm 1$ and $b= \pm 1$.  This means that in going from four  to six
dimensions the six-dimensional Lagrangian contains the parameters $a$ and $b$.
Using the previous formulas and the
six-dimensional $\Gamma$-matrices given in Eqs. (\ref{Gamma}) we can compute:
\begin{eqnarray}
{\bar{\zeta}} \, \Gamma^{\hat{\mu}} D_{\hat{\mu}} \zeta = {\bar{\beta}}
\left[  \gamma^{\mu} D_{\mu} + i a \gamma_5 D_4 - ab D_5  \right] \beta  \,\, .
\label{DIRAC}
\end{eqnarray}
In the final part of this Appendix we discuss the embedding of the six-dimensional
Dirac matrices in the ten dimensional ones that will be useful for uplifting the fermions from six to ten dimensions.

It is convenient to use the following  representation of the ten-dimensional Dirac matrices
in terms of the four dimensional $\gamma$-matrices and the Pauli matrices $\vec{\tau}$:
\begin{eqnarray}
\Gamma^0=\gamma^0_{4}\otimes \mathbb{I}\otimes\mathbb{I}\otimes \mathbb{I}~~&;&~~\Gamma^5=\gamma^5\otimes i\tau^2\otimes\mathbb{I}\otimes \mathbb{I}\nonumber\\
\Gamma^1=\gamma^1_{4}\otimes \mathbb{I}\otimes\mathbb{I}\otimes \mathbb{I}~~&;&~~\Gamma^6=\gamma^5\otimes \tau^3\otimes - i\tau^1\otimes \mathbb{I}\nonumber\\
\Gamma^2=\gamma^2_{4}\otimes \mathbb{I}\otimes\mathbb{I}\otimes \mathbb{I}~~&;&~~\Gamma^7=\gamma^5\otimes \tau^3\otimes i\tau^2\otimes \mathbb{I}\nonumber\\
\Gamma^3=\gamma^3_{4}\otimes \mathbb{I}\otimes\mathbb{I}\otimes \mathbb{I}~~&;&~~\Gamma^8=\gamma^5\otimes \tau^3\otimes\tau^3\otimes i\tau^1\nonumber\\
\Gamma^4=\gamma^5\otimes i\tau^1\otimes\mathbb{I}\otimes \mathbb{I}~~&;&~~\Gamma^9=\gamma^5\otimes \tau^3\otimes\tau^3\otimes i\tau^2  \,\, .
\label{10dim}
\end{eqnarray}
They  satisfy the ten-dimensional   Clifford algebra:
\begin{eqnarray}
\left\{\Gamma^M,\,\Gamma^N\right\}=-2\eta^{MN}\qquad \eta^{MN}=\left(-,\,+\dots\right)  \,\, .
\label{Clifford}
\end{eqnarray}
They can be also conveniently expressed in terms of the six dimensional $\Gamma$-matrices
 introduced in Eq. (\ref{Gamma}) as follows:
\begin{eqnarray}
\Gamma^0=\Gamma^0_{(6)} \otimes\mathbb{I}\otimes \mathbb{I}~~
&;&~~\Gamma^5=\Gamma^5_{(6)} \otimes\mathbb{I}\otimes \mathbb{I}\nonumber\\
\Gamma^1=\Gamma^1_{(6)}\otimes\mathbb{I}\otimes \mathbb{I}~~&;&~~\Gamma^6=\Gamma^7_{(6)} \otimes - i\tau^1\otimes \mathbb{I}\nonumber\\
\Gamma^2=\Gamma^2_{(6)}  \otimes\mathbb{I}\otimes \mathbb{I}~~&;&~~\Gamma^7=\Gamma^7_{(6)} \otimes i\tau^2\otimes \mathbb{I}\nonumber\\
\Gamma^3=\Gamma^3_{(6)}\otimes\mathbb{I}\otimes \mathbb{I}~~&;&~~\Gamma^8=\Gamma^7_{(6)} \otimes\tau^3\otimes i\tau^1\nonumber\\
\Gamma^4=\Gamma^4_{(6)} \otimes\mathbb{I}\otimes \mathbb{I}~~&;&~~\Gamma^9=\Gamma^7_{(6)} \otimes\tau^3\otimes i\tau^2
\label{10dim6}
\end{eqnarray}
where we have added an index $6$ to the six dimensional $\Gamma$-matrices to distinguish them from the ten dimensional ones and
\begin{eqnarray}
\Gamma^{11}  \equiv
\Gamma^0\dots \Gamma^9= \Gamma^7_{6} \otimes\tau^3\otimes\tau^3 ~~~;~~~
\Gamma^{7}_{6} = \Gamma^{0}_{(6)}  \Gamma^{1}_{(6)}  \Gamma^{2}_{(6)}  \Gamma^{3}_{6}  \Gamma^{4}_{(6)}  \Gamma^{5}_{(6)} .
\label{ga11ga7}
\end{eqnarray}
The spinor $\lambda$ in ${\cal N}=1$ ten-dimensional SYM theory is  Majorana-Weyl and therefore satisfies the Weyl-condition:
 \begin{eqnarray}
 \left(\frac{1+ b\Gamma^{11}}{2}\right)\lambda=\left(\frac{1+b\Gamma^{7}_{6} \otimes\tau^3\otimes\tau^3 }{2}\right)\lambda_6 \otimes\lambda_4=0  \,\, .
 \label{Weyl10}
\end{eqnarray}
The Majorana condition, instead, is encoded in the constraint:
\begin{eqnarray}
\xi=B^{-1}\xi^*
\end{eqnarray}
where $B$ allows one  to connect the ten-dimensional Dirac matrices with  their complex conjugates~\cite{GSO}:
\begin{eqnarray}
B\,\Gamma^M \,B^{-1}={\Gamma^M}^* \nonumber
\end{eqnarray}
which implies $B\,B^*=\mathbb{I}$.
In our ten-dimensional representation $B$ is given by:
\begin{eqnarray}
B=c\Gamma^2\,\Gamma^4\,\Gamma^6\, \Gamma^8 \nonumber
\end{eqnarray}
where  $c$ is  an arbitrary constant such that $|c|=1$.  In this paper we take it to be $\pm 1$.
 In the six-dimensional representation of the Dirac matrices $B$ becomes:
\begin{eqnarray}
B = c \Gamma^{2}_{(6)} \Gamma^{4}_{(6)} \otimes i \tau_2  \otimes  \tau_1
\equiv B_6 \otimes i \tau_2 \otimes  \tau_1
\label{B6}
\end{eqnarray}
with $B_6$ being the operator that allows to connect the six-dimensional Dirac matrices with their conjugates. It satifies:  $B_6\,B_6^*=-\mathbb{I}$.

We have now all the ingredients to decompose the ten-dimensional Majorana-Weyl spinor in terms of the six-dimensional components. It is given by the following expression:
\begin{eqnarray}
\lambda&=& \Psi\otimes \left(\begin{array}{c}1\\0\end{array}\right)\otimes \left(\begin{array}{c}1\\0\end{array}\right)-   (B_6\Psi)^*\otimes \left(\begin{array}{c}0\\1\end{array}\right)\otimes \left(\begin{array}{c}0\\1\end{array}\right) \nonumber\\
&+&\left.  \Lambda^2\otimes \left(\begin{array}{c} 1\\0\end{array}\right)\otimes \left(\begin{array}{c}0\\1\end{array}\right) +  c^*\, a \Lambda^1\otimes \left(\begin{array}{c}0\\1\end{array}\right)\otimes \left(\begin{array}{c}1\\0\end{array}\right)\right]
\label{ans1}
\end{eqnarray}
where $a= \pm 1$ to be fixed later. It satisfies the Weyl condition in Eq. (\ref{Weyl10}).
if $\Psi$ and $\Lambda^i$ satisfy the two conditions:
\begin{eqnarray}
( 1 + b \Gamma^{7}_{6} ) \Psi = ( 1 - b \Gamma^{7}_{6} ) \Lambda^i =0  \,\, .
\label{weylcondi}
\end{eqnarray}
It satisfies the Majorana condition $\lambda^{*} = B \lambda$ if  the six-dimensional spinor $\Lambda^i$ satisfies the relation:
\begin{eqnarray}
B_6 \Lambda^i = -  ac  \epsilon_{ij} (\Lambda^j)^*~~~;~~~\epsilon_{12} =-1
\label{Majoco}
\end{eqnarray}
The spinor  ${\bar{\lambda}}$ is then given by:
\begin{eqnarray}
\bar{\lambda}&=& \bar{\Psi}\otimes \left(1,\,0\right)\otimes \left(1,\,0\right) -  (B_6\Psi)^T\Gamma^0_{(6)}\otimes \left(0,\,1\right)\otimes \left(0,\,1 \right) \nonumber\\
&+&   \bar{\Lambda}_2\otimes \left( 1,\,0\right)\otimes \left(0,\,1\right) + c\, a \bar{\Lambda}_1\otimes \left(0,\,1\right)\otimes \left(1,\,0\right)
\label{ans2}
\end{eqnarray}
where, in analogy with two-dimensional spinors (see Eq. (\ref{lamb})), we are following the convention $ (\Lambda^{i})^{*} = \Lambda^{*}_{i} $ and lowering and rising indices through the antisymmetric tensor $\epsilon$. The same rule is adopted for the field $\Psi^\alpha$, in detail $(\Psi^\alpha)^*=\Psi^*_\alpha$ which determines:
\begin{eqnarray}
\bar{\Psi}^\alpha=\left(\begin{array}{c}
                             0\\-\bar{\Psi}\end{array}\right).\label{105}
\end{eqnarray}
The ten dimensional spinors $\lambda$ and ${\bar{\lambda}}$ will be used in
Appendix \ref{Uplift99} for the uplift from six to ten dimensions.

\section{Magnetized D branes}
\label{magne}

In Eq. (\ref{M2b})  of Sect. \ref{2} we have given the mass spectrum of  the open
strings attached to two magnetized D branes in terms of the number operators.
It   is valid in the entire interval $- \frac{1}{2} \leq \nu_r \leq \frac{1}{2}$, but the explicit expression of the number operators  is in general different for positive and negative values of  $\nu_r$
$(r=1,2,3)$.   In  Sect. \ref{2} we have given their form when $0  \leq \nu_r \leq \frac{1}{2}$.  Here
we extend it to the case $- \frac{1}{2} \leq \nu_r \leq  0$.

The interval  $- \frac{1}{2} \leq \nu_r \leq \frac{1}{2}$ is the natural range for the fermions in the
NS  sector and therefore, in this case, the number operator given in Eq. (\ref{NPsi}), for $ x =1$
is valid in the entire interval  $- \frac{1}{2} \leq \nu_r \leq \frac{1}{2}$. This range is also natural
according to the first  equation in (\ref{tana}). For  the bosonic coordinate and for
the R sector we have to change the number operators in Eqs. (\ref{NcalZ})  and (\ref{NPsi})
for $x=0$ as follows:
\begin{eqnarray}
N^{\cal{Z}}_{r} =  \sum_{n=0}^{\infty} \left[
(n+1+ \nu_r ) a^{(r)\dagger}_{n+1+\nu_r}  a_{n+1+ \nu_r}^{(r)} +
 (n- \nu_r ) \bar{a}^{(r) \dagger}_{n- \nu_r} \bar{a}_{n- \nu_r}^{(r)}
 \right]
\label{NcalZb}
\end{eqnarray}
and
\begin{eqnarray}
N^{{\Psi}}_{r} =  \sum_{n=0}^{\infty} \left[
(n+1+ \nu_r ) \Psi^{(r)\dagger}_{n+1+\nu_r}  \Psi_{n+1+ \nu_r}^{(r)} +
 (n- \nu_r ) \bar{\Psi}^{(r) \dagger}_{n- \nu_r} \bar{\Psi}_{n- \nu_r}^{(r)}
 \right]  \,\, .
\label{NPsibx}
\end{eqnarray}
We see that, in going from the interval $0 \leq \nu_r \leq \frac{1}{2}$ to the interval
$- \frac{1}{2} \leq \nu_r \leq 0$, we exchange the role of  the oscillators $a^{(r)}$
and ${\bar{a}}^{(r)}$ and $\Psi^{(r)}$ and ${\bar{\Psi}}^{(r)}$ in the R sector.
This can be seen more directly by writing Eqs. (\ref{NcalZ}) and (\ref{NcalZb}) as follows:
\begin{eqnarray}
N^{\cal{Z}}_{r} =
\sum_{n=0}^{\infty} \left[ (n + |\nu_r | ) a^{(r)\dagger}_{n+ |\nu_r |}  a_{n+ |\nu_r |}^{(r)} +
 (n+1 - |\nu_r | ) \bar{a}^{(r) \dagger}_{n+1 - |\nu_r |} \bar{a}_{n+1- |\nu_r |}^{(r)}\right]
\label{NcalZbcx}
\end{eqnarray}
and
\begin{eqnarray}
N^{\cal{Z}}_{r} =  \sum_{n=0}^{\infty} \left[
(n+1- |\nu_r | ) a^{(r)\dagger}_{n+1- |\nu_r |}  a_{n+1- |\nu_r| }^{(r)} +
 (n+  |\nu_r | ) \bar{a}^{(r) \dagger}_{n+  |\nu_r |} \bar{a}_{n+ | \nu_r |}^{(r)}
 \right]
\label{NcalZbd}
\end{eqnarray}
and analogously in the R sector. In the NS  sector, as well as for the four-dimensional
non-compact directions, instead nothing changes.

In the second part of this Appendix we give the number operators for
the second and third torus for the open strings D5/D9.  They are
given by:
 \begin{eqnarray}
N^{\cal{Z}}_{r=2,3}&=&
\sum_{n=0}^{\infty}\left[  (n + \frac{1}{2} - |\nu_{r}^{(9)}| )
a^{(r)\dagger}_{n+  \frac{1}{2} -  |\nu_{r}^{(9)}|}  a_{n+  \frac{1}{2} - |\nu_{r}^{(9)}|}^{(r) }\right.\nonumber\\
&&\left. ~+
 (n +  \frac{1}{2} + |\nu_{r}^{(9)}| ) \bar{a}^{(r) \dagger}_{n +  \frac{1}{2} +
|\nu_{r}^{(9)} |} \bar{a}_{n+  \frac{1}{2} + |\nu_{r}^{(9)}|}^{(r)} \right]
\label{NcalZf1}
\end{eqnarray}
for the bosonic coordinate,
\begin{eqnarray}
(N^{{\Psi}}_{r=2,3})_{R} &=&\sum_{n=0}^{\infty}\left[ (n + \frac{1}{2} - |\nu_{r}^{(9)}| )
{\Psi}^{(r)\dagger}_{n+  \frac{1}{2} - | \nu_{r}^{(9)}|}
\Psi_{n+  \frac{1}{2} - |\nu_{r}^{(9)}|}^{(r)}\right.\nonumber\\
&&\left. ~+
 (n +  \frac{1}{2} + |\nu_{r}^{(9)}| ) \bar{\Psi}^{(r)\dagger}_{n +  \frac{1}{2} +
|\nu_{r}^{(9)}|} \bar{\Psi}_{n+  \frac{1}{2} + |\nu_{r}^{(9)}|}^{(r)} \right]
\label{NcalZf2}
\end{eqnarray}
for the fermionic coordinate in the  Ramond sector and
\begin{eqnarray}
(N^{\Psi}_{r=2,3})_{NS} &=&
\sum_{n=\frac{1}{2} }^\infty \left[ \!(n+  \frac{1}{2} - |\nu_{r}^{(9)}|)
\Psi^{\,\dagger \,(r)}_{n+  \frac{1}{2} - |\nu_{r}^{(9)}|}\!\! \Psi^{\, (r)}_{n+  \frac{1}{2} - |\nu_{r}^{(9)}|}
\right.\nonumber\\
&&\left.~+      (n-  \frac{1}{2} + |\nu_{r}^{(9)}|)
{\overline \Psi}^{\,\dagger \,(r)}_{n-  \frac{1}{2} + |\nu_{r}^{(9)}|}{
\overline \Psi}^{\, (r)}_{n-  \frac{1}{2} + |\nu_{r}^{(9)}|}\right]\label{NPsif1}
\end{eqnarray}
for the fermionic coordinate in the NS sector.

The previous expressions (\ref{NcalZf1}), (\ref{NcalZf2}) and (\ref{NPsif1})
are valid if $0 \leq \nu_{2,3}^{(9)} <\frac{1}{2}$. If instead $ -\frac{1}{2}\leq \nu_{2,3}^{(9)} <0$
we have the following expressions:
\begin{eqnarray*}
N^{\cal{Z}}_{r=2,3} &=&
\sum_{n=0}^{\infty}\left[  (n + \frac{1}{2} + | \nu_{r}^{(9)}| )
a^{(r)\dagger}_{n+  \frac{1}{2} + | \nu_{r}^{(9)}|}  a_{n+  \frac{1}{2} + |\nu_{r}^{(9)}|}^{(r)}\right.\nonumber\\
&&\left.~ +
 (n +  \frac{1}{2} - |\nu_{r}^{(9)}| ) \bar{a}^{(r) \dagger}_{n +  \frac{1}{2} -
|\nu_{r}^{(9)} |} \bar{a}_{n+  \frac{1}{2} - |\nu_{r}^{(9)}|}^{(r)} \right]
\label{NcalZf1a}
\end{eqnarray*}
and for the Ramond sector:
\begin{eqnarray*}
(N^{{\Psi}}_{r=2,3})_{R} &=&
\sum_{n=0}^{\infty}\left[  (n + \frac{1}{2} + |\nu_{r}^{(9)}| )
{\Psi}^{(r)\dagger}_{n+  \frac{1}{2} + | \nu_{r}^{(9)}|}
\Psi_{n-  \frac{1}{2} + |\nu_{r}^{(9)}|}^{(r)}\right.\nonumber\\
&&\left.~ +
 (n +  \frac{1}{2} - |\nu_{r}^{(9)}| ) \bar{\Psi}^{(r)\dagger}_{n +  \frac{1}{2} -
|\nu_{r}^{(9)}|} \bar{\Psi}_{n+  \frac{1}{2} -| \nu_{r}^{(9)}|}^{(r)} \right]
\label{NcalZf2a}
\end{eqnarray*}
while for the NS sector we have:
\begin{eqnarray*}
(N^{\Psi}_{r=2,3})_{NS} &=&
\sum_{n=\frac{1}{2} }^\infty \left[ (n-  \frac{1}{2} + |\nu_{r}^{(9)}|)
\Psi^{\,\dagger \,(r)}_{n-  \frac{1}{2} + |\nu_{r}^{(9)}|} \Psi^{\, (r)}_{n-  \frac{1}{2} + |\nu_{r}^{(9)}|}\right.\nonumber\\
&&\left. ~+      (n + \frac{1}{2} - |\nu_{r}^{(9)}|)
{\overline \Psi}^{\,\dagger \,(r)}_{n+  \frac{1}{2} -  |\nu_{r}^{(9)}|}{
\overline \Psi}^{\, (r)}_{n+  \frac{1}{2} - |\nu_{r}^{(9)}|}\right]
\label{NPsif1a}
\end{eqnarray*}

\section{Uplift from four to six for the strings 55}
\label{uplift55}

We start from the Lagrangian of ${\cal{N}}=4$ super Yang-Mills in four dimensions given in the
${\cal{N}}=1$ superfield formalism by:
\begin{eqnarray}
&&\!\!\!\!\!\!\! L_4 = 2  {  \rm Tr }   \left[  \int d^2  \theta d^2 { \bar{ \theta}}  \sum_{i=1}^{3}
{ \bar{\Phi}}_i { \rm e}^{2 g V}  \Phi_i   +  i  \sqrt{2} g    \left( \int d^2  \theta
 \Phi_1 [ \Phi_2 ,  \Phi_3 ] +  \int d^2   {\bar{\theta}}
 { \bar{\Phi}}_1 [ { \bar{\Phi}}_2 ,  { \bar{\Phi}}_3 ]  \right) \right]
 \nonumber \\
 && \!\!\!+2   {\rm Tr} \left( \frac{1}{4} \left[ \int d^2 \theta \,W^{\alpha} W_{\alpha} +
 \int d^2 {\bar{\theta}} \, {\bar{W}}_{\dot{\alpha}} {\bar{W}}^{\dot{\alpha}}
 \right] \right)
\label{supfi}
\end{eqnarray}
where $V$ is the vector superfield in the Wess-Zumino gauge:
\begin{eqnarray}
V (x , \theta, {\bar{\theta}})  & = & - \theta \sigma^{\mu} \bar{\theta} A_{\mu} + i {\theta}^{2} \bar{\theta} \bar{\lambda} - i {\bar{\theta}}^{2} \theta \lambda
+ \frac{1}{2} {\theta}^{2} {\bar{\theta}}^{2} D \,\, ,
\label{V}
\end{eqnarray}
$W_{\alpha}$ is its superfield strength:
 \begin{eqnarray}
W_{\alpha} (x, \theta) = - i \lambda_{\alpha} + \left[ \delta^{~\beta}_{\alpha} D - i \left( \sigma^{\mu \nu} \right)_{\alpha}^{~\beta} F_{\mu \nu} \right] \theta_{\beta} + \theta^{2} \sigma^{\mu}_{\alpha \dot{\alpha} } D_{\mu} \bar{\lambda}^{\dot{\alpha}}
\label{W}
\end{eqnarray}
and  the  $\Phi_i$'s  are three chiral superfields given by:
\begin{eqnarray}
\Phi_i (x , \theta) = A_i (x) + \sqrt{2}\,\, \theta^{\alpha} \psi_{i \,\alpha} (x)  + \theta^2 F_i (x) \,\, .
\label{chir}
\end{eqnarray}
Any of the previous fields that we denote  collectively by $\phi$  is a matrix:
\begin{eqnarray}
\phi \equiv \phi^{A} ( T^{A} )^{a}_{\,\,b}~~~;~~{\rm Tr} ( T^A T^B ) = \frac{1}{2} \delta_{AB}~~~;~~~~A,B= 1 \dots N^2
\label{ma63}
\end{eqnarray}
 where $T^A$ are the matrices of $U(N)$ in the fundamental representation.

In terms of the component fields the previous Lagrangian is equal to:
\begin{eqnarray}
L_{4}\! &=& \! \,\,\, 2  {\rm Tr} \left[  - \frac{1}{4} F^{\mu\nu}  F_{\mu \nu}  - i {\bar{\lambda}}
{\bar{\sigma}}^{\mu} D_{\mu} \lambda
+ \frac{1}{2}   \sum_{c=1}^{3}   \left(    ( D_{c+5} )^2  +  i  g D_{c+5}
\eta_{cmn} [A^m , A^n ]  \right)  \right.
\nonumber \\
&& + \,  \sum_{i=1}^{3} \left( - {( D^{\mu}A_i )^{\dagger}}  ( D_{\mu} A_i )
- i {\bar{\psi}}_i {\bar{\sigma}}^{\mu} D_{\mu} \psi_i  \right) + \bar{F}_2  F_2  +
\bar{F}_3  F_3 - g D_{5}  [A_1 , {\bar{A}}_1]
 \nonumber \\
&&  +\,  i \,  \sqrt{2}   g  \left( F_2 [ A_3 , A_1] + F_3 [ A_1 , A_2]  +
{\bar{F}}_2 [ {\bar{A}}_3 , {\bar{A}}_1] +
{\bar{F}}_3 [ {\bar{A}}_1 , {\bar{A}}_2]    \right)   \nonumber \\
&& + \,  i  \, \sqrt{2} g
\left( \psi_{1}^{\alpha} [\psi_{3\alpha}, A_2] +
\psi_{1}^{\alpha} [A_3 ,  \psi_{2\alpha}]
+ \psi_{3}^{\alpha} [ \psi_{2\alpha}, A_1  ]
\right. \nonumber \\
&& +   \left.
{\bar{\psi}}_{1 {\dot{\alpha}}} [{\bar{\psi}}_{3}^{\dot{\alpha}}, {\bar{A}}_2]  +
{\bar{\psi}}_{1{\dot{\alpha}}} [{\bar{A}}_3 ,  {\bar{\psi}}_{2}^{\dot{\alpha}}]   +
{\bar{\psi}}_{3 {\dot{\alpha}}} [ {\bar{\psi}}_{2 }^{\dot{\alpha}}, {\bar{A}}_1  ]     \right)
\nonumber \\
&&  -  \left.
 \sqrt{2} ig \sum_{i=1}^{3}   \left( \psi^{\alpha }_{i}  [ \lambda_{\alpha}  ,   \bar{A}_i ] +
 {\bar{\psi}}_{i{\dot{\alpha}}}
[ {\bar{\lambda}}^{\dot{\alpha}}, A_i ]  \right)
 \right]
\label{N4b}
\end{eqnarray}
where we have introduced a triplet of auxiliary fields $D_{c+5}$ with $c=1,2,3$:
\begin{eqnarray}
D_8 \equiv -D~~~;~~~i F_1 = \frac{D_6 - i D_7}{\sqrt{2}}
\label{D345}
\end{eqnarray}
the real fields $A_m  (m=6,7,8,9)$ given by:
\begin{eqnarray}
A_2 \equiv \frac{A_6 + i A_7}{\sqrt{2}}~~;~~A_3 \equiv \frac{A_8 + i A_9}{\sqrt{2}}
\label{A2A3A4}
\end{eqnarray}
and the {}'t Hooft symbols:
\begin{eqnarray}
\eta_{c mn} &=& \epsilon_{c mn9} + \delta_{c m} \delta_{n9} - \delta_{c n} \delta_{m9}
\nonumber \\
{\bar{\eta}}_{c mn} &=& \epsilon_{c mn9} - \delta_{c m} \delta_{n9} + \delta_{c n} \delta_{m9}
\label{eta}
\end{eqnarray}
where $m,n =6,7,8,9$,  $c=6,7,8$ and $\epsilon_{6789}=1$.

In the following we split the previous Lagrangian in a part $L_1$, corresponding to
${\cal{N}}=2$ super Yang-Mills,  and in a part  $L_2$, corresponding to its interaction with an  hypermultiplet in the adjoint,
we write them in a formalism where the $SU(2)$ R-invariance is manifest and finally we
uplift them to six dimensions.

The Lagrangian corresponding to ${\cal{N}}=2$ super Yang-Mills is given by :
\begin{eqnarray}
L_{1} &=& 2  {\rm Tr} \left[
- \frac{1}{4} F^{\mu \nu}  F_{\mu \nu}  - i {\bar{\lambda}}
{\bar{\sigma}}^{\mu} D_{\mu} \lambda
+ \frac{1}{2} \sum_{c=1}^{3} {\cal{D}}_{c+5}^{2}  -
\frac{1}{2} \sum_{i=4}^{5} {( D^{\mu}A_i )^\dagger}  ( D_{\mu} A_i )
\right.\nonumber \\
&+& \left. \frac{g^2}{2}  [A_4 , {{A}}_5]^2   -i {\bar{\psi}}_1
{\bar{\sigma}}^{\mu} D_{\mu} \psi_1
 - \sqrt{2} ig    \left( \psi^{\alpha }_{1}  [ \lambda_{\alpha}  ,   \bar{A}_1 ] +
 {\bar{\psi}}_{1{\dot{\alpha}}}
[ {\bar{\lambda}}^{\dot{\alpha}}, A_1 ]  \right)
   \right]
\label{sN=2b}
\end{eqnarray}
where we have redefined  one of  the three auxiliary fields:
\begin{eqnarray}
{\cal{D}}_{c+5} = D_{c+5} ~~for~ c=1,2~~~~;~~~~{\cal{D}}_8 = D_8 - g [\phi, \phi^{*} ]
\label{DcalD}
\end{eqnarray}
and we have introduced  the following fields $A_4$ and $A_5$:
\begin{eqnarray}
A_1 \equiv \phi= a \frac{ A_4 - ib A_5}{\sqrt{2}}~~~;~~~{\bar{A}}_1 \equiv \phi^{*}
= a\frac{ A_4  +ibA_5}{\sqrt{2}}
\label{A1barA1}
\end{eqnarray}
where we have written the complex scalar field $A_1$ belonging to
the first chiral multiplet  in terms of the two real scalar fields $A_4$ and $A_5$  that in the uplift from four to six dimensions will provide the extra two components of the six-dimensional gauge field.  The presence of the phases  $a = \pm 1$ and $b = \pm 1$ will be become clear later on.

The Lagrangian corresponding to the interaction with an hypermultiplet is given by:
\begin{eqnarray}
L_{2}\!\! &=&\!\! 2  {\rm Tr} \left[  \frac{i}{2} g \sum_{c=1}^{3} {\cal{D}}_{c+5}
\eta_{cmn} [A^m , A^n ]
+  \sum_{i=2}^{3}\! \left( - {( D^{\mu}A_i )^{\dagger}} ( D_{\mu} A_i )
- i {\bar{\psi}}_i {\bar{\sigma}}^{\mu} D_{\mu} \psi_i   +\bar{F}_i  F_i \right)
  \right.
 \nonumber \\
&-& \frac{b}{2} g^2 [A_4 , A_5]  \eta_{8mn} [A^m , A^n ]
 \nonumber \\
&+& i \sqrt{2}  g   \left( F_2 [ A_3 , A_1] + F_3 [ A_1 , A_2]  +
{\bar{F}}_2 [ {\bar{A}}_3 , {\bar{A}}_1] +
{\bar{F}}_3 [ {\bar{A}}_1 , {\bar{A}}_2]    \right)   \nonumber \\
&+&  i \sqrt{2} g
\left( \psi_{1}^{\alpha} [\psi_{3\alpha}, A_2] +
\psi_{1}^{\alpha} [A_3 ,  \psi_{2\alpha}]
+ \psi_{3}^{\alpha} [ \psi_{2\alpha}, A_1  ]
\right.  \nonumber \\
&+&  \left.
{\bar{\psi}}_{1 {\dot{\alpha}}} [{\bar{\psi}}_{3}^{\dot{\alpha}}, {\bar{A}}_2]  +
{\bar{\psi}}_{1{\dot{\alpha}}} [{\bar{A}}_3 ,  {\bar{\psi}}_{2}^{\dot{\alpha}}]   +
{\bar{\psi}}_{3 {\dot{\alpha}}} [ {\bar{\psi}}_{2 }^{\dot{\alpha}}, {\bar{A}}_1  ]     \right)
\nonumber \\
&-&  \left.
 \sqrt{2} ig \sum_{i=2}^{3}   \left( \psi^{\alpha }_{i}  [ \lambda_{\alpha}  ,   \bar{A}_i ] +
 {\bar{\psi}}_{i{\dot{\alpha}}}
[ {\bar{\lambda}}^{\dot{\alpha}}, A_i ]  \right)
\right]
\label{inte}
\end{eqnarray}
Let us now write the two Lagrangians in a way that the $SU(2)$ R-symmetry is manifest.
Let us introduce the following $SU(2)$ doublets \cite{DMNP}:
\begin{eqnarray}
\lambda_{\alpha}^{i} = \left( \begin{array}{c} - \psi_{1\alpha} \\
\lambda_{\alpha} \end{array} \right)~;~ \lambda_{\alpha i} =
\left( \begin{array}{c} - \lambda_{\alpha} \\
- \psi_{1\alpha} \end{array} \right)~~;~~{\bar{\lambda}}^{\dot{\alpha}}_{i} =
\left( \begin{array}{c} - {\bar{\psi}}^{\dot{\alpha}}_{1} \\
{\bar{\lambda}}^{\dot{\alpha}} \end{array} \right)~;~
{\bar{\lambda}}^{\dot{\alpha i}} =
\left( \begin{array}{c}  {\bar{\lambda}}^{\dot{\alpha}} \\
{\bar{\psi}}^{\dot{\alpha}} _{1}\end{array} \right)
\label{lamb}
\end{eqnarray}
In order to show the consistency of the previous equations we have to use the following convention:
\begin{eqnarray}
( \lambda_{\alpha}^{i} )^{*} \equiv {\bar{\lambda}}_{\dot{\alpha} i} =
\epsilon_{ij} {\bar{\lambda}}_{\dot{\alpha}}^{j}~;~
{\bar{\lambda}}^{\dot{\alpha}}_{i} =
\epsilon_{ij}  \epsilon^{\dot{\alpha} \dot{\beta}} {\bar{\lambda}}_{\dot{\beta} }^{j}~;~
\lambda^{i}_{\alpha} = \epsilon^{ij} \lambda_{\alpha j}
\label{conve}
\end{eqnarray}
where
\begin{eqnarray}
\epsilon^{ij} = \left( \begin{array}{cc}   0 & 1 \\
-1 & 0 \end{array}  \right)~;~\epsilon_{ij}   = \left( \begin{array}{cc}   0 & -1 \\
1 & 0 \end{array}  \right)
\label{epsil}
\end{eqnarray}
It is easy to see that, using the first  equation in Eq. (\ref{conve}), one can
show that the last two equations in (\ref{lamb}) are consistent. The consistency of
the first two equations in Eq. (\ref{lamb}) can be shown using the last equation in
(\ref{conve}).

Using the previous formulas we can rewrite Eq. (\ref{sN=2b}) as follows
\begin{eqnarray}
L_{1} &=& 2  {\rm Tr} \left[
- \frac{1}{4} F^{\mu \nu}  F_{\mu \nu}   -
\frac{1}{2} \sum_{i=4}^{5} {( D^{\mu}A_i )^\dagger}  ( D_{\mu} A_i ) +
\frac{g^2}{2}  [A_4 , {{A}}_5]^2
\right.\nonumber \\
&-& \left.
 i {\bar{\lambda}}_{i}
{\bar{\sigma}}^{\mu} D_{\mu} \lambda^i
+ \frac{1}{2} \sum_{c=1}^{3} {\cal{D}}_{c+5}^{2}  + i \frac{\sqrt{2} g}{2}
 \left( \lambda^{\alpha }_{i}  [ \lambda_{\alpha}^{i}   ,   \bar{A}_1 ] +
 {\bar{\lambda}}_{ {\dot{\alpha}} i}
[ {\bar{\lambda}}^{\dot{\alpha} i}, A_1 ]  \right)
   \right]
\label{sN=2bc}
\end{eqnarray}
where now the invariance under $SU(2)$ is manifest.

Introducing the following $SU(2)$ doublets:
\begin{eqnarray}
Z^i=\left( \begin{array}{c}
          \bar{A}_3\\
          A_2
          \end{array}\right) ~~;~~\bar{Z}_i\equiv(Z^i)^{\dagger} =
          \left( A_3 ,\bar{A}_2\right)
          \label{bdoub}
\end{eqnarray}
together with
\begin{eqnarray}
Z_i\equiv \left( \epsilon_{12}Z^2= - A_2,\epsilon_{21}Z^1=\bar{A}_3\right)~~;~~\bar{Z}^i\equiv \left(\begin{array}{c}
\epsilon^{12}\bar{Z}_2= \bar{A}_2\\
\epsilon^{21}\bar{Z}_1= -A_3
\end{array}\right)
\label{Zboub4}
\end{eqnarray}
and eliminating the auxiliary fields $F_2$ and $F_3$,
we can write Eq. (\ref{inte}) as follows:
\begin{eqnarray}
L_{2} &=& 2  {\rm Tr} \left[
 -  {( D^{\mu} Z^i )^{\dagger}} ( D_{\mu} Z^i )
- i {\bar{\psi}}_2 {\bar{\sigma}}^{\mu} D_{\mu} \psi_2   -
i {\bar{\psi}}_3 {\bar{\sigma}}^{\mu} D_{\mu} \psi_3
  \right. \nonumber \\
&+&
g\bar{Z}_i\sum_{c=1}^{3} ({\cal D}_{c+5}\tau^{c})^i_{\,j}Z^j-g Z_i
\sum_{c=1}^{3}({\cal D}_{c+5}\tau^{c})^i_{\,j}\bar{Z}^j
 \nonumber \\
&+& g^2 \left(
 [{{A}}_4 , {\bar{Z}}_i ] [ {{A}}_4 , Z^i] +   [{{A}}_5 , {\bar{Z}}_i ] [ {{A}}_5 , Z^i]
\right) \nonumber \\
&+&
i \sqrt{2} g\left( \lambda^{\alpha i} [\psi_{3\alpha},\,Z^j]\epsilon_{ij}
-\lambda^\alpha_i[\psi_{2\alpha},\,\bar{Z}_j]\epsilon^{ij}-\bar{\lambda}_{\dot{\alpha}i}[\bar{\psi}_3^{\dot{\alpha}},\,\bar{Z}_j]\epsilon^{ij}
-\bar{\lambda}_{\dot{\alpha}}^i[\bar{\psi_2}^{\dot{\alpha}},\,Z^j]\epsilon_{ij}\right.
\nonumber\\
&&\left. \left.+ \,\psi_{3}^{\alpha} [ \psi_{2\alpha}, A_1  ] +
{\bar{\psi}}_{3 {\dot{\alpha}}} [ {\bar{\psi}}_{2 }^{\dot{\alpha}}, {\bar{A}}_1  ]\right)\right]
\label{intene}
\end{eqnarray}
that is manifestly $SU(2)$ invariant. In deriving the previous equation we have used the following identity:
\begin{eqnarray*}
{\rm Tr} \left[  \frac{i}{2} g
\sum_{c=1}^{3} {{\cal D}}_{c+5}   \eta_{(c+5)mn} [A^m , A^n ]\right]\!\!=\!
{\rm Tr}\! \left[
 g\bar{Z}_i \sum_{c=1}^{3} ({\cal D}_{c+5}\tau^{c})^i_{\,j}Z^j
 -g Z_i \sum_{c=1}^{3} ({\cal D}_{c+5}\tau^{c})^i_{\,j}\bar{Z}^j\right]
 \label{DDD}
\end{eqnarray*}
that follows from Eqs. (\ref{A2A3A4}) and (\ref{bdoub}).
The next step is to write both Eq.s. (\ref{sN=2bc}) and (\ref{intene}) using
four-dimensional Dirac fermions:
\begin{eqnarray}
\xi \equiv \left( \begin{array}{c} \psi_{2\alpha} \\
          { \bar{\psi}}_{3}^{ \dot{\alpha}} \end{array} \right)~~~;~~~{\bar{\xi}} \equiv
         \left( \begin{array}{cc} {{\psi}}_{3}^{\alpha}  &
          {\bar{\psi}}_{2\dot{\alpha}} \end{array} \right)
~~~;~~~
\eta^i = \left( \begin{array}{c}
             \lambda_{\alpha}^{i} \\
             {\bar{\lambda}}^{\dot{\alpha} i} \end{array} \right) ~;~
{\bar{\eta}}_{i} = \left( \begin{array}{cc}
- \lambda^{\alpha}_{i} & {\bar{\lambda}}_{\dot{\alpha} i} \end{array}\right)
\label{fermi77}
\end{eqnarray}
We get:
\begin{eqnarray}
L_{1} &=& 2  {\rm Tr} \left[
- \frac{1}{4} F_{\mu \nu}  F_{\mu \nu}   -
\frac{1}{2} \sum_{i=4}^{5} {( D_{\mu}A_i )}  ( D_{\mu} A_i ) +
\frac{g^2}{2}  [A_4 , {{A}}_5]^2
\right.\nonumber \\
&+& \left.
 \frac{1}{2} \sum_{c=3}^{5} {\cal{D}}_{c+5}^{2}  - \frac{i}{2} {\bar{\eta}}_i \left[  \gamma^{\mu} D_{\mu}
 \eta^i - a \gamma_5 g [ A_4, \eta^i] - ab ig [A_5, \eta^i]  \right]
\right]
\label{sN=2bcx}
\end{eqnarray}
and
\begin{eqnarray}
L_{2} &=& 2  {\rm Tr} \left[
 + g^2 \left(
 [{{A}}_4 , {\bar{Z}}_i ] [ {{A}}_4 , Z^i] +   [{{A}}_5 , {\bar{Z}}_i ] [ {{A}}_5 , Z^i]
\right)+ g\bar{Z}_i\sum_{c=1}^{3} ({\cal D}_{c+5}\tau^{c})^i_{\,j}Z^j \right.
\nonumber \\
&-&g Z_i
\sum_{c=1}^{3}({\cal D}_{c+5}\tau^{c})^i_{\,j}\bar{Z}^j -
i  {\bar{\xi}} \left(   \gamma^{\mu} D_{\mu} \xi +  g a b   [ {{A}}_5, \xi] +
\gamma_5 g a [{{A}}_4 , \xi] \right)
 \nonumber \\
&-&\left.
 \sqrt{2} g \left( [ {\bar{\xi}} , Z^j] \epsilon_{ij}
\gamma_5 \eta^i + {\bar{\eta}}_i \gamma_5 [ \xi, {\bar{Z}}_j ] \epsilon^{ij}   \right)
\right]
\label{intenex}
\end{eqnarray}
We are now ready to uplift the two previous Lagrangians  to six dimensions
by introducing the two
six-dimensional chiral  spinors:
\begin{eqnarray}
( 1 - b \Gamma_7) \Lambda^i =0 ~~;~~\Lambda^i = \left( \begin{array}{c}
\frac{1+ b\gamma_5}{2} \eta^i \\
a \frac{1- b\gamma_5}{2} \eta^i  \end{array} \right)~~;~~ {\bar{\Lambda}}_i =
\left(  \begin{array}{cc} {\bar{\eta}}_i \frac{1-b\gamma_5}{2}  &
 a {\bar{\eta}}_i \frac{1+b\gamma_5}{2}     \end{array}
  \right)
\label{Lambda}
\end{eqnarray}
and
\begin{eqnarray}
( 1 - b_{\Psi}  \Gamma_7) \Psi =0 ~~;~~
\Psi = \left( \begin{array}{c} \frac{1 + b_{\Psi} \gamma_5}{2} \xi \\
a_{\Psi} \frac{ 1 - b_{\Psi} \gamma_5}{2} \xi \end{array} \right)~~;~~~
{\bar{\Psi}} = \left( \begin{array}{cc}  {\bar{\xi}}  \frac{ 1 - b_{\Psi} \gamma_5}{2} &
a_{\Psi} {\bar{\xi}}  \frac{ 1 + b_{\Psi} \gamma_5}{2} \end{array} \right)
\label{PSI}
\end{eqnarray}
Eq. (\ref{sN=2bcx}) becomes:
\begin{eqnarray}
L_{{\cal{N}}=2\,sYM} &=& 2  {\rm Tr} \left[
- \frac{1}{4} F_{ {\hat{\mu}}  {\hat{\nu}} }  F_{ {\hat{\mu}}   {\hat{\nu}}}
 + \frac{1}{2} \sum_{c=3}^{5} {\cal{D}}_{c+5}^{2} -
 \frac{i}{2}   {\bar{\Lambda}}_i \Gamma^{\hat{\mu}}D_{\hat{\mu}} \Lambda^i
\right]
\label{superN=2}
\end{eqnarray}
while Eq. (\ref{intenex}) becomes:
\begin{eqnarray}
L_{2} &=& 2  {\rm Tr} \left[   - {( D_{\hat{\mu}} {{Z}}^i )}^{\dagger} ( D_{\hat{\mu}} Z^i )
+ g\bar{Z}_i\sum_{c=1}^{3} ({\cal D}_{c+5}\tau^{c})^i_{\,j}Z^j-g Z_i
\sum_{c=1}^{3}({\cal D}_{c+5}\tau^{c})^i_{\,j}\bar{Z}^j
   \right. \nonumber \\
 &-&\left.  i   {\bar{\Psi}}\Gamma^{\hat{\mu}} D_{\hat{\mu}} \Psi  +
  i \sqrt{2} g b \left( [{\bar{\Psi}}, Z^j] \epsilon_{ij} \Lambda^i -
  {\bar{\Lambda}}_i [ \Psi , {\bar{Z}}_j] \epsilon^{ij} \right) \right]
\label{final43x}
\end{eqnarray}
where ${\hat{\mu}}= (\mu , 4, 5)$. The previous equations have been obtained by imposing
that:
\begin{eqnarray}
a_{\Psi} = - a ~~~~;~~~~b_{\Psi} = - b
\label{ababpsi}
\end{eqnarray}
This means that the two chiral spinors in six dimensions have opposite chirality.

\section{Uplift from four to six for the strings 59}
\label{uplift59}

An ${\cal{N}}=2$  hypermultiplet consists of two ${\cal{N}}=1$ chiral superfields
$R_1$ and $R_2$ coupled
to the gauge superfield $V$   and with a very special superpotential. Its
Lagrangian is  given by:
\begin{eqnarray}
L_{hyper} &=&  \int d^2 \theta \,  d^2 {\bar{\theta}} \left[ {{R}}_{1a}^{*}
\left(  {\rm e}^{2 gV} \right)^{a}_{\,\,\,b}
R_{1}^{b} + {{R}}_{2a}  \left(  {\rm e}^{-2 gV} \right)^{a}_{\,\,\,b} R_{2}^{*b}
 \right]  \nonumber \\
 && +  \sqrt{2} g \left[  \int d^2 \theta R_{2a} \Phi^{a}_{\,\,\,b} R_{1}^{b} + \int d^2 {\bar{\theta}}
 R_{1a}^{*} (\Phi^{*})^{a}_{\,\,\,b} R_{2}^{*b} \right]
\label{hyp}
\end{eqnarray}
where
\begin{eqnarray}
R_{i} = z_i + \sqrt{2} \theta \psi_i + \theta^2 G_i~~;~~
\Phi= \phi + \sqrt{2} \theta \psi + \theta^2 F
\label{sf}
\end{eqnarray}
Here * stands for a complex conjugation and the superfield $\Phi$ has to be identified with $\Phi_1$ of the previous section. In terms of the component fields  the first two terms of Eq. (\ref{hyp}) are equal to:
\begin{eqnarray}
&&\int d^2 \theta \,  d^2 {\bar{\theta}} \left[ {{R}}_{1a}^{*}
\left(  {\rm e}^{2 gV} \right)^{a}_{\,\,\,b}
R_{1}^{b} + {{R}}_{2a}  \left(  {\rm e}^{-2 gV} \right)^{a}_{\,\,\,b} R_{2}^{{*}b}
 \right]
\nonumber \\
 &&=
 \left[ - (D_{\mu} z_1)^{*}_{a} (D_{\mu} z_{1})^{a} + G_{1}^{a}  G_{1a}^{*} -
 i {\bar{\psi}}_{1 a} {\bar{\sigma}}^{\mu} (D_{\mu} \psi_1 )^b \right]
 \nonumber \\
 &&+\left[  - (D_{\mu} z_2)_{a} (D_{\mu} z_{2})^{*a} + G_{2a}  G_{2}^{*a}  -
 i {\bar{\psi}}_{2}^{ a} {\bar{\sigma}}^{\mu} (D_{\mu} \psi_2 )^a  \right] +
 g \left( z_{1a}^{*} D^{a}_{\,\,\,b} z_{1}^{b}
-  z_{2a}  D^{a}_{\,\,\,b}  z_{2}^{{*b}} \right) \nonumber \\
&&+ i \sqrt{2} g \left( z_{1a}^{*} \lambda^{a}_{\,\,\,b}  \psi_{1}^{b} +  z_{2a}
({\bar{\lambda}})^{a}_{\,\,\,b}  {\bar{\psi}}_{2}^{b}   -
{\bar{\psi}}_{1a}  ({\bar{\lambda}})^{a}_{\,\,\,b}  z_{1}^{b}    - {{\psi}}_{2a}
( {{\lambda}})^{a}_{\,\,\,b}  z_{2}^{*b} \right)
\label{kinte}
\end{eqnarray}
where the covariant derivatives are given by
\begin{eqnarray}
(D_{\mu} z_{1} )^{a} = \partial_{\mu} z_{1}^{a} + i g (A_{\mu})^{a}_{\,\,\,b} z_{1}^{b}~~;~~
(D_{\mu} z_{2}^{*} )^{a} = (\partial_{\mu} z_{2}^{*}) ^{a} + i g (A_{\mu})^{a}_{\,\,\,b} (z_{2}^{*})^{b}
\label{Dmu}
\end{eqnarray}
\begin{eqnarray}
(D_{\mu} z_{1}^{*}  )_{a} = (\partial_{\mu} z_{1}^{*})_{a} - i g (A_{\mu}^{*})_{a}^{\,\,\,b}
( z_{1}^{*})_{b} = (\partial_{\mu} z_{1}^{*})_{a} - i g
(A_{\mu} )_{\,\,\,a}^{b}  (z_{1}^{*})_{b}
\label{z1}
\end{eqnarray}
\begin{eqnarray}
(D_{\mu} z_{2}   )_{a} = (\partial_{\mu} z_{2})_{a} - i g
(A_{\mu} )_{\,\,\,a}^{b}  (z_{2})_{b}
\label{z2}
\end{eqnarray}
\begin{eqnarray}
D_{\mu} \psi_{1}^{a} = \partial_{\mu} \psi_{1}^{a}  +i g (A_{\mu})^{a}_{\,\,\,b} \psi_{1}^{b}~~;~~
D_{\mu} \psi_{2a} = \partial_{\mu} \psi_{2a}  - i g (A_{\mu})^{b}_{\,\,\,a} \psi_{2b}
\label{psi12}
\end{eqnarray}
with
\begin{eqnarray}
(A_{\mu})^{a}_{\,\,\,b} \equiv (T^A)^{a}_{\,\,\,b} A^{A}_{\mu}~~;'~~
(\phi)^{a}_{\,\,\,b} \equiv (T^A)^{a}_{\,\,\,b} \phi^{A}~~;~~
(\phi^{*})^{a}_{\,\,\,b} \equiv (T^A)^{a}_{\,\,\,b} (\phi^{*})^{A}
\label{Aphib}
\end{eqnarray}
We can now write the superpotential in terms of the component fields. We get:
\begin{eqnarray}
&&\sqrt{2} g \left[  \int d^2 \theta R_{2a} \Phi^{a}_{\,\,\,b} R_{1}^{b} + \int d^2 {\bar{\theta}}
 R_{1a}^{*} (\Phi^{*})^{a}_{\,\,\,b} R_{2}^{*b} \right]  = \sqrt{2} g \left[ z_{2a} F^{a}_{\,\,\,b} z_{1}^{b}
 + z_{1a}^{*}( F^{*})^{a}_{\,\,\,b} z_{2}^{*b}\right.
 \nonumber \\
&&     + G_{2a} \phi^{a}_{\,\,\,b} z_{1}^{b} +
 z_{2a} \phi^{a}_{\,\,\,b} G_{1}^{b}   +  G_{1a}^{*} (\phi^{*})^{a}_{\,\,\,b}  z_{2}^{*b} +
 z_{1a}^{*} (\phi^{*})^{a}_{\,\,\,b} G_{2}^{*b} -\psi_{2a} \phi^{a}_{\,\,\,b} \psi_{1}^{b }
 \nonumber \\
 && \left.  -
 {\bar{\psi}}_{1a}  (\phi^{*})^{a}_{\,\,\,b} {\bar{\psi}}_{2}^{b}   -
 z_{2a} ( \psi)^{a}_{\,\,\,b} \psi_{1}^{b} -
 z_{1a}^{*} ({\bar{\psi}})^{a}_{\,\,\,b} {\bar{\psi}}_{2}^{b} - \psi_{2a} ( \psi)^{a}_{\,\,\,b}
 z_{1}^{b}
 - {\bar{\psi}}_{1a}  ({\bar{\psi}})^{a}_{\,\,\,b} z_{2}^{*b}   \right]
\label{supo}
\end{eqnarray}
We want to rewrite the previous Lagrangian in a way where the $SU(2)$
R symmetry  is manifest.
We introduce the following $SU(2)$ doublets \cite{DMNP}:
\begin{eqnarray}
{\bar{w}}_i \equiv  \left(  \begin{array}{cc} {\bar{w}}_1 &  {\bar{w}}_2 \end{array} \right) =
 \left(  \begin{array}{cc} - i z_2  &  z_{1}^{*} \end{array} \right) ~~;~~
 w^i \equiv \left( \begin{array}{c} w^1 \\  w^2 \end{array} \right) =
- \left( \begin{array}{c} i z_{2}^{*}  \\  z_1 \end{array} \right)
\label{wz}
\end{eqnarray}
The index $i$ is raised or lowered by the antisymmetric tensor $\epsilon$
given in Eq. (\ref{epsil}).
The complex variables $w, \bar{w}$ are not independent, but satisfy the relation:
\begin{eqnarray}
 \epsilon^{ij } {\bar{w}}_{j} = ( w_{i} )^{*}
\label{rel74}
\end{eqnarray}
The previous relations imply:
\begin{eqnarray}
{\bar{w}}^{\dot{2}} = - {\bar{w}}_{\dot{1}} =   ( w_{\dot{2}} )^{*} =  ( w^{\dot{1}} )^{*} ~~;~~
{\bar{w}}^{\dot{1}} =  {\bar{w}}_{\dot{2}} =   ( w_{\dot{1}} )^{*} = - ( w^{\dot{2}} )^{*}
\label{rel76}
\end{eqnarray}
In terms of the previous doublets for the scalar fields and of those connected with the gauginos
given in Eq. (\ref{lamb}), the total Lagrangian, that is the sum of Eqs. (\ref{kinte})
and   (\ref{supo}),
 can be written in the following equivalent forms:
\begin{eqnarray}
&&L_{hyper}=   (D_{\mu} {\bar{w}})_{i a} (D^{\mu} w)^{ia} -
g {\bar{w}}_{ia} \sum_{c=1}^{3} (\tau^c)^{i}_{\,\,\,j} {{D}}_{c+5}   w^{aj}
\nonumber \\
&&- i {\bar{\psi}}_{1a} {\bar{\sigma}}^{\mu} (D_{\mu} \psi_{1})^a -
 i {\bar{\psi}}_{2}^{a} {\bar{\sigma}}^{\mu} (D_{\mu} \psi_{2})_a
  - \sqrt{2}g \left( \psi_{2a} \phi^{a}_{\,\,\,b} \psi_{1}^{b }
  +  {\bar{\psi}}_{1a} ( \phi^{*})^{a}_{\,\,\,b} {\bar{\psi}}_{2}^{b} \right)
\nonumber \\
&&+
\sqrt{2} g  \left[ i {\bar{w}}_{i} \lambda^{i\alpha } \psi_{1\alpha}  -
{\bar{w}}_{i} {\bar{\lambda}}^{i}_{\dot{\alpha}} {\bar{\psi}}_{2}^{\dot{\alpha}}
 - \psi_{2}^{\alpha} \lambda_{\alpha i} w^{i} +i  {\bar{\psi}}_{1\dot{\alpha}}
 {\bar{\lambda}}^{\dot{\alpha}}_{i} w^i \right]  - {\bar{G}}_{i} G^i
\nonumber \\
&&-
 \sqrt{2} g \left(    {\bar{G}}_{ia}  \left( \begin{array}{cc} ( \phi^{*})^{a}_{\,\,\,b}  & 0 \\
                        0  &  (\phi)^{a}_{\,\,\,b}   \end{array} \right)^{i}_{\,\,\,j}  w^{jb} +
 {\bar{w}}_{ia}
 \left( \begin{array}{cc} (\phi)^{a}_{\,\,\,b} & 0 \\
                                       0 & (\phi^{*})^{a}_{\,\,\,b} \end{array} \right)^{i}_{\,\,\,j}
 G^{jb} \right)
 \label{Lyper}
\end{eqnarray}
The equations of motion for the auxiliary fields $G$ and ${\bar{G}}$ are:
\begin{eqnarray}
&&G^{ia} + \sqrt{2}g \left( \begin{array}{cc} (\phi^{*})^{a}_{\,\,\,b} & 0 \\
                                                                0 & (\phi)^{a}_{\,\,\,b} \end{array} \right)^{i}_{\,\,\,j} w^{jb} =0
\nonumber \\
&&{\bar{G}}_{ia} + \sqrt{2}g {\bar{w}}_{jb} \left( \begin{array}{cc} (\phi)^{b}_{\,\,\,a} & 0 \\
                                                                0 & (\phi^{*} )^{b}_{\,\,\,a} \end{array} \right)^{j}_{\,\,\,i}  =0
\label{GGc}
\end{eqnarray}
Inserting them  in Eq. (\ref{Lyper}) we get:
\begin{eqnarray}
L_{hyper}  &=& \epsilon^{i j} ( D_{\mu} {\bar{w}}_{i })_a
(D_{\mu}  {{w}}_{j})^{a}  -
g {\bar{w}}_{ia} \sum_{c=1}^{3} (\tau^c)^{i}_{\,\,\,j} ({{D}}_{c+2})^{a}_{\,\,\,b}   w^{bj}
\nonumber \\
&-& i {\bar{\psi}}_{1a} {\bar{\sigma}}^{\mu} (D_{\mu} \psi_{1})^a -
 i {\bar{\psi}}_{2}^{a} {\bar{\sigma}}^{\mu} (D_{\mu} \psi_{2})_a
  - \sqrt{2}g \left(  \psi_{2a} \phi^{a}_{\,\,\,b} \psi_{1}^{b }
  +   {\bar{\psi}}_{1a} ( \phi^{*})^{a}_{\,\,\,b} {\bar{\psi}}_{2}^{b} \right)
\nonumber \\
&+&
\sqrt{2} g  \left[ i {\bar{w}}_{i} \lambda^{i\alpha } \psi_{1\alpha}  -
{\bar{w}}_{i} {\bar{\lambda}}^{i}_{\dot{\alpha}} {\bar{\psi}}_{2}^{\dot{\alpha}}
 - \psi_{2}^{\alpha} \lambda_{\alpha i} w^{i} +i  {\bar{\psi}}_{1\dot{\alpha}}
 {\bar{\lambda}}^{\dot{\alpha}}_{i} w^i \right]
\nonumber \\
&+&
g^2 {\bar{w}}_{ia} (\{\phi , \phi^{*} \})^{a}_{\,\,\,b} w^{bi} + g^2  {\bar{w}}_{ia} (\tau^3 )^{i}_{\,\,\,j}
 ([\phi, \phi^* ])^{a}_{\,\,\,b} w^{jb}
\label{Lyperb}
\end{eqnarray}
Using the redefinition of the auxiliary fields given in Eq. (\ref{DcalD}) we can rewrite the previous Lagrangian as follows:
\begin{eqnarray}
{\cal{L}}_{hyper} &=&   (D_{\mu} {\bar{w}})_{i a} (D^{\mu} w)^{ia}-
g {\bar{w}}_{ia} \sum_{c=1}^{3} (\tau^c)^{i}_{\,\,\,j} {\cal{D}}_{c+5}   w^{aj}+
g^2  {\bar{w}}_{ia}(\{ \phi , \phi^{*} \}^{i}_{\,\,\,j})^{a}_{\,\,\,b}  w^{bj}
\nonumber \\
&-& i ({\bar{\psi}}_{1\dot{\alpha}} )_{a}
({\bar{\sigma}}^{\mu} )^{\dot{\alpha} \beta} (D_{\mu}{{\psi}}_{1 \beta})^a
 - i ({\bar{\psi}}_{2 \dot{\gamma}})^{a}
({\bar{\sigma}}^{\mu} )^{\dot{\gamma} \delta} (D_{\mu}
\psi_{2\delta})_a  \nonumber \\
&-&  g  \sqrt{2} \left( \psi^{\alpha}_{2a} \phi^{a}_{\,\,\,b} \psi_{1\alpha}^{b} +
{\bar{\psi}}_{1\dot{\alpha}a} \phi^{*a}_{\,\,\,\,b} {\bar{\psi}}_{2}^{\dot{\alpha}b} \right)
\nonumber \\
&-& \sqrt{2}i g \left[   \left( ({\bar{\psi}}_{1\dot{\alpha}})_a
({\bar{\lambda}}^{\dot{\alpha}i} )^{a}_{\,\,\,b}
+ i (\psi_{2}^{\alpha})_a ( \lambda_{\alpha}^{i})^{a}_{\,\,\,b} \right)  \epsilon_{ij} w^j  \right.
\nonumber \\
&-& \left.
{\bar{w}}_{i} \epsilon^{ij} \left(  (\lambda_{j}^{\alpha})^{a}_{\,\,\,b} (\psi_{1\alpha })^{b} + i
({\bar{\lambda}}_{\dot{\alpha} j })^{a}_{\,\,\,b}
({\bar{\psi}}_{2}^{\dot{\alpha}} )^{b}  \right) \right]
\label{Lfin}
\end{eqnarray}
where
\begin{eqnarray}
(D_{\mu} w^i)^{a}  = \partial_{\mu} w^{ia} + ig (A_{\mu})^{a}_{\,\,\,b} w^{ib}~~;~~
(D_{\mu} {\bar{w}}_i )_a = \partial_{\mu} {\bar{w}}_{ia} -ig (A_{\mu})^{b}_{\,\,\,a}
{\bar{w}}_{i}^{b}
\label{DDe}
\end{eqnarray}
Using the notations given in Appendix \ref{nota} we can write the previous four-dimensional
Lagrangian
in terms of Dirac spinors. One gets:
\begin{eqnarray}
{{L}}_{hyper} &=& \epsilon^{i j} ( D_{ {{\mu}} } {\bar{w}}_{i })_a
(D^{{{\mu}}}  {{w}}_{j})^{a} -
g {\bar{w}}_{ia}  \sum_{c=1}^{3}(\tau^c)^{i}_{\,\,\,j} ({\cal{D}}_{c+5})^{a}_{\,\,\,b}   w^{bj}
+  g^2  {\bar{w}}_{ia}(\{ \phi , \phi^{*} \}^{i}_{\,\,\,j})^{a}_{\,\,\,b}  w^{bj}
\nonumber \\
&-& i {\bar{\psi}}_{a} \left[  \gamma^{\mu} ( {D }_{\mu} )^{a}_{\,\,\,b}
+ \sqrt{2}g \left(  \phi^{a}_{\,\,\,b} \frac{1 +\gamma_5}{2} - (\phi^{*})^{a}_{\,\,\,b}
\frac{1- \gamma_5}{2}    \right)
 \right] \psi^b
\nonumber \\
&-& \sqrt{2} g i \left[ {\bar{\psi}}_{a} \gamma_5
(\eta^{i})^{a}_{\,\,\,b} \epsilon_{ij} (z^j )^{b}  +
(z^{\dagger}_{i} )_{a}
\epsilon^{ij} ({\bar{\eta}}_{j})^{a}_{\,\,\,b} \gamma_5  \psi^b
 \right]
\label{dmnpg}
\end{eqnarray}
where in the first line we have written what is in the first and last line of Eq. (\ref{Lyperb}). The Dirac field $\psi$ is defined in terms of the Weyl fields by:
\begin{eqnarray}
\psi = \left(\begin{array}{c}      \psi_{1 \alpha} \\
i {\bar{\psi}}_{2}^{\dot{\alpha}} \end{array}  \right)
\label{psidi}
\end{eqnarray}
and the covariant derivative is given by:
\begin{eqnarray}
(D_{\mu} \psi)^a = \partial_{\mu}  \psi^a + i g (A_{\mu})^{a}_{\,\,\,b} \psi^b
\label{covderfe}
\end{eqnarray}
It remains now to rewrite the last two lines of Eq. (\ref{dmnpg}) in six-dimensional notations.

This can be done by introducing the following six-dimensional Weyl spinor:
\begin{eqnarray}
(1-b_{\mu} \Gamma_7) \mu =0~;~
\mu = \left(\begin{array}{c}  \frac{1+ b_{\mu} \gamma_5}{2} \psi \\
       a_{\chi} \frac{1- b_{\mu} \gamma_5}{2} \psi \end{array}                  \right)~;~
        \bar{\mu}  = \mu^{\dagger} \Gamma^0 = \left( \begin{array}{cc}
       \bar{\psi}  \frac{1- b_{\mu} \gamma_5}{2}  & a_{\mu} \bar{\psi}  \frac{1+ b_{\mu}
  \gamma_5}{2}
        \end{array} \right)
\label{chila}
\end{eqnarray}
If the following conditions are satisfied
\begin{eqnarray}
a_{\mu} = -a~~~;~~~ b_{\mu} = - b
\label{abchi}
\end{eqnarray}
 we can uplift the previous Lagrangian to the following six-dimensional one:
\begin{eqnarray}
L_{hyper} &=&  \epsilon^{i j} ( D_{ {\hat{\mu}} } {\bar{w}}_{i })_a
(D^{{\hat{\mu}}}  {{w}}_{j})^{a} -
g {\bar{w}}_{ia}  \sum_{c=1}^{3}(\tau^c)^{i}_{\,\,\,j} ({\cal{D}}_{c+5})^{a}_{\,\,\,b}   w^{bj}
\nonumber \\
&-&
i  {\bar{\mu}}_{a} \Gamma^{\hat{\mu}} (D_{\hat{\mu}} \mu)^{a}+
 \sqrt{2} gi b  \left[  {\bar{\mu}}_{a} (\Lambda^i)^{a}_{~b} \epsilon_{ij}w^{jb}    +
{\bar{w}}_{ia} \epsilon^{ij}   ({\bar{\Lambda}}_j)^{a}_{~b} \mu^{b} \right]
\label{Lfi}
\end{eqnarray}
where $\mu$ and $\Lambda$ are six-dimensional Weyl spinors with opposite chirality.

\section{Uplift from six to ten dimensions for the   strings 99} \label{Uplift99}

In this Appendix we uplift the Lagrangian in Eq. (\ref{L99a}) to ten dimensions.

The uplift of the purely bosonic part is easy if we observe that  the term with the double commutator can be written as follows:
\begin{eqnarray}
\frac{\tilde{g}_9^2}{4} \sum_{c=1}^3 \eta_{(c+5)mn} \eta_{(c+5)pq}
[ {\hat{A}}^{m(9)},  {\hat{A}}^{n(9)}]    [ {\hat{A}}^{p(9)},  {\hat{A}}^{q(9)}]  =
\frac{\tilde{g}_{9}^{2}}{4} \sum_{m,n} [ {\hat{A}}_m , {\hat{A}}_{n} ]^2
\label{etaeta}
\end{eqnarray}
that can be uplifted to ten dimensions to become:
\begin{eqnarray}
\frac{\tilde{g}_{9}^{2}}{4} \sum_{m,n} [ {\hat{A}}_m , {\hat{A}}_{n} ]^2  \Longrightarrow
- \frac{1}{4} {\hat{F}}^{mn} {\hat{F}}_{mn}
\label{FmnFmn}
\end{eqnarray}
Then it is easy to see that together with the two other purely bosonic terms gives rise to
the Lagrangian of  pure Yang-Mills theory in ten dimensions:
\begin{eqnarray}
{\cal{L}}_{99} = - \frac{1}{4} F_{MN} F_{MN} +  \,\,\, fermions
\label{L99+}
\end{eqnarray}
In the following we want to uplift the terms with the fermions present in Eq. (\ref{L99a}).

It is easier to start from the ten dimensional expression that is given by:
\begin{eqnarray}
- i {\rm Tr} \left( {\bar{\lambda}} \Gamma^{M} D_M  \lambda  \right)
\label{10ferm}
\end{eqnarray}
and use in it the representation for the ten dimensional spinors given in Eqs. (\ref{ans1})
and (\ref{ans2}).
For $M=0 \dots 5$  we get
\begin{eqnarray}
-2i {\rm Tr} \left( {\bar{\Psi}} \Gamma^{\hat{\mu}} D_{\hat{\mu}} \Psi  + \frac{1}{2}
{\bar{\Lambda}}_i
\Gamma^{\hat{\mu}} D_{\hat{\mu}} \Lambda^i  \right)
\label{05}
\end{eqnarray}
while for the other four components we get respectively:
\begin{eqnarray}
- i {\rm Tr} \left( {\bar{\lambda}} \Gamma^{6} D_6  \lambda  \right) &=&-
2b \left[ c^* a {\bar{\Psi}}D_6 \Lambda^1 - ca {\bar{\Lambda}}_1 D_6 \Psi  \right]
\nonumber \\
- i {\rm Tr} \left( {\bar{\lambda}} \Gamma^{7} D_7  \lambda  \right) &=&
-2bi \left[ c^* a {\bar{\Psi}}D_7 \Lambda^1 +  ca {\bar{\Lambda}}_1 D_7 \Psi  \right]
\nonumber \\
- i {\rm Tr} \left( {\bar{\lambda}} \Gamma^{8} D_8  \lambda  \right) &=&
2b \left[   {\bar{\Psi}}D_8 \Lambda^2 -   {\bar{\Lambda}}_2 D_8 \Psi  \right]
\nonumber \\
- i {\rm Tr} \left( {\bar{\lambda}} \Gamma^{9} D_9  \lambda  \right) &=&
-2i b \left[  {\bar{\Psi}}D_9 \Lambda^2 +  {\bar{\Lambda}}_2 D_9 \Psi  \right]
\label{6789}
\end{eqnarray}
In deriving the previous equations we have used the following relations:
\begin{eqnarray}
&&B_{6}^{T }= - B_6~~:~~B_6 \Gamma^{0}_{(6)} =  \Gamma^{0}_{(6)} B_6~~;~~ (\Gamma^{0}_{(6)})^T =  \Gamma^{0}_{(6)} \nonumber \\
&&\Gamma^{0\dagger}_{(6)} = \Gamma^{0}_{(6)}~~~;~~~\Gamma^{i\dagger}_{(6)} = - \Gamma^{i}_{(6)}~~~;~~~\Gamma^{{\hat{\mu}}\dagger}_{(6)} \Gamma_{(6)}^{0} = \Gamma^{0}_{(6)}
\Gamma_{(6)}^{\hat{\mu}}
\label{relazi78}
\end{eqnarray}
The terms in Eq. (\ref{05}) reproduce the kinetic terms of the fermions in Eq. (\ref{L99a}), while
the four terms in the previous equation reproduce the Yukawa terms in the last line
of Eq. (\ref{L99a}) provided that we make the following identifications:
\begin{eqnarray}
 c^* a \left(A_6 + i A_7 \right) = \sqrt{2} A_2~~~;~~~A_8 + i A_9 = \sqrt{2} A_3
\label{6to10}
\end{eqnarray}
with $ac^* =1$.  $A_2$ and $A_3$ are connected to the variables $Z^i$ and ${\bar{Z}}_i$ through
Eqs. (\ref{bdoub}).

In conclusion, if we start from the following ten dimensional Lagrangian:
\begin{eqnarray}
L_{10} = 2 {\rm Tr} \left(  - \frac{1}{4} F_{MN} F^{MN} - \frac{i}{2} {\bar{\lambda}}
\Gamma^{M} D_M \lambda  \right)
\label{9910}
\end{eqnarray}
we reproduce the six-dimensional Lagrangian in Eq. (\ref{L99a}).

\section{${\cal N}=1$ supersymmetry  transformations}
\label{susyap}

In this appendix we prove that the following six dimensional actions:
\begin{eqnarray}
&&{\cal S}_{g}= 2 \int d^6x {\rm Tr}\left[ -\frac{1}{4}F_{\hat{\mu}\hat{\nu}}^2+\frac{1}{2}\sum_{c=1}^3 {\cal D}_c^2-\frac{i}{2}\bar{\Lambda}_i \Gamma^{\hat{\mu}}D_{\hat{\mu}}\Lambda^i\right]
\label{as1}
\end{eqnarray}
for the gauge sector living on the D5 branes, and
\begin{eqnarray}
{\cal S}_{m}&=& \int d^6x \left[\epsilon^{i j} ( D_{ {\hat{\mu}} }  {\bar{w}}_{i })_{a}
(D^{{\hat{\mu}}}  {{w}}_{j})^{a}-
i  {\bar{\mu}}_{a} \Gamma^{\hat{\mu}} (D_{\hat{\mu}}  \mu)^{a}
-
g {\bar{w}}_{ia}  \sum_{c=1}^{3}(\tau^c)^{i}_{\,\,\,j}
({\hat{{\cal{D}}}}_{c+2})^{a}_{\,\,\,\,\,b}   w^{jb}\right.\nonumber\\
&+&\left.
 \sqrt{2} g i b  \left(  {\bar{\mu}}_{a} ( {{ \Lambda}}^i )^{a}_{~b}
  \epsilon_{ij}(w^j)^{b} +
({\bar{w}}_i)_{a} \epsilon^{ij}   ({{{\bar{{{\Lambda}}}}}}_{j})^{a}_{~b} \mu^{b} \right)\right]
\label{as2}
\end{eqnarray}
for the twisted matter, are invariant under the following ${\cal N}=1$  supersymmetry transformations:
\begin{eqnarray}
&&\delta A^{\hat{\mu}}=\frac{i}{2}\left( \bar{\epsilon}_i \Gamma^{\hat{\mu}} \Lambda^i-\bar{\Lambda}_i \Gamma^{\hat{\mu}} \epsilon^i \right)~~;~~\delta {\cal D}^c=
\frac{1}{2}(\tau^c)^i_{~j} \left( D_{\hat{\mu}}\bar{\Lambda}_i\Gamma^{\hat{\mu}}\epsilon^j+
\bar{\epsilon}_i\Gamma^{\hat{\mu}}D_{\hat{\mu}}\Lambda^j\right)
\nonumber\\
&&\delta \Lambda^i= \frac{1}{2} \Gamma^{{ \hat{\mu}}  {\hat{\nu}} } F_{\hat{\mu}\hat{\nu}}
\,\,
\epsilon^i + i { \cal{D}}^i_{~j}\epsilon^j~~;~~\delta\bar{\Lambda}_i= -\bar{\epsilon}_iF_{\hat{\mu}\hat{\nu}} \frac{1}{2}  \Gamma^{ {\hat{\mu}} {\hat{\nu}}}
- i\bar{\epsilon}_j{ {\cal{D}}}^j_{~i}
\label{as3}
\end{eqnarray}
and:
\begin{eqnarray}
\delta w^{ia}=- \sqrt{2}\,  b\,  \epsilon^{ij} \bar{\epsilon}_j\mu^a~~&;&~~
\delta \bar{w}_{ia}= - \sqrt{2} \,b\, \bar{\mu}_a\epsilon_{ij} {\epsilon}^j\nonumber\\
\delta\mu^a =-i \sqrt{2} \, b\,  \Gamma^{\hat{\mu}}\epsilon^i\epsilon_{ij}(D_{\hat{\mu}} w^j)^a~~&;&\delta\bar{\mu}_a = -i \sqrt{2} \, b \,(D_{\hat{\mu}} \bar{w}_j)_a \epsilon^{ji}\bar{\epsilon}_i\Gamma^{\hat{\mu}}
\label{susy1}
\end{eqnarray}
where ${ \cal{D}}^i_{~j} \equiv {\cal{D}}^c  ( \tau^{c} )^{i}_{\,\,j}$ and $\Gamma^{\hat{\mu} \hat{\nu}}
\equiv \frac{1}{2} [ \Gamma^{\hat{\mu}}, \Gamma^{\hat{\nu}}]$.

Before showing the supersymmetry of the previous actions let us discuss some properties
of the  parameter of the supersymmetry transformation $\epsilon^i$. It  is a chiral fermion
with the same chirality of the gaugino. This property follows from  the requirement that, if the gaugino is a chiral fermion, also its supersymmetry variation must be a chiral fermion.
This means that:
\begin{eqnarray}
0=  \delta \left(\frac{1-b\Gamma_7}{2}\right)\Lambda^i
=
\frac{1}{4}  F_{\hat{\mu}\hat{\nu}}\Gamma^{\hat{\mu}\hat{\nu}}  \left(\frac{1-b\Gamma_7}{2}\right)
\epsilon^i  + i {\cal{D}}^i_{~j}\left(\frac{1- b\Gamma_7}{2}\right)\epsilon^j
\label{susych}
\end{eqnarray}
which implies:
\begin{eqnarray}
\left(\frac{1-b\Gamma_7}{2}\right)
\epsilon^i=0\label{eps}
\end{eqnarray}
 that is the same chirality condition imposed on the  $\Lambda^i$.  For the same reason as before it  must also satisfy   the relation (\ref{Majoco}) as the gaugino. We get in fact:
\begin{eqnarray}
&&0= \delta \left( B_6 \Lambda^i  + ac \epsilon_{ij} ( \Lambda^j)^* \right) =
\frac{1}{2} F_{\hat{\mu}\hat{\nu}} ( \Gamma^{\hat{\mu}\hat{\nu}})^*
\left( B_6 \epsilon^i  + ac \epsilon_{ij} (\epsilon^j)^*  \right) \nonumber \\
&&+ i {\cal{D}}^{i}_{\,\,j} \left( B_6 \epsilon^j
+ ac \epsilon_{jk} (\epsilon^k)^* \right)
\label{as5}
\end{eqnarray}
that implies
\begin{eqnarray}
B_6\epsilon^i=-ac \epsilon_{ij} (\epsilon^j)^*
\label{as6}
\end{eqnarray}
In deriving Eq. (\ref{as5}) we have used the transformation properties of $\Lambda^*$:
\begin{eqnarray}
\delta (\Lambda^i)^*= \frac{1}{2} F_{\hat{\mu}\hat{\nu}}(\Gamma^{\hat{\mu}\hat{\nu}})^*(\epsilon^i)^*- i { {\cal{D}}}^{j}_{~i}(\epsilon^j)^*
\end{eqnarray}
where we have used that ${\cal{D}}^\dag={\cal{D}}$ which implies
$({\cal{D}}^*)^i_{~j}={\cal{D}}^{j}_{~i}$,
the relation:
\begin{eqnarray}
B_6\Gamma^{\hat{\mu}\hat{\nu}}=
(\Gamma^{\hat{\mu}\hat{\nu}})^*B_6
\end{eqnarray}
which is valid because in our representation of the Dirac matrices
$\Gamma^{2,4}$ are purely imaginary, while the other $\Gamma$-matrices are real,
and the identity:
\begin{eqnarray}
(\tau^c)^i_{~j}=-\epsilon_{jm}(\tau^c)^m_{~~l}\epsilon^{li}\label{nr}
\end{eqnarray}
Let us now start by analysing the susy-invariance of the gauge
action in Eq. (\ref{as1}). The invariance under the supersymmetry transformations of the
action in  Eq. (\ref{as1}) without the auxiliary field has been
shown in Ref. \cite{NPB121}~\footnote{See also Section 9 of Ref. \cite{9803026}  for more details.} .
Here we consider only the additional contribution given by the presence of the auxiliary fields.
We get the following extra contributions to the variation of the action in Eq. (\ref{as1}):
\begin{eqnarray}
&& \delta S_{g}^{D} =
2 \int d^6 x  {\rm Tr} \left[  \frac{1}{2}
\partial_{\hat{\mu}} \left( \bar{\Lambda}_i \Gamma^{\hat{\mu}}D^i_{~j} \epsilon^j\right)  - \frac{1}{2}  \left( D_{\hat{\mu}}
\bar{\Lambda}_i\Gamma^{\hat{\mu}}\epsilon^j +
 \bar{\epsilon}_i \Gamma^{\hat{\mu}}D_{\hat{\mu}}\Lambda^j\right){\cal{D}}^i_{~j}
 \right. \nonumber \\
&& \left.   + {\cal{D}}^{c}  \delta {\cal{D}}^c \right]  = 2 \int d^6 x  {\rm Tr} \left[ \frac{1}{2}
\partial_{\hat{\mu}} \left( \bar{\Lambda}_i \Gamma^{\hat{\mu}}D^i_{~j} \epsilon^j\right) \right]
  \nonumber\\
\label{Dte}
\end{eqnarray}
that is a  total derivative. This implies that the action
in Eq. (\ref{as1}) is invariant under the supersymmetry transformations in Eq. (\ref{as3}).
In the last step in Eq. (\ref{Dte}) we have used the supersymmetry transformation of the auxiliary field in Eq. (\ref{as3}).

Let us consider now the matter sector
that  is described by the following Lagrangian:
\begin{eqnarray}
{\cal{L}}_{59}   &=&  \epsilon^{i j} ( D_{ {\hat{\mu}} }  {\bar{w}}_{i })_{a}
(D^{{\hat{\mu}}}  {{w}}_{j})^{a}
-
i  {\bar{\mu}}_{a} \Gamma^{\hat{\mu}} (D_{\hat{\mu}}  \mu)^{a}
-
g {\bar{w}}_{ia}  \sum_{c=1}^{3}(\tau^c)^{i}_{\,\,\,j}
({\hat{{\cal{D}}}}_{c+2})^{a}_{\,\,\,\,\,b}   w^{jb}
\nonumber \\
&+&
 \sqrt{2} g i b  \left[  {\bar{\mu}}_{a} ( {\hat{ \Lambda}}^i )^{a}_{~b}
  \epsilon_{ij}(w^j)^{b} +
({\bar{w}}_i)_{a} \epsilon^{ij}   ({{{\bar{{\hat{\Lambda}}}}}}_{j})^{a}_{~b} \mu^{b} \right]
\label{59a}
\end{eqnarray}
Let us start evaluating  the variation of the various terms present in the previous Lagrangian.
The variation of the kinetic term for the scalars is given by:
\begin{eqnarray}
&&\delta\left[\epsilon^{i j} ( D_{ {\hat{\mu}} }  {\bar{w}}_{i })_{a}
(D^{{\hat{\mu}}}  {{w}}_{j})^{a} \right] =-ig \epsilon^{i j}{\bar{w}}_{ia }\, \delta (A^{\hat{\mu}})^a_{~b}(D^{{\hat{\mu}}}  {{w}}_{j})^{b}+\epsilon^{i j}( D_{ {\hat{\mu}} } \delta {\bar{w}}_{i })_{a}
(D^{{\hat{\mu}}}  {{w}}_{j})^{a}\nonumber\\
&&+i\,g \epsilon^{i j}( D_{ {\hat{\mu}} }  {\bar{w}}_{i })_{a}\delta (A^{\hat{\mu}})^a_{~b}{{w}}_{jb}
+ \epsilon^{i j} ( D_{ {\hat{\mu}} }  {\bar{w}}_{i })_{a}
(D^{{\hat{\mu}}}  \delta{{w}}_{j})^{a} \nonumber\\
&&=\frac{g}{2} {\bar{w}}_{ia }\left( \bar{\epsilon}_j \Gamma^{\hat{\mu}}\Lambda^j-\bar{\Lambda}_j \Gamma^{\hat{\mu}} \epsilon^j\right)^a_{~b}(D_{{\hat{\mu}}}  {{w}}^{i})^{b} -\frac{g}{2} ( D_{ {\hat{\mu}} }  {\bar{w}}_{i })_{a}\left( \bar{\epsilon}_j \Gamma^{\hat{\mu}}\Lambda^j-\bar{\Lambda}_j \Gamma^{\hat{\mu}} \epsilon^j\right)^a_{~b} {{w}}^{ib}\nonumber\\
&& + \sqrt{2} \, b \, (D_{\hat{\mu}}\bar{\mu})_a\epsilon^j \epsilon_{ji}(D^{\hat{\mu}}w^i)^a - \sqrt{2}\,
b \,  (D_{\hat{\mu}} \bar{w}_i )_a \epsilon^{ij}\bar{\epsilon}_j (D^{\hat{\mu}} \mu)^a
\label{vktb}
\end{eqnarray}
while the variation of the kinetic term for the fermions is equal to:
\begin{eqnarray}
&&\delta \left[-i  {\bar{\mu}}_{a} \Gamma^{\hat{\mu}} (D_{\hat{\mu}}  \mu)^{a}\right]=-i \delta{\bar{\mu}}_{a}\Gamma^{\hat{\mu}} (D_{\hat{\mu}}  \mu)^{a}+g {\bar{\mu}}_{a} \Gamma^{\hat{\mu}}\delta (A_{\hat \mu})^a_{~b}\mu^b-i  {\bar{\mu}}_{a} \Gamma^{\hat{\mu}} (D_{\hat{\mu}}  \delta\mu)^{a}\nonumber\\
&&= - \sqrt{2} \, b \, (D_{\hat{\mu}}\bar{w}_j)_a\epsilon^{ji}\bar{\epsilon}_i
\left(\frac{1}{2}\left\{\Gamma^{\hat{\mu}},\,\Gamma^{\hat{\nu}}\right\}
+\frac{1}{2}\left[\Gamma^{\hat{\mu}},\,\Gamma^{\hat{\nu}}\right]\right)(D_{\hat{\nu}}\mu)^a\nonumber\\
&&- \sqrt{2} \, b\,
{\bar{\mu}}_{a} \left(\frac{1}{2}\left\{\Gamma^{\hat{\mu}},\,\Gamma^{\hat{\nu}}\right\}
+\frac{1}{2}\left[\Gamma^{\hat{\mu}},\,\Gamma^{\hat{\nu}}\right]\right)\epsilon^i\epsilon_{ij}(D_{\hat \mu}D_{\hat \nu}w^j)^a
\nonumber\\
&&+\frac{g\,i}{2}{\bar{\mu}}_{a}\Gamma_{\hat{\mu}}\left(\bar{\epsilon}_i \Gamma^{\hat{\mu}}\Lambda^i-\bar{\Lambda}_i \Gamma^{\hat{\mu}} \epsilon^i\right)^a_{~b}\mu^b\nonumber\\
&&= \sqrt{2} \, b\, (D_{\hat{\mu}}\bar{w}_j)_a\epsilon^{ji}\bar{\epsilon}_i(D^{\hat{\mu}}\mu)^a
+\frac{ig\, b}{ \sqrt{2}}\bar{w}_{ja}\epsilon^{ji}\bar{\epsilon}_i \Gamma^{\hat{\mu}\hat{\nu}}(F_{\hat{\mu}\hat{\nu}})^a_{~b}\mu^b- \sqrt{2} \, b\,
(D_{\hat{\mu}}{\bar{\mu}})_{a} \epsilon^i\epsilon_{ij}(D^{\hat \mu}w^j)^a\nonumber\\
&&-\frac{ig \, b }{\sqrt{2}}\bar{\mu}_a \Gamma^{\hat{\mu}\hat{\nu}}\epsilon^i\epsilon_{ij}(F_{\hat{\mu}\hat{\nu}})^a_{~b}w^{jb}+\frac{g\,i}{2}{\bar{\mu}}_{a}\Gamma_{\hat{\mu}}\left(\bar{\epsilon}_i \Gamma^{\hat{\mu}}\Lambda^i-\bar{\Lambda}_i \Gamma^{\hat{\mu}} \epsilon^i\right)^a_{~b}\mu^b \nonumber\\
&&- \sqrt{2} \, b \,  \partial_{\hat{\mu}} \left[ \bar{w}_{ja}\epsilon^{ji}\bar{\epsilon}_i [\Gamma^{\hat{\mu}},\,\Gamma^{\hat{\nu}}](D_{\hat{\nu}} \mu)^a\right]
+ \sqrt{2} \, b\,
 \partial_{\hat{\mu}} \left[ {\bar{\mu}}_a \epsilon^i \epsilon_{ij} (D^{\hat{\mu}} w^j)^a  \right]
\label{vktf}
\end{eqnarray}
where $ [ D_{\hat{\mu}},  D_{\hat{\nu}}] = ig F_{\hat{\mu} \hat{\nu}}$.
Summing the two  previous variations we see that the two last terms in Eq. (\ref{vktb}) cancel
with two equal terms in Eq. (\ref{vktf}) and  we get:
\begin{eqnarray}
&&\delta\left[\epsilon^{i j} ( D_{ {\hat{\mu}} }  {\bar{w}}_{i })_{a}
(D^{{\hat{\mu}}}  {{w}}_{j})^{a}-i  {\bar{\mu}}_{a} \Gamma^{\hat{\mu}} (D_{\hat{\mu}}  \mu)^{a}\right]=\frac{g}{2} {\bar{w}}_{ia }\left( \bar{\epsilon}_j \Gamma^{\hat{\mu}}\Lambda^j-\bar{\Lambda}_j \Gamma^{\hat{\mu}} \epsilon^j\right)^a_{~b}(D_{{\hat{\mu}}}  {{w}}^{i})^{b}\nonumber\\
 &&-\frac{g}{2} ( D_{ {\hat{\mu}} }  {\bar{w}}_{i })_{a}\left( \bar{\epsilon}_j \Gamma^{\hat{\mu}}\Lambda^j -\bar{\Lambda}_j \Gamma^{\hat{\mu}} \epsilon^j\right)^a_{~b} {{w}}^{ib}+
 \frac{ig\, b}{\sqrt{2}}\bar{w}_{ja}\epsilon^{ji}\bar{\epsilon}_i \Gamma^{\hat{\mu}\hat{\nu}}(F_{\hat{\mu}\hat{\nu}})^a_{~b}\mu^b\nonumber\\
&&- \frac{ig\, b}{\sqrt{2}}\bar{\mu}_a \Gamma^{\hat{\mu}\hat{\nu}}\epsilon^i\epsilon_{ij}(F_{\hat{\mu}\hat{\nu}})^a_{~b}w^{jb}+\frac{i\,g}{2}{\bar{\mu}}_{a}\Gamma_{\hat \mu}\left(\bar{\epsilon}_i \Gamma^{\hat{\mu}}\Lambda^i-\bar{\Lambda}_i \Gamma^{\hat{\mu}} \epsilon^i\right)^a_{~b}\mu^b\nonumber\\
&&- \sqrt{2} \, b\,  \partial_{\hat{\mu}} \left[ \bar{w}_{ja}\epsilon^{ji}\bar{\epsilon}_i [\Gamma^{\hat{\mu}},\,\Gamma^{\hat{\nu}}](D_{\hat{\nu}} \mu)^a\right]
+\sqrt{2} \, b \, \partial_{\hat{\mu}} \left[ {\bar{\mu}}_a \epsilon^i \epsilon_{ij} (D^{\hat{\mu}} w^j)^a  \right]
\label{sum1a}
\end{eqnarray}
The variation of the Yukawa couplings  is equal to
\begin{eqnarray}
&&\sqrt{2} g i b\delta  \left[  {\bar{\mu}}_{a} ( {{ \Lambda}}^i )^{a}_{~b}
  \epsilon_{ij}(w^j)^{b} +
({\bar{w}}_i)_{a} \epsilon^{ij}   ({{{\bar{{{\Lambda}}}}}}_{j})^{a}_{~b} \mu^{b} \right]\nonumber\\
&&= 2 \, g \,(D_{\hat{\mu}}\bar{w}_j)_a\epsilon^{ji}\bar{\epsilon}_i\Gamma^{\hat{\mu}}\left({\Lambda}^h\right)^a_{~b}\epsilon_{hk}w^{kb}
+\frac{i gb}{\sqrt{2}} \bar{\mu}_a \Gamma^{\hat{\mu}\hat{\nu}}(F_{\hat{\mu}\hat{\nu}})^a_{~b}\epsilon^i\epsilon_{ij}w^{jb}\nonumber\\
&&- \sqrt{2}g b \bar{\mu}_a ({\cal{D}}^i_{~j})^a_{~b}\epsilon^j\epsilon_{ik}w^{kb}-  2 i g  \bar{\mu}_a\left(  ({ \Lambda}^i)^a_{~b} \bar{\epsilon}_i -
 \epsilon^i(\bar{ \Lambda}_i)^a_{~b}\right)\mu^b\nonumber\\
&&- \frac{igb}{\sqrt{2}}
\bar{w}_{ia}\epsilon^{ij}\bar{\epsilon}_j\Gamma^{\hat{\mu}\hat{\nu}}(F_{\hat{\mu}\hat{\nu}})^a_{~b}\mu^b +\sqrt{2}\, g \, b\,
\bar{w}_{ia}{\epsilon}^{ih}\bar{\epsilon}_j({\cal{D}}^j_{~h})^a_{~b}\mu^b\nonumber\\
&&+ 2 \, g \,  \bar{w}_{ia}\epsilon^{ij}(\bar{  \Lambda}_j)^a_{~b}\Gamma^{\hat{\mu}}\epsilon^k\epsilon_{kh}(D_{\hat{\mu}}w^h)^b
\label{yuka3}
\end{eqnarray}
The first and the last term of the previous expression  can be written as:
\begin{eqnarray}
&& 2\, g \, \left[ (D_{\hat{\mu}}\bar{w}_j)_a\epsilon^{ji}\bar{\epsilon}_i\Gamma^{\hat{\mu}}
\left(\Lambda^h\right)^a_{~b}\epsilon_{hk}w^{kb}
+
\bar{w}_{ja}\epsilon^{ji}(\bar{\Lambda}_i)^a_{~b}\Gamma^{\hat{\mu}}\epsilon^h\epsilon_{hk}
(D_{\hat{\mu}}w^k)^b \right]\nonumber\\
&&= 2g  \left[ (D_{\hat{\mu}}\bar{w}_j)_a\frac{1}{2}\left( \epsilon^{ji}\bar{\epsilon}_i\Gamma^{\hat{\mu}}\left(\Lambda^h\right)^a_{~b}\epsilon_{hk} - \epsilon^{ji}
(\bar{\Lambda}_i)^a_{~b}\Gamma^{\hat{\mu}}\epsilon^h\epsilon_{hk}\right)w^{kb}
\right. \nonumber\\
&&
+\bar{w}_{ja}\frac{1}{2}\left(\epsilon^{ji}(\bar{\Lambda}_i)^a_{~b}\Gamma^{\hat{\mu}}\epsilon^h\epsilon_{hk}
-\epsilon^{ji}\bar{\epsilon}_i\Gamma^{\hat{\mu}}
\left(\Lambda^h\right)^a_{~b}\epsilon_{hk}\right)
(D_{\hat{\mu}}w^k)^b\nonumber\\
&& -
\bar{w}_{ja}\frac{1}{2}\left(\epsilon^{ji}D_{\hat{\mu}}(\bar{\Lambda}_i)^a_{~b}\Gamma^{\hat{\mu}}\epsilon^h\epsilon_{hk}
+\epsilon^{ji}\bar{\epsilon}_i\Gamma^{\hat{\mu}}D_{\hat{\mu}}
\left(\Lambda^h\right)^a_{~b}\epsilon_{hk}\right)w^{kb}\nonumber\\
&& \left. +
\frac{1}{2}  \partial_{\hat{\mu}}\left[\bar{w}_{ja}\epsilon^{jk}\left( \bar{\epsilon}_k \Gamma^{\hat{\mu}}(\Lambda^i)^a_{~b}+(\bar{\Lambda}_k)^a_{~b}\Gamma^{\hat{\mu}}\epsilon^i\right)\epsilon_{ik}w^{kb}\right] \right]
\label{yuka4}
\end{eqnarray}
Using the following identities  which are proved at the end of this section:
\begin{eqnarray}
\bar{\epsilon}_i\Gamma^{\hat{\mu}}\Lambda^h- \bar{\Lambda}_i\Gamma^{\hat{\mu}}\epsilon^h=\delta^h_i\frac{1}{2}\left(\bar{\epsilon}_j\Gamma^{\hat{\mu}}\Lambda^j- \bar{\Lambda}_j\Gamma^{\hat{\mu}}\epsilon^j\right)~~;~~\bar{\epsilon}_i\Gamma^{\hat{\mu}}\Lambda^i+ \bar{\Lambda}_i\Gamma^{\hat{\mu}}\epsilon^i=0\label{wi}
\end{eqnarray}
we can write the variation of the Yukawa couplings as follows:
\begin{eqnarray}
&&\sqrt{2} g i b\delta  \left[  {\bar{\mu}}_{a} ( {\hat{ \Lambda}}^i )^{a}_{~b}
  \epsilon_{ij}(w^j)^{b} +
({\bar{w}}_i)_{a} \epsilon^{ij}   ({{{\bar{{\hat{\Lambda}}}}}}_{j})^{a}_{~b} \mu^{b} \right]
=+\sqrt{2}gb \bar{w}_{ia}{\epsilon}^{ih}\bar{\epsilon}_j({\cal{D}}^j_{~h})^a_{~b}\mu^b
\nonumber\\
&&-\sqrt{2}g b \bar{\mu}_a ({\cal{D}}^i_{~j})^a_{~b}\epsilon^j\epsilon_{ik}w^{kb}+\frac{g }{2} (D_{\hat{\mu}}\bar{w}_j)_a\left(\bar{\epsilon}_i\Gamma^{\hat{\mu}}\left(\Lambda^i\right)^a_{~b}-
(\bar{\Lambda}_i)^a_{~b}\Gamma^{\hat{\mu}}\epsilon^i\right)w^{jb}\nonumber\\
&&-\frac{g }{2} \bar{w}_{ja}\left(\bar{\epsilon}_i\Gamma^{\hat{\mu}}\left(\Lambda^i\right)^a_{~b}-
(\bar{\Lambda}_i)^a_{~b}\Gamma^{\hat{\mu}}\epsilon^i\right)(D_{\hat{\mu}}w^j)^b+ i g b \bar{\mu}_a\left( a_w (\Lambda^i)^a_{~b} \bar{\epsilon}_i -a^*_w \epsilon^i(\bar{\Lambda}_i)^a_{~b}\right)\mu^b\nonumber\\
&&- g  \bar{w}_{ja} \left(\epsilon^{ji}D_{\hat{\mu}}(\bar{\Lambda}_i)^a_{~b}\Gamma^{\hat{\mu}}\epsilon^h\epsilon_{hk}
+\epsilon^{ji}\bar{\epsilon}_i\Gamma^{\hat{\mu}}D_{\hat{\mu}}
\left(\Lambda^h\right)^a_{~b}\epsilon_{hk}\right)w^{hb}\nonumber\\
&&+\frac{igb }{\sqrt{2}} \bar{\mu}_a \Gamma^{\hat{\mu}\hat{\nu}}(F_{\hat{\mu}\hat{\nu}})^a_{~b}\epsilon^i\epsilon_{ij}w^{jb}
- \frac{igb}{\sqrt{2}}   \bar{w}_{ia}\epsilon^{ij}\bar{\epsilon}_j\Gamma^{\hat{\mu}\hat{\nu}}(F_{\hat{\mu}\hat{\nu}})^a_{~b}\mu^b\nonumber\\
&&+  {g}
\partial_{\hat{\mu}}\left[\bar{w}_{ja}\epsilon^{jk}\left( \bar{\epsilon}_k \Gamma^{\hat{\mu}}(\Lambda^i)^a_{~b}+(\bar{\Lambda}_k)^a_{~b}\Gamma^{\hat{\mu}}\epsilon^i\right)\epsilon_{ik}w^{kb}\right]\label{189}
\end{eqnarray}
Summing Eqs.  (\ref{sum1a}) and (\ref{189}), we see that  the first four terms of Eq. (\ref{sum1a})
cancel with the corresponding terms in Eq. (\ref{189}) and  we get:
\begin{eqnarray}
&&\!\!\!\!\!\!\!\!\!\delta\left[\epsilon^{i j} ( D_{ {\hat{\mu}} }  {\bar{w}}_{i })_{a}
(D^{{\hat{\mu}}}  {{w}}_{j})^{a}-i  {\bar{\mu}}_{a} \Gamma^{\hat{\mu}} (D_{\hat{\mu}}  \mu)^{a}
+\sqrt{2} g i b\left({\bar{\mu}}_{a} ( {{ \Lambda}}^i )^{a}_{~b}
  \epsilon_{ij}(w^j)^{b} +
({\bar{w}}_i)_{a} \epsilon^{ij}   ({{{\bar{{{\Lambda}}}}}}_{j})^{a}_{~b} \mu^{b}\right)
\right]\nonumber\\
 &&\!\!\!\!\!=\frac{i}{2}g{\bar{\mu}}_{a}\Gamma_{\hat{\mu}}\left(\bar{\epsilon}_i \Gamma^{\hat{\mu}}\Lambda^i-\bar{\Lambda}_i \Gamma^{\hat{\mu}} \epsilon^i\right)^a_{~b}\mu^b -
 2 i g  \bar{\mu}_a\left(  (\Lambda^i)^a_{~b} \bar{\epsilon}_i -  \epsilon^i(\bar{\Lambda}_i)^a_{~b}\right)\mu^b\nonumber\\
 &&\!\!\!\!\!- \sqrt{2}g b \bar{\mu}_a ({\cal{D}}^i_{~j})^a_{~b}\epsilon^j\epsilon_{ik}w^{kb} +  \sqrt{2}gb \bar{w}_{ia}{\epsilon}^{ih}\bar{\epsilon}_j({\cal{D}}^j_{~h})^a_{~b}\mu^b\nonumber\\
 &&\!\!\!\!\!- g  \bar{w}_{ja} \epsilon^{ji}
 \left(D_{\hat{\mu}}(\bar{\Lambda}_i)^a_{~b}\Gamma^{\hat{\mu}}\epsilon^h
+  \bar{\epsilon}_i\Gamma^{\hat{\mu}}D_{\hat{\mu}}
\left(\Lambda^h\right)^a_{~b} \right) \epsilon_{hk}   w^{kb}\nonumber\\
&&\!\!\!\!\!- \sqrt{2} \, b\,  \partial_{\hat{\mu}} \left[ \bar{w}_{ja}\epsilon^{ji}\bar{\epsilon}_i [\Gamma^{\hat{\mu}},\,\Gamma^{\hat{\nu}}](D_{\hat{\nu}} \mu)^a\right]   +
\sqrt{2} \, b \, \partial_{\hat{\mu}} \left[ {\bar{\mu}}_a \epsilon^i \epsilon_{ij} (D^{\hat{\mu}} w^j)^a  \right]\nonumber\\
&&\!\!\!\!\!+ {g}  \partial_{\hat{\mu}}\left[\bar{w}_{ja}\epsilon^{jk}\left( \bar{\epsilon}_k \Gamma^{\hat{\mu}}(\Lambda^i)^a_{~b}+(\bar{\Lambda}_k)^a_{~b}\Gamma^{\hat{\mu}}\epsilon^i\right)\epsilon_{ik}w^{kb}\right]
 \label{sum2}
\end{eqnarray}
The last term of the action to consider is the variation of the term with the auxiliary fields given by:
\begin{eqnarray}
&&-g \delta\left[\bar{w}_{ia}( {\cal{D}}^i_{~j})^a_{~b}w^{jb}\right]=+ \sqrt{2} g b \bar{\mu}_a\epsilon_{ik}\epsilon^k({\cal{D}}^i_{~j})^a_{~b}w^{jb} + \sqrt{2} g  b
\bar{w}_{ia}({\cal{D}}^i_{~j})^a_{~b}\epsilon^{jk}\bar{\epsilon}_k\mu^b\nonumber\\
&&- g \bar{w}_{ia}\frac{1}{2}(\tau^c)^h_{~k}(\tau^c)^i_{~j} D_{\hat{\mu}}\left( \bar{\Lambda}_h\Gamma^{\hat{\mu}}\epsilon^k+\bar{\epsilon}_h\Gamma^{\hat{\mu}}\Lambda^k\right)w^{jb}\label{fv}
\end{eqnarray}
Using the identity in Eq. 12.21 of Ref. \cite{Westbook}
\begin{eqnarray}
\frac{1}{2} (\tau^c)^h_{~k}(\tau^c)^i_{~j}=\delta^h_j\delta^i_k-\frac{1}{2}\delta^h_k\delta^i_j
=\left(\delta^h_j\delta^i_k-\delta^h_k\delta^i_j\right)+\frac{1}{2}\delta^h_k\delta^i_j=\epsilon^{hi}\epsilon_{kj}
+\frac{1}{2}\delta^h_k\delta^i_j
\end{eqnarray}
the last term of eq. (\ref{fv}) becomes:
\begin{eqnarray}
&&\!\!\!\!\!\!\!\!\!\!\!\!\!\!\!-g \bar{w}_{ia}\frac{1}{2}(\tau^c)^h_{~k}(\tau^c)^i_{~j} D_{\hat{\mu}}
\left( \bar{\Lambda}_h\Gamma^{\hat{\mu}}\epsilon^k+\bar{\epsilon}_h\Gamma^{\hat{\mu}}\Lambda^k\right)^a_{~b}w^{jb}\nonumber\\
&&\!\!\!\!\!\!\!\!\!\!\!\!\!\!\!=
 g \bar{w}_{ia}\epsilon^{ih} D_{\hat{\mu}}\left( \bar{\Lambda}_h\Gamma^{\hat{\mu}}\epsilon^k+
\bar{\epsilon}_h\Gamma^{\hat{\mu}}\Lambda^k\right)^a_{~b}\epsilon_{kj}w^{jb}-\frac{1}{2} \bar{w}_{ia}D_{\hat{\mu}}\left( \bar{\Lambda}_j\Gamma^{\hat{\mu}}\epsilon^j+\bar{\epsilon}_j\Gamma^{\hat{\mu}}\Lambda^j\right)^a_{~b}w^{ib}\label{3var}
\end{eqnarray}
The first term of the previous equation cancels the fourth line of  Eq. (\ref{sum2}),
while the second term of Eq. (\ref{3var}) is zero according to the second identity written in Eq.(\ref{wi}).
Using Eq. (\ref{nr})
we can write the first two terms of eq. (\ref{fv}) as follows:
\begin{eqnarray}
&&+ \sqrt{2} g\,b  \bar{\mu}_a\epsilon_{ik}\epsilon^k({\cal{D}}^i_{~j})^a_{~b}w^{jb}+ \sqrt{2} g b
\bar{w}_{ia}({\cal{D}}^i_{~j})^a_{~b}\epsilon^{jk}\bar{\epsilon}_k\mu^b= + \sqrt{2} g b
\bar{\mu}_a \,({\cal{D}}^i_{~j})^a_{~b}\epsilon^j \epsilon_{ik}w^{kb}\nonumber\\
&&- \sqrt{2} g b \bar{w}_{ia}\epsilon^{ih}\bar{\epsilon}_j (D^j_{~h})^a_{~b}\mu^b
\end{eqnarray}
These two terms cancel the terms in the third line of Eq. (\ref{sum2}).

In the last part of this Appendix we first prove the identities written in Eq. (\ref{wi}) and finally that the second line of Eq. (\ref{sum2}) is identically zero.

The starting point to prove the Eq.s (\ref{wi}) is the identity:
\begin{eqnarray*}
&&\bar{\epsilon}_i \Gamma^{\hat{\mu}_1\dots \hat{\mu}_{n}}\Lambda^j=-\bar{\epsilon}_iB_6^*B_6 \Gamma^{\hat{\mu}_1\dots \hat{\mu}_{n}}\Lambda^j= -(-)^{\sum_{i=1}^n(\delta_{\hat{\mu}_i0}+1)}\bar{\epsilon}_iB_6^*{\Gamma^T}^{\hat{\mu}_1\dots \hat{\mu}_{n}}B_6\Lambda^j
\end{eqnarray*}
where we have denoted with ${\Gamma^T}^{\hat{\mu}_1\dots \hat{\mu}_{n}}$ the completely antisymmetrized product of the transposed Dirac-matrices and  used the identity $B_6\Gamma^{\hat{\mu}}B_6^{-1}={\Gamma^{\hat{\mu}}}^*$ together with:
\begin{eqnarray}
{\Gamma^{\hat{\mu}}}^\dagger=(-)^{\delta_{\hat{\mu}0}+1}\Gamma^{\hat{\mu}}\Rightarrow
{\Gamma^{\hat{\mu}}}^*=(-)^{\delta_{\hat{\mu}0}+1}{\Gamma^{\hat{\mu}}}^T
\end{eqnarray}
Furthermore, observing that $\Gamma^0B_6^*=B_6^*{\Gamma^0}^T$ and using, both for the gaugino end the susy parameters, the Eq.s (\ref{Lambdai}) and (\ref{as6}) we get:
\begin{eqnarray}
\bar{\epsilon}_i \Gamma^{\hat{\mu}_1\dots \hat{\mu}_{2n-1}}\Lambda^j&=& (-)^{n-2}\epsilon_{ik} \epsilon^{jl} \bar{\Lambda}_l \Gamma^{\hat{\mu}_1\dots \hat{\mu}_{2n+1}}\epsilon^k
\nonumber\\
\bar{\epsilon}_i \Gamma^{\hat{\mu}_1\dots \hat{\mu}_{2n}}\Lambda^j&=& (-)^{n-1} \epsilon_{ik}\epsilon^{jl} \bar{\Lambda}_l \Gamma^{\hat{\mu}_1\dots \hat{\mu}_{2n}+2}\epsilon^j
\label{i=0}
\end{eqnarray}
with $n=0\dots 4$. Along the same lines we can prove an analogous relation where the role of the gaugino and susy parameters are exchanged. Eq.s (\ref{wi}) can be easily derived by considering the case  $n=0$ of the first of the two previous identities.

Let us now consider the following Fierz identity valid for two generic six-dimensional spinors, here denoted with $\chi$ and $\psi$, having the same chirality\cite{0203157}:
\begin{eqnarray}
\psi^A\bar{\chi}_B=-\frac{1}{4} \left( \bar{\chi}\Gamma^{\hat{\mu}}\psi\right){\Gamma_{\hat{\mu}}}^A_{~B}
+\frac{1}{48} \left( \bar{\chi}\Gamma^{\hat{\mu}\hat{\nu}\hat{\rho}}\psi\right){\Gamma_{\hat{\mu}\hat{\nu}\hat{\rho}}}_{~B}^A
\end{eqnarray}
This identity allow us to write:
\begin{eqnarray}
\bar{\mu}_A{\Lambda^i}^A \bar{\epsilon}_{i\,B}\mu^B=- \frac{1}{16} (\bar{\mu}\Gamma^{\hat{\mu}}\mu)
(\bar{\epsilon}_i \Gamma^{\hat{\nu}}\Lambda^i) {\rm Tr}[\Gamma_{\hat{\mu}}\Gamma_{\hat{\nu}}]- \frac{1}{48}
(\bar{\mu}\Gamma^{\hat{\mu}\hat{\nu}\hat{\rho}}\mu) (\bar{\epsilon}_i \Gamma^{\hat{\sigma}\hat{\tau}\hat{\delta}} \Lambda^i){\rm Tr}[\Gamma_{\hat{\mu}\hat{\nu}\hat{\rho}}\Gamma_{\hat{\sigma}\hat{\tau}\hat{\delta}}]\nonumber\\
&&\label{i=1}
\end{eqnarray}
where we have used the condition $ {\rm Tr}[\Gamma_{\hat{\mu}}\Gamma_{\hat{\sigma}\hat{\tau}\hat{\delta}}]=0$, and
\begin{eqnarray}
\bar{\mu}_A{\epsilon^i}^A \bar{\Lambda}_{i\,B}\mu^B= -\frac{1}{16} (\bar{\mu}\Gamma^{\hat{\mu}}\mu)
(\bar{\Lambda}_i \Gamma^{\hat{\nu}}\epsilon^i) {\rm Tr}[\Gamma_{\hat{\mu}}\Gamma_{\hat{\nu}}]- \frac{1}{48}
(\bar{\mu}\Gamma^{\hat{\mu}\hat{\nu}\hat{\rho}}\mu) (\bar{\Lambda}_i \Gamma^{\hat{\sigma}\hat{\tau}\hat{\delta}} \epsilon^i){\rm Tr}[\Gamma_{\hat{\mu}\hat{\nu}\hat{\rho}}\Gamma_{\hat{\sigma}\hat{\tau}\hat{\delta}}]\nonumber\\
&&\label{i=2}
\end{eqnarray}
Subtracting Eq.s (\ref{i=1}) and (\ref{i=2}) and using  the first identity in Eq.(\ref{i=0}) with $n=0$ and $n=1$, we get:
\begin{eqnarray}
 \bar{\mu}_A{\Lambda^i}^A \bar{\epsilon}_{i\,B}\mu^B-\bar{\chi}_A{\epsilon^i}^A \bar{\Lambda}_{i\,B}\chi^B=- (\bar{\mu}\Gamma^{\hat{\mu}}\mu)
(\bar{\Lambda}_i \Gamma^{\hat{\nu}}\epsilon^i)
\end{eqnarray}
which prove the vanishing of the second line of Eq. (\ref{sum2}).

\section{Calculation of  correlators}
\label{correla}

In this Appendix we compute in detail various three-point functions in order to check some of
the terms of the Lagrangians written in Sect. (\ref{595559}).
\vskip 0.5cm

\noindent
{\bf Two fermions in the adjoint and a gauge field}

The first amplitude that we consider is the one involving two fermions in the adjoint representation of the gauge group $U(N_5)$ and the gauge field $A_{\hat{\mu}}^{(5)}$:
\begin{eqnarray}
&& {\cal A}^{\bar{\Psi}^{(5)} A_{\hat{\mu}}^{(5)} \Psi^{(5)}}= C_0\int \frac{ \prod_{i=1}^{3} dx_i}{dV}
\langle V_{\bar{\Psi}^{(5)}} ^{(-1/2)}(x_1)\,V_{A_{\hat{\mu}}}^{(-1)}(x_2)\, V_{  {\Psi}^{(5) }}^{(-1/2)}(x_3)\rangle
\label{am1}
\end{eqnarray}
where $C_0$ is defined after eq.(\ref{corre}) and:
\begin{eqnarray}
dV=\frac{dx_adx_bdx_c}{x_{ab}x_{ac}x_{bc}}~~;~~x_{ab}=x_a-x_b.
\label{vol8}
\end{eqnarray}
The amplitude (\ref{am1}) can easily be computed using for the
fields $\Psi^{(5)}$ and $A_{\hat{\mu}}$  the vertex operators
defined in sec.[\ref{D9D9}]  and observing that the vertex for
$\bar{\Psi}^{(5)}$ is given by the Eq.( \ref{Vlab}) with
$\Psi^\alpha$ replaced by the doublet given in the  Eq. (\ref{105}).
This latter definition follows from the observation that, in our
conventions, the complex-conjugation of the $SU(2)$ doublets
doesn't change the four dimensional chirality associated to the
compact directions transverse to the D5-brane and, being the
ten-dimensional chirality fixed by the GSO, it doesn't change also
the six dimensional one. It follows, that the spin fields
associated to the vertex $\bar{\Psi}^\alpha$ must have the same
chirality  as the ones associated to the field
 $\Psi^\alpha$ .

The result is:
\begin{eqnarray}
&& A^{\bar{\Psi}^{(5)} A_{\hat{\mu}}^{(5)} \Psi^{(5)}}
=i2\sqrt{2}g_{5} \left( \bar{\Psi}^{(5)\,{A} \alpha}   \right)^{c}_{~a} (p)
(A_{{\hat{\mu}}}^{(5)} )^a_{~b}  (k) \left( \Psi^{(5)\,{B}  \beta }
\right)^{b}_{~c}   (q) [x_{12}x_{13}x_{23}]
\nonumber\\
&&\times [\langle S_{{A}}(x_1)\psi^{\hat{\mu}}(x_2)S_{{B}}(x_3)\rangle][\langle S_{{\alpha}}(x_1)S_{{\beta}}(x_3)\rangle] [\langle e^{-\frac{1}{2}\phi(x_1)} e^{-\phi(x_2)}e^{-\frac{1}{2} \phi(x_3)}\rangle]\nonumber\\
&&\times
 \langle {\rm e}^{i \sqrt{2 \pi \alpha' } p \cdot X (x_1)} {\rm e}^{i\sqrt{2 \pi \alpha' }  k \cdot  X (x_2)}
 {\rm e}^{i \sqrt{2 \pi \alpha' }  q \cdot  X (x_3)} \rangle
\nonumber \\
\label{cor2}
\end{eqnarray}
Using the following correlators\cite{Billo:2002hm, 0911.5168}:
\begin{eqnarray}
&\langle S_{{A}}(x_1)\psi^{\hat{\mu}}(x_2)S_{{B}}(x_3)\rangle=-\frac{i}{\sqrt{2}}
(\bar{\Sigma}^{\hat{\mu}})_{AB}  x_{12}^{-1/2}x_{13}^{-1/4}x_{23}^{-1/2} \nonumber &\\
&\langle e^{-\frac{1}{2}\phi(x_1)} e^{-\phi(x_2)}e^{-\frac{1}{2} \phi(x_3)}\rangle=x_{12}^{-1/2}x_{13}^{-1/4}x_{23}^{-1/2}~~;~~\langle S_{{\alpha}}(x_1)S_{{\beta}}(x_3)\rangle=-
\epsilon_{\alpha\beta} x_{13}^{-1/2}& \nonumber
\end{eqnarray}
and taking into account that the correlator involving the bosonic coordinate $x$ gives 1
because the momentum is conserved and the three states are massless,
we get \footnote{Here and in the following we omit writing the  delta-function of momentum conservation.}:
\begin{eqnarray}
&&\!\!\!\! A^{\bar{\Psi}^{(5)} A_{\hat{\mu}}^{(5)} \Psi^{(5)}}
=-2g_{5}\,\left( \bar{ {\Psi}}^{(5)\,{A}{\alpha}}   \right)^{c}_{~a} (p)
(A_{\hat{\mu}}^{(5)}  )^a_{~b} (k) \left({ \Psi}^{(5)\,{B}{\beta}}  \right)^{b}_{~c} (q)
(\bar{\Sigma}^{\hat{\mu}})_{\hat{A}\hat{B}}\epsilon_{\alpha\beta}\nonumber\\
&&\!\!\!\!=-2 g_{5}\left(0 \,\,\, \,(\bar{{\Psi}}^{(5)\,{A}{\alpha}} )^{c}_{~a} (p) \right)
(A_{\hat{\mu}}^{(5)})^a_{~b} (k)
\left(\begin{array}{cc}
           0& (\Sigma^{\hat{\mu}})^{{A}{B}}\\
           (\bar{\Sigma}^{\hat{\mu}})_{{A}{B}}&0\end{array}\right)
           \left(\begin{array}{c}(\Psi^{(5)\,{B}{\beta}})^{b}_{~c} (q)
           \\0\end{array}\right)\epsilon_{\alpha\beta}\nonumber\\
&&\!\!\!\!=-2 g_{5}{\rm Tr}[ \bar{\Psi}^{(5)\,{\alpha}} (p) A_{\hat{\mu}}^{(5)} (k) \Gamma^{\hat{\mu}}_{(6)} {\Psi}^{(5)}_{\alpha} (q)]=2g_{5}{\rm Tr}[ \bar{\Psi} (p) A_{\hat{\mu}}^{(5)} (k) \Gamma^{\hat{\mu}}_{(6)} {\Psi}^{(5)}]
\label{ppl}
\end{eqnarray}
where we have introduced the six dimensional  Weyl-spinors and  Dirac matrices so defined\cite{1011.1249}:
\begin{eqnarray}
{\Psi}^{(5)\,\beta}\equiv \left(\begin{array}{c}\Psi^{(5)\,{B}{\beta}}\\0\end{array}\right)\nonumber~~;~~
\Gamma^{\hat{\mu}}=\left( \begin{array}{cc}
                             0&\Sigma^{\hat{\mu}}\\
                             \bar{\Sigma}^{\hat{\mu}}&0
                          \end{array}\right)
\end{eqnarray}
being $\Sigma^{\hat{\mu}}$ and $\bar{\Sigma}^{\hat{\mu}}$  $4\times 4$ matrices satisfying the anticommutation algebra
$\left\{ \Sigma^{\hat{\mu}}, {\bar{\Sigma}}^{\hat{\nu}} \right\} = -2\eta^{\hat{\mu} \hat{\nu}}$ and $A,B=1,\dots 4$.
Notice that the representation of Dirac matrices which we are using in this section differs from the one adopted in the field theory calculation of the effective action of our system. However, this difference will not generate any confusion because the final expression of the string amplitudes will be always written in a form which is independent of the chosen representation of the Dirac matrices.

Eq. (\ref{ppl}) reproduces the correct coupling of two adjoint fermions  with  a gauge field,
written for example in Eq. (\ref{L55}).
\vskip 0.5 cm

\noindent
{\bf A  twisted scalar, a  twisted fermion and a  D5 gaugino}

The next string amplitude which we are going to consider is the one involving the interaction between a twisted scalar, a  twisted fermion and the gaugino of the gauge theory
living  on the D5-brane:
\begin{equation}
A^{\bar{\mu} \Lambda^{(5)} w }=  C_{0} \int    \frac{ \prod_{i=1}^{3} dx_{i}}{dV}  < V_{\bar{\mu}}^{-(\frac{1}{2})}(x_{1}) V_{{\Lambda}^{(5)}}^{- (\frac{1}{2})} (x_{2}) V_{w}^{(-1)}(x_{3}) >
\end{equation}
We need to compute:
\begin{eqnarray}
A^{\bar{\mu} \Lambda w } \!\!\!\!& = &\!\!\! \!\sqrt{2} g_{5} x_{12} x_{13} x_{23} (\bar{\mu}_{\hat{A}})^u_a  ({p})
\left( {\Lambda}^{(5)\,\hat{B}}_{~~~~ \dot{\alpha}} \right)^{a}_{~b} (k)  ({w}^{\dot{\gamma}})^b_u
(q)\epsilon_{\dot{\gamma}\dot{\beta}}
<e^{-\frac{1}{2} \phi(x_1) }  e^{-\phi(x_2)} e^{-\frac{1}{2} \phi(x_{3}) }>\nonumber\\
&\times& <S^{\hat{A}} (x_{1}) S_{\hat{B}} (x_{2})>   <S^{\dot{\alpha}} (x_2) S^{\dot{\beta}} (x_3)><\bar{\Delta}(x_2) \Delta(x_3)> \nonumber\\
&\times&\,< e^{ i \sqrt{2 \pi \alpha'}  p_{ \hat{\mu}} X^{\hat{\mu}} (x_{1}) }   e^{ i \sqrt{2 \pi \alpha'}
k_{ \hat{\mu}} X^{\hat{\mu}} (x_{2}) } e^{i \sqrt{2 \pi \alpha'} q_{\hat{\mu}} X^{\hat{\mu}} (x_{3}) }>
\label{2twi1gau}
\end{eqnarray}
Using the following correlators:
\begin{eqnarray}
<S^{\hat{A}} (x_{1}) S_{\hat{B}} (x_{3})> \sim \frac{i \delta^{A}_{B}}{(x_{13})^{1/4} }~~;~~
<S^{\dot{\alpha}} (x_1) S^{\dot{\beta}} (x_2)>  \sim - \frac{ \epsilon^{\dot{\alpha} \dot{\beta} }}{(x_{12})^{1/2}}\nonumber\\
<e^{ -\frac{1}{2} \phi (x_1) }  e^{ -\phi(x_2)} e^{-\frac{1}{2} \phi (x_3)} > = \frac{1}{ x_{12}^{ {\Delta}_{1}+{\Delta}_{2}-{\Delta}_3} x_{13}^{{\Delta}_{1}+{\Delta}_{3}-{\Delta}_{2}} z_{23}^{{\Delta}_{2}+{\Delta}_{3}-{\Delta}_{1}} }
\end{eqnarray}
where  ${\Delta}_{1}= {\Delta}_{3} = \frac{3}{8} $  is the conformal dimension of $e^{ -\frac{1}{2} \phi (z) }$  and ${\Delta}_{2}= \frac{1}{2}$  is the conformal dimension of  $e^{ -\phi(z)}$,
we get
\begin{eqnarray}
A^{\bar{\mu} \Lambda w }  =  \sqrt{2}g_{5}  i \left[ \bar{\mu}^{u}_{a} (p)
\left( {\Lambda}^{(5)i} \right)^a_{~b} (k)  \epsilon_{ij} (w^{j})^{b}_{u} (q) \right]  \nonumber
\label{}
\end{eqnarray}

\vskip 0.5 cm
\noindent
{\bf A twisted scalar, a twisted fermion and a D9 gaugino}

It is also enlightening to repeat  the previous calculation in the case of
the gaugino  living in the world-volume of the $D9$-brane. The result is:
\begin{eqnarray}
&&A^{\bar{\mu} \Lambda^{(9)} w }=  C_{0} \int \prod_{i=1}^{3}  \frac{dx_{i}}{dV}  < V_{w}^{(59)}(x_{3})  V_{{\Lambda}^{(9)}} (x_{2}) V_{\bar{\mu}}^{95}(x_{1}) >\nonumber\\
&&=- i\sqrt{2}\,g_{9} e^{i k_{{m}} y_0^{{m}}}\, \bar{\mu}^{u}_{a} (p)
\left( {\Lambda}^{(9)i}(k^{\hat{\mu}},\,k^{\hat m}) \right)^v_{~u} \epsilon_{ij} (w^{j})^{a}_{v} (q) \nonumber
\end{eqnarray}
where  $y_0^m$ is the position of the $D5$ in the last two tori and we have used the correlator:
\begin{eqnarray}
\langle  \Delta(x_1)e^{i k_{\hat m}X^{\hat m}}\bar{\Delta}(x_3) \rangle= x_{12}^{-\pi\alpha' k_m k^m} x_{13}^{-1/2+\pi\alpha' k_m k^m}
x_{23}^{-\pi\alpha' k_m k^m} e^{i k_m y^m_0}
\label{part43}
\end{eqnarray}
that is consistent with
the general formula  for the correlator of three operators $A, B,C$ with conformal
dimension $\Delta_A , \Delta_B, \Delta_C$ respectively:
\begin{eqnarray}
\langle A(x_1) B(x_2) C( x_3) \rangle =
\frac{1}{x_{12}^{\Delta_A + \Delta_B - \Delta_C}
x_{13}^{\Delta_A + \Delta_C - \Delta_B} x_{23}^{\Delta_B + \Delta_C - \Delta_A} }
\label{cor67}
\end{eqnarray}
Remember  that in our case $\Delta_A =
\Delta_C =\frac{1}{4}$ and $ \Delta_B =\pi \alpha'
k_{m}  k^{{m}}$.
By performing the Fourier transformations along the compact momenta, according to the relation:
\begin{eqnarray}
{\Lambda}^{(9)i}(k^{\hat{\mu}},y_0) =\sum_{k_{\hat m}} {\Lambda}^{(9)i}(k^{\hat{\mu}},\,k^{\hat m}) e^{i k_{\hat m} y_0^{\hat m}}
\label{four}
\end{eqnarray}
we get:
\begin{eqnarray}
&&A^{\bar{\mu} \Lambda^{(9)} w }=-\sqrt{2}g_{9} i \, \bar{\mu}^{u}_{a} (p)
\left( {\Lambda}^{(9)i}(k^{\hat{\mu}},\,y_0) \right)^v_{~u} \epsilon_{ij} (w^{j})^{a}_{v} (q)
\label{amp83}
\end{eqnarray}
Finally by observing that $g_9 \Lambda^{(9)}_{d=10}=\tilde{g}_9 \Lambda^{(9)}_{d=6}$ we reproduce, from a string calculus, the interaction term written in the fourth line of
Eq. (\ref{59c})

\vskip 0.5 cm
\noindent
{\bf Two twisted scalars and a D9 gauge field}

The last amplitude that we  consider is the three-point correlator involving the gauge field living on the ten dimensional world-volume of the $D9$ and two twisted scalar fields living on the six-dimensional world-volume
of the D5 brane.
\begin{eqnarray}
&& {\cal A}^{\bar{w} A_{M}^{(9)} w}= C_0\int \frac{\prod_{i=1}^{3} dx_i }{dV}
\langle V_{\bar{w}} ^{(-1)}(x_1)\,V_{A_{M}^{(9)}}^{(0)}(x_2)\, V_{w }^{(-1)}(x_3)\rangle
\label{cor62}
\end{eqnarray}
Using the vertex operators introduced in Sect. \ref{vertexope}) we have to
compute the following quantity:
\begin{eqnarray}
&&{\cal A}^{\bar{w} A_{M}^{(9)} w}=g_{9}\sqrt{\frac{2}{\pi\alpha'}}\,{\bar {w}^u}_{\dot\alpha a }(p) ~{A^a}_{Mb} (k) ~{w}^{b}_{{\dot{\beta}} u}(q)
\times \langle
\left({\bar{\Delta}} S^{\dot\alpha}\,{\rm e}^{-\varphi}
{\rm e}^{i\sqrt{2\pi\alpha'} p_\hmu  X^\hmu }\right)(x_1)
\nonumber\\
&&
\times \left(
\left[ \partial  X^{M} + i \sqrt{2\pi \alpha'} k_N \psi^N  \,  \psi^M  \right]\,
{\rm e}^{i\sqrt{2\pi\alpha'}  k_N  X^N }
\right)(x_2)
\left(
{{\Delta}} S^{\dot\beta}\,{\rm e}^{-\varphi}  {\rm e}^{i \sqrt{2\pi\alpha'}q_\hmu  X^\hmu }
\right)(x_3)
\rangle\nonumber\\
&&\equiv g_{9}\sqrt{\frac{2}{\pi\alpha'}}\,{\bar {w}^u}_{\dot\alpha a }(p) ~{A^a}_{Mb} (k) ~{w}^{b}_{{\dot{\beta}} u}(q)\,(x_{12}\,x_{13}\,x_{23})
\times {\cal C}^{M\dot{\alpha}\dot{\beta}}
\label{corre62}
\end{eqnarray}

We start with the gauge polarization  $M=\hmu$. In this case  the
previous correlator reduces to
\begin{eqnarray}
{\cal C}^{\hat{\mu}\dot{\alpha}\dot{\beta}}\!\!\!\!&=&\!\!\!\!\langle
\left({\bar{\Delta}} S^{\dot\alpha}\,{\rm e}^{-\varphi}
{\rm e}^{i\sqrt{2\pi\alpha'} p_\hnu  X^\hnu }\right)\!\!(x_1)\!
\left(\!\partial  X^{\hmu} {\rm e}^{i \sqrt{2\pi\alpha'} k_m  X^m } \right)\!\!(x_2)\!
\left(\!
{{\Delta}} S^{\dot\beta}\,{\rm e}^{-\varphi}  {\rm e}^{i \sqrt{2\pi\alpha'}q_\hnu  X^\hnu }
\right)\!(x_3)
\rangle
\nonumber\\
&=&
\langle {\rm e}^{-\varphi (x_1) }  ~{\rm e}^{-\varphi (x_3)}  \rangle
\times
\langle {\bar{\Delta}}(x_1)  ~{\rm e}^{i \sqrt{2\pi\alpha'} k_m  X^m (x_2)}~{{\Delta}} (x_3) \rangle
\times
\langle S^{\dot\alpha} (x_1)  S^{\dot\beta} (x_3) \rangle
\nonumber \\
&\times&
\langle {\rm e}^{i\sqrt{2\pi\alpha'}  p_\hnu  X^\hnu  (x_1)}
~\partial  X^{\hmu} (x_2) {\rm e}^{i \sqrt{2\pi\alpha'}  k_\hlambda  X^\hlambda (x_2)}
{\rm e}^{i\sqrt{2\pi\alpha'}  q_\hnu  X^\hnu (x_3) }\rangle
\nonumber\\
&=&
x_{13}^{-1}
\times x_{12}^{-\pi\alpha' k_m k^m} x_{13}^{-1/2+\pi\alpha' k_m k^m}
x_{23}^{-\pi\alpha' k_m k^m} e^{i k_m y^m_0}
\,  x_{13}^{-1/2} \epsilon^{\dot \alpha  \dot \beta}
\nonumber\\
&\times & x_{12}^{2\pi\alpha' p_\hmu k^\hmu} x_{13}^{2\pi\alpha' p_\hmu q^\hmu}
x_{23}^{2\pi\alpha' k_\hmu q^\hmu}
( i\sqrt{2\pi\alpha'}  ) \left[\frac{p^\hmu}{x_{12}} - \frac{q^\hmu}{x_{23}}\right]
\label{c56}
\end{eqnarray}
We can  rewrite Eq. (\ref{c56}) as follows
\begin{eqnarray}
{\cal C}^{\hat{\mu}\dot{\alpha}\dot{\beta}}&=&i
\sqrt{2\pi \alpha' } \,
~e^{i k_m y^m_0} \epsilon^{\dot \alpha  \dot \beta}\,x_{12}^{-\pi\alpha' k_m k^m+2\pi\alpha' p_\hmu k^\hmu}
\nonumber\\
&\times &x_{13}^{-2+\pi\alpha' k_m k^m+2\pi\alpha' p_\hmu q^\hmu}\,
x_{23}^{-\pi\alpha' k_m k^m+2\pi\alpha' k_\hmu q^\hmu}
\left[\frac{p^\hmu}{x_{12}} - \frac{q^\hmu}{x_{23}}\right]
\nonumber\\
&=&  i\sqrt{2\pi\alpha'}
~e^{i k_m y^m_0} \epsilon^{\dot \alpha  \dot \beta}
\left[\frac{p^\hmu}{x_{12}x_{13}x_{23}} + \frac{k^\hmu}{x_{13}^{2}x_{23}}\right]
\label{uno}
\end{eqnarray}
where we have used the  momentum conservation  $ p_{\hat{\mu}} + q_{\hat{\mu}} +k_{\hat{\mu}}
=0$ that implies, together with the mass shell conditions $p_{\hat{\mu}}p^{\hat{\mu}} =
q_{\hat{\mu}} q^{\hat{\mu}} =0$, the following conditions:
$2\alpha' p_\hmu q^\hmu
= -2\alpha' p_\hmu k^\hmu
= - 2\alpha' q_\hmu k^\hmu
= \alpha' k_\hmu k^\hmu
$. We have also used the mass-shell condition for the vector field: $k_\hmu k^\hmu +
k_m k^m =0$.  Notice that the piece proportional to $k_\hmu$ does not have the right
dependence on the Koba-Nielsen variables and  therefore it must  be cancelled  out in the final result.

We now consider the  polarization of the vector field to be  $M=m$. In this case the
amplitude written in Eq. (\ref{corre62})  is the sum of  two pieces.
The first one is
\begin{eqnarray}
{\cal C}^{{ m}\dot{\alpha}\dot{\beta}}_1\!\!\!\!&=&\!\!\!\!
\langle
\left({\bar{\Delta}} S^{\dot\alpha}\,{\rm e}^{-\varphi}
{\rm e}^{i \sqrt{2\pi\alpha'} p_\hnu  X^\hnu }\right)\!\!(x_1)\!
\left(\!\partial  X^{m} {\rm e}^{i \sqrt{2\pi\alpha'}  k_n  X^n } \right)\!\!(x_2)\!
\left(\!
{{\Delta}} S^{\dot\beta}\,{\rm e}^{-\varphi}  {\rm e}^{i\sqrt{2\pi\alpha'}  q_\hnu  X^\hnu }
\right)\!\!(x_3)
\rangle
\nonumber\\
&=&\!\!\!\!
\langle {\rm e}^{-\varphi (x_1)}  ~{\rm e}^{-\varphi (x_3)}  \rangle
~
\langle {\bar{\Delta}} (x_1)  ~\partial  X^{m} (x_2)  {\rm e}^{i \sqrt{2\pi\alpha'}  k_n  X^n (x_2) }
~{{\Delta}} (x_3) \rangle
~
\langle S^{\dot\alpha} (x_1) S^{\dot\beta} (x_3) \rangle
\nonumber \\
& \times&\!\!\!\! \langle {\rm e}^{i\sqrt{2\pi\alpha'}  p_\hnu  X^\hnu  (x_1 )}
~ {\rm e}^{i \sqrt{2\pi\alpha'}  k_\hlambda  X^\hlambda (x_2) }
{\rm e}^{i\sqrt{2\pi\alpha'}  q_\hnu  X^\hnu (x_3)}\rangle
\nonumber\\
&=&\!\!\!\!
x_{13}^{-1}
\,
\frac{ i\sqrt{\pi \alpha'} k^m e^{i k_\hmm y^m_0} (x_{12}+x_{32})}
{\sqrt{2} x_{12}^{\pi\alpha' k_m k^m+1} x_{13}^{1/2-\pi\alpha' k_m k^m}
x_{23}^{\pi\alpha' k_m k^m+1}
}
\,  x_{13}^{-1/2} \epsilon^{\dot \alpha  \dot \beta}
\, x_{12}^{2\pi\alpha' p_\hmu k^\hmu} x_{13}^{2\pi\alpha' p_\hmu q^\hmu} x_{23}^{2\pi\alpha' k_\hmu q^\hmu}
\nonumber\\
\nonumber\\
&=&  \frac{i \sqrt{\pi\alpha'} k^m e^{i k_m y^m_0} (x_{12}+x_{32}) \epsilon^{\dot \alpha  \dot \beta}  }{\sqrt{2}x_{12} x_{13}^{2} x_{23}}
\label{due}
\end{eqnarray}
The only correlator that needs some explanation is the one  containing the twist fields and
that is equal to:
\begin{eqnarray}
\langle {\bar{\Delta}}(x_1)\!
\left(\partial X^n {\rm e}^{i \sqrt{2\pi\alpha'}  k_m X^m }\right)\!\!(x_2)
{{\Delta}}(x_3) \rangle =
\frac{( x_{12}-x_{23})  \left( i \sqrt{\pi\alpha'} k^n e^{i k_m y^m_0} \right)  }{\sqrt{2}
x_{12}^{\pi\alpha' k_m k^m+1} x_{13}^{1/2-\pi\alpha' k_m k^m}
x_{23}^{\pi\alpha' k_m k^m+1}
}
\label{twi}
\end{eqnarray}
in order to get the proper conformal dependence of the total
correlator. There are several reasons why we think that this choice
is the correct one. If we assume the correlator in Eq.
(\ref{part43}), then the previous one can be obtained from it by
taking the derivative with respect to $x_2$ of Eq. (\ref{part43}).
The operator $\partial X^n {\rm e}^{i \sqrt{2\pi\alpha'} k_m  X^m }$
is {\sl not} a good conformal operator, but the operator $A_M
\partial X^M {\rm e}^{i \sqrt{2\pi\alpha'}  k_M  X^M }$ with $A_M
k^M=0$ is a good conformal operator and therefore, when used in a
correlator, one  must obtain the proper dependence on the
Koba-Nielsen variables. The choice in Eq. (\ref{twi}) is symmetric
under the exchange of $1$ and $3$.

The second one is
\begin{eqnarray}
{\cal C}^{ { m} \dot{\alpha} \dot{\beta} }_2&=&
\langle
\left({\bar{\Delta}} S^{\dot\alpha}\,{\rm e}^{-\varphi}
{\rm e}^{i \sqrt{2\pi\alpha'} p_\hnu  X^\hnu }\right)(x_1)
\left( i \sqrt{2\pi\alpha'} k_n \psi^n \psi^m {\rm e}^{i \sqrt{2\pi\alpha'}  k_\hll  X^\hll } \right)(x_2)
\nonumber\\~
&\times&\left(
{{\Delta}} S^{\dot\beta}\,{\rm e}^{-\varphi}  {\rm e}^{i\sqrt{2\pi\alpha'}  q_\hnu  X^\hnu }
\right)(x_3)
\rangle
\nonumber\\
&=&
i \sqrt{2\pi\alpha'} k_n
\langle {\rm e}^{-\varphi (x_1) }  ~{\rm e}^{-\varphi (x_3)}  \rangle
~
\langle {\bar{\Delta}} (x_1)  ~{\rm e}^{i\sqrt{2\pi\alpha'}   k_\hll  X^\hll (x_2) }~{{\Delta}} (x_3) \rangle
\nonumber \\
&\times&\langle S^{\dot\alpha} (x_1) ~\psi^n (x_2 )  \psi^m (x_2)
~S^{\dot\beta} (x_3) \rangle \langle {\rm e}^{i\sqrt{2\pi\alpha'}
p_\hnu  X^\hnu (x_1) } {\rm e}^{i \sqrt{2\pi\alpha'}  k_\hlambda
X^\hlambda (x_2)} {\rm e}^{i \sqrt{2\pi\alpha'} q_\hnu  X^\hnu (x_3)
}\rangle
\nonumber\\
&=&
i \sqrt{2\pi\alpha'} k_n~
x_{13}^{-1}
\times x_{12}^{-\pi\alpha' k_m k^m} x_{13}^{-1/2+\pi\alpha' k_m k^m}
x_{23}^{-\pi\alpha' k_m k^m} e^{i k_m y^m_0}
\times  x_{13}^{1/2} x_{12}^{-1} x_{23}^{-1}
\nonumber\\
&\times&[-\frac{1}{2}(\bar\sigma^{n m} ){}^{\dot \alpha  \dot \beta}]\,
x_{12}^{2\pi\alpha' p_\hmu k^\hmu} x_{13}^{2\pi\alpha' p_\hmu q^\hmu}
x_{23}^{2\pi\alpha' k_\hmu q^\hmu}
\nonumber\\
&=&-
\frac{ i}{2}\sqrt{2\pi\alpha'}\,k_n\,(\bar\sigma^{ n m} ){}^{\dot \alpha  \dot \beta}
x_{12}^{-1} x_{13}^{-1} x_{23}^{-1}
\label{tre}
\end{eqnarray}
where we have again used the momentum conservation and the mass shell conditions.

The final result is the sum of the three contributions in Eqs. (\ref{uno}), (\ref{due})
and (\ref{tre}) that is equal to:
\begin{eqnarray}
{\cal A}^{\bar{w} A_{M}^{(9)} w}&=&2\,i\,g_{9}  {\bar {w}}^{u}_{\dot\alpha a }(p)
{w}^{b}_{{\dot{\beta}} u}(q)
e^{i k_m y^m_0}\,(x_{12}\,x_{13}\,x_{23}) \nonumber \\
& \times&\left\{  \epsilon^{ \dot{\alpha} \dot{\beta}}
\left[ A^{a}_{ {\hat{\mu}} b} (k) \left( \frac{p^{\hat{\mu} }}{x_{12} x_{13} x_{23}} +
\frac{k^{\hat{\mu}}}{x_{13}^{2} x_{23}} \right) + A^{a}_{ {{m}} b} (k) k^{{m}}
\left(  \frac{ x_{12} - x_{23}}{
2 x_{12} x_{13}^{2} x_{23}} \right) \right]  \right. \nonumber \\
&-& \left. A^{a}_{ {{m}} b} (k)
k_{{n}} \frac{ (\bar\sigma^{n m} ){}^{\dot \alpha  \dot \beta}}{
2x_{12} x_{13} x_{23}}   \right\}
\label{1+2+3}
\end{eqnarray}
It can be checked that the previous expression is gauge invariant. In fact, if we make
the substitutions:
\begin{eqnarray}
A_{\hat{\mu}} \rightarrow k_{\hat{\mu}}~~;~~
A_{{m}} \rightarrow k_{{m}}
\label{Ak}
\end{eqnarray}
it is easy to check that the last term in Eq. (\ref{1+2+3}) vanishes, while the term
between square brackets becomes:
\begin{eqnarray}
k_{  {\hat{\mu}} }
\left( \frac{p^{\hat{\mu} }}{x_{12} x_{13} x_{23}} +
\frac{k^{\hat{\mu}}}{x_{13}^{2} x_{23}} \right) +
k_{ {{m}} }  k^{{m}}
\left(  \frac{ x_{12} - x_{23}}{ 2 x_{12} x_{13}^{2} x_{23}} \right)
\label{sb}
\end{eqnarray}
Using the mass shell condition $k_{\hat{\mu}} k^{\hat{\mu}} + k_{{m}} k^{{m}}
=0$ and  $k_{\hat{\mu}} p^{\hat{\mu}} = - \frac{1}{2} k_{\hat{\mu}} k^{\hat{\mu}}$
we can rewrite the previous equation  as follows:
\begin{eqnarray}
&&k_{\hat{\mu}} k^{\hat{\mu}} \left[  - \frac{1}{2 x_{12} x_{13} x_{23}}  +
\frac{1}{x_{13}^{2} x_{23}} -   \frac{ x_{12} - x_{23}}{ 2 x_{12} x_{13}^{2} x_{23}}
\right]  \nonumber \\
&&= k_{\hat{\mu}} k^{\hat{\mu}} \left[ - \frac{1}{2 x_{12} x_{13} x_{23}}
+ \frac{1}{2 x_{13}^{2} x_{23}} +  \frac{1}{2 x_{13}^{2} x_{12}} \right]
\nonumber \\
&&=  k_{\hat{\mu}} k^{\hat{\mu}} \frac{ - x_{13}+ x_{12}+ x_{23 }
}{2 x_{13}^{2} x_{23} x_{12}}=0
\label{sb6}
\end{eqnarray}
that proves the gauge invariance of the correlator.

Using the condition:
\begin{eqnarray*}
k^M A_{M} = k^{\hat{\mu}} A_{\hat{\mu}} + k^{{m}} A_{{m}} =0
\label{k2=0}
\end{eqnarray*}
we can rewrite Eq. (\ref{1+2+3}) as follows:
\begin{eqnarray*}
{\cal A}^{\bar{w} A^{(9)}_{M} w}= i \,{\bar {w}}^{u}_{\dot\alpha a }(p)
{w}^{b}_{{\dot{\beta}} u}(q)
e^{i k_m y^m_0} \left[  \epsilon^{ \dot{\alpha} \dot{\beta}}
(A^{(9)}_{\hat{\mu}})^{a}_{~  b} (k)  ( p^{\hat{\mu} }  -q^{\hat{\mu} } )
-  (A^{(9)}_{{m}})^{a}_{~  b} (k)
k_{{n}} (\bar\sigma^{n m} ){}^{\dot \alpha  \dot \beta}
 \right]
\label{fin3}
\end{eqnarray*}
that has the correct dependence on the Koba-Nielsen variables. Finally, by performing the Fourier transformation as shown after
Eq. (\ref{cor67}), we get  an amplitude which is  evaluated at  the position of the $D5$-brane in the last two tori :
\begin{eqnarray}
{\cal A}^{\bar{w} A^{(9)}_{M} w}\!\!=\! i \,{\bar {w}}^{u}_{\dot\alpha a }(p)
{w}^{b}_{{\dot{\beta}} u}\!(q)\!\!
\left[\!  \epsilon^{ \dot{\alpha} \dot{\beta}}
(A^{(9)}_{\hat{\mu}})^{a}_{~  b} (k_{\hat \mu},y_0)  ( p^{\hat{\mu} } \!\! -\!q^{\hat{\mu} } )\!
-\!  (A^{(9)}_{{m}})^{a}_{~  b} (k_{\hat \mu},y_0)
k_{{n}} (\bar\sigma^{ n m} ){}^{\dot \alpha  \dot \beta}
\! \right]
\label{fin3a}
\end{eqnarray}

The previous calculation can be easily extended to the interaction of two twisted scalars with  the gauge field living in the world-volume of the $D5$-brane. Since the vertex of the  gauge field
$A_{\hat{\mu}}^{(5)}$ depends only on the six dimensional momentum, one has that the corresponding amplitude contains only one term which  is equal and opposite in sign, due to the different ordering of the twisted fields, to the one in Eq. (\ref{uno}) taken
at $k_{ m}=0$. Using the
 trasversality condition $k^{\hat\mu}A_{\hat \mu}^{(5)}=0$ we
 arrive at  the following expression:
\begin{eqnarray*}
{\cal A}^{\bar{w} A_{M}^{(5)} w}= -i \,{\bar {w}}^{u}_{\dot\alpha a }(p)
{w}^{b}_{{\dot{\beta}} u}(q)(A^{(5)}_ {\hat{\mu}})^{a}_{~b} (k)  ( p^{\hat{\mu} }  -q^{\hat{\mu} } )\epsilon^{ \dot{\alpha} \dot{\beta}}
\end{eqnarray*}
which is in agreement with the one obtained from the kinetic term of the twisted scalar  in the field theoretical approach.

\end{document}